\documentclass[10pt,twocolumn,a4paper]{aastex631}

\newcommand{\kms}{$\mathrm{km \, s}^{-1}$}
\shorttitle{Molecular gas kinematics in jellyfish galaxies}
\shortauthors{Bacchini et al.}

\begin{document}

\title{3D modeling of the molecular gas kinematics in optically-selected jellyfish galaxies}

\correspondingauthor{Cecilia Bacchini}
\email{cecilia.bacchini@inaf.it}

\author[0000-0002-8372-3428]{Cecilia Bacchini}
\affiliation{INAF - Osservatorio Astronomico di Padova, vicolo dell'Osservatorio 5, IT-35122 Padova, Italy}

\author[0000-0003-2589-762X]{Matilde Mingozzi}
\affiliation{Space Telescope Science Institute, 3700 San Martin Drive, Baltimore, MD 21218, USA}

\author[0000-0001-8751-8360]{Bianca M. Poggianti}
\affiliation{INAF - Osservatorio Astronomico di Padova, vicolo dell'Osservatorio 5, IT-35122 Padova, Italy}

\author[0000-0002-1688-482X]{Alessia Moretti}
\affiliation{INAF - Osservatorio Astronomico di Padova, vicolo dell'Osservatorio 5, IT-35122 Padova, Italy}

\author[0000-0002-7296-9780]{Marco Gullieuszik}
\affiliation{INAF - Osservatorio Astronomico di Padova, vicolo dell'Osservatorio 5, IT-35122 Padova, Italy}

\author[0000-0002-5655-6054]{Antonino Marasco}
\affiliation{INAF - Osservatorio Astronomico di Padova, vicolo dell'Osservatorio 5, IT-35122 Padova, Italy}

\author[0000-0002-2897-9121]{Bernardo Cervantes Sodi}
\affiliation{Istituto de Radioastronom\'{i}a y Astrof\'{i}sica, Universidad Nacional Aut\'{o}noma de M\'{e}xico, Campus Morelia, A.P. 3-72, C.P. 58089, Michoac\'{a}n, M\'{e}xico}

\author[0000-0002-2808-1223]{Osbaldo S\'{a}nchez-Garc\'{i}a}
\affiliation{Istituto de Radioastronom\'{i}a y Astrof\'{i}sica, Universidad Nacional Aut\'{o}noma de M\'{e}xico, Campus Morelia, A.P. 3-72, C.P. 58089, Michoac\'{a}n, M\'{e}xico}

\author[0000-0003-0980-1499]{Benedetta Vulcani}
\affiliation{INAF - Osservatorio Astronomico di Padova, vicolo dell'Osservatorio 5, IT-35122 Padova, Italy}

\author[0000-0002-4382-8081]{Ariel Werle}
\affiliation{INAF - Osservatorio Astronomico di Padova, vicolo dell'Osservatorio 5, IT-35122 Padova, Italy}

\author[0000-0001-9143-6026]{Rosita Paladino}
\affiliation{INAF - Istituto di Radioastronomia, via P. Gobetti 101, I-40129 Bologna, Italy}

\author[0000-0002-3585-866X]{Mario Radovich}
\affiliation{INAF - Osservatorio Astronomico di Padova, vicolo dell'Osservatorio 5, IT-35122 Padova, Italy}



\begin{abstract}
Cluster galaxies are subject to the ram pressure exerted by the intracluster medium, which can perturb or even strip away their gas while leaving the stars undisturbed. 
We model the distribution and kinematics of the stars and the molecular gas in four late-type cluster galaxies (JO201, JO204, JO206, and JW100), which show tails of atomic and ionized gas indicative of ongoing ram pressure stripping. 
We analyze MUSE@VLT data and CO data from ALMA searching for signatures of radial gas flows, ram pressure stripping, and other perturbations. 
We find that all galaxies, with the possible exception of JW100, host stellar bars. 
Signatures of ram pressure are found in JO201 and JO206, which also shows clear indications of ongoing stripping in the molecular disk outskirts. 
The stripping affects the whole molecular gas disk of JW100. 
The molecular gas kinematics in JO204 is instead dominated by rotation rather than ram pressure.  
We also find indications of enhanced turbulence of the molecular gas compared to field galaxies. 
Large-scale radial flows of molecular gas are present in JO204 and JW100, but more uncertain in JO201 and JO206.
We show that our sample follows the molecular gas mass-size relation, confirming that it is essentially independent of environment even for the most extreme cases of stripping. 
Our findings are consistent with the molecular gas being affected by the ram pressure on different timescales and less severely than the atomic and ionized gas phases, likely because the molecular gas is denser and more gravitationally bound to the galaxy.
\end{abstract}


%

\section{Introduction} \label{sec:intro}
In dense environments, such as groups and clusters, galaxies are affected by various physical mechanisms that can significantly influence their properties and evolution \citep{1982Nulsen,2006BoselliGavazzi,2021Cortese}. 
These processes are usually divided into gravitational and hydrodynamical interactions \citep{2022Boselli}. 
Gravitational perturbations can be induced by tidal forces due to the potential of other cluster/group members \citep{1983Merritt} or the large scale structure itself \citep{1990Byrd}, but also by fly-by encounters and mergers \citep{1992BarnesHernquist,2006Kronberger}. 
Gravitational interactions affect both the stellar and the gaseous components of galaxies. 
Instead, the hydrodynamical interactions between the galactic interstellar medium (ISM) and the intracluster medium (ICM) are expected to influence only the gaseous components of galaxies. 
These mechanisms are the thermal evaporation of the cold and warm ($T \lesssim 10^4$~K) ISM due to the interaction of the hot ($T \approx 10^7-10^8$~K) ICM \citep{1977Cowie}, the removal of the outer ISM layer due to the viscosity momentum transfer with the ICM (viscous stripping) or instabilities \citep[][]{1982Nulsen,2008Roediger}, and the ram pressure stripping, that is the removal of the ISM due to the pressure exerted by the ICM while a galaxy is moving through the cluster \citep{1972GunnGott}. 
In addition, the interaction with the ICM can heat up or strip away the hot ($T\approx 10^6$~K) gas corona surrounding galaxies and prevent them from accreting new gas, finally quenching star formation \citep[starvation;][]{1980Larson}. 

Ram pressure is often considered among the dominant mechanisms affecting the ISM in cluster galaxies. 
The ram pressure can be calculated as $P_\mathrm{ram} = \rho_\mathrm{ICM} V_\mathrm{gal}^2$, where $\rho_\mathrm{ICM}$ and $V_\mathrm{gal}$ are the ICM density and the galaxy velocity relative to the cluster \citep{1972GunnGott}. 
Hence, this mechanism is expected to be particularly strong for galaxies with high $V_\mathrm{gal}$ located close to the cluster center, where $\rho_\mathrm{ICM}$ is the highest. 
The ram pressure can have different effects on the gas distribution and kinematics in galaxies. 
The compression of the gas disk can make it morphologically lopsided and asymmetric \citep[e.g.][]{2008Mapelli,2008Kronberger_b}. 
Depending on its direction with respect to the galaxy rotation, the ram pressure can decelerate one side of the gas disk and accelerate the other, resulting in a kinematically lopsided disk, or also shift the kinematic center of the gas disk with respect to the optical center of the galaxy \citep[e.g.][]{2008Kronberger_b}. 
Typical signatures of ram pressure stripping are one-sided tails of gas extending outside the stellar disk and gas clouds that are spatially detached and kinematically decoupled from the galaxy \citep[e.g.][]{2007Chung,2013Merluzzi,2017Lee}. 
Moreover, gas disks in cluster galaxies are sometimes less extended and less massive that those in field galaxies \citep{1980Chamaraux,1984Haynes,1990Cayatte,2001Schroder,2009Chung}. 
Both truncation and gas deficiency are properties ascribed to ram pressure stripping, being relatively common in both low- \citep[e.g.][]{2007Chung,2018Gavazzi} and intermediate-redshift cluster galaxies \citep[e.g.][]{2007Cortese,2019Boselli,2022Moretti}. 
The efficiency of ram pressure stripping depends on the gas properties, being more effective on a diffuse medium than on dense gas clumps, and on the gravitational pull caused by the galactic potential, that weakens with increasing galactocentric distance and height above the midplane \citep[e.g.][]{1972GunnGott,2009Tonnesen,2018Koppen}. 

The atomic gas in galaxies is relatively diffuse and typically distributed in a disk that is very extended \citep[up to twice the stellar disk diameter; see  e.g.][]{2001VerheijenSancisi,2016Wang,2016Lelli_c} and thick \citep[up to $\approx$1~kpc; see e.g.][]{1996Olling,2011Yim,2014Yim,2017Marasco,2019Bacchini_a,2019Bacchini_b,2020Bacchini_b}, being very susceptible to ram pressure. 
Indeed, cluster galaxies often contain less atomic gas than expected from their optical size or stellar mass and have truncated and/or asymmetric HI discs \citep{1980Chamaraux,1984Haynes,1985Giovanelli,1990Cayatte,2001Solanes,2001Schroder,2002Waugh,2009Chung,2021Loni,2022Zabel}. 
Morever, long tails of atomic gas are commonly observed in cluster galaxies \citep{2000BravoAlfaro,2004Kenney,2007Chung,2010Scott,2017Sorgho,2019Ramatsoku,2020Ramatsoku,2020Deb,2021HealyDeb,2022Deb,2022Hess}. 

The molecular gas is typically denser and clumpier than the atomic gas \citep[e.g.][]{2008Leroy} and its distribution is also less extended in both the radial \citep[up to the stellar disk diameter;][]{2013Davis,2021Brown,2022Zabel} and vertical \citep[up to $\approx 0.5$~kpc;][]{2011Yim,2014Yim,2017Marasco,2019Bacchini_a,2019Bacchini_b} directions. 
Hence, it is expected that the molecular gas is more resilient to ram pressure than the atomic gas \citep[e.g.][]{2017Lee,2021Brown,2022Zabel,2022Boselli}. 
Nevertheless, there is growing observational evidence that the ram pressure actually influences the molecular gas in cluster galaxies, as indicated by signatures of compression, kinematic lopsidedness, and shifts between the optical and kinematic center \citep{2017Lee,2019Zabel,2020Cramer}. 
Direct observations of molecular gas stripping by ram pressure are limited, but tails and blobs of molecular gas far from the stellar disk have been observed in some cluster galaxies \citep{2008Vollmer,2014Jachym,2017Jachym,2017Lee,2018Moretti,2020Moretti}, as well as truncated molecular gas disks and H$_2$-deficient galaxies \citep{2009Fumagalli,2014Boselli_c,2019Zabel,2022Zabel,2022Lee}. 
However, a few authors have found that cluster galaxies can also host a normal (both in size and mass) or even enhanced reservoir of molecular gas \citep{2009Fumagalli,2020MorettiLetter,2021Brown,2022Zabel}, possible indication that ram pressure can increase the efficiency of the HI-to-H$_2$ conversion \citep{2020Moretti}. 

In this work, we analyze the molecular gas distribution and kinematics in four cluster galaxies observed with the Atacama Large Millimeter Array (ALMA). 
These objects are part of the sample of 114 galaxies observed within the Large Program "GAs Stripping Phenomena in galaxies with MUSE" (GASP), which is a survey carried out with integral-field Multi Unit Spectroscopic Explorer (MUSE) at the Very Large Telescope (VLT). 
The GASP survey aims at understanding the impact of environment on the evolution of galaxies by studying their stellar and ionized gas emission. 
Recently, follow-up programs have provided multi-wavelength observations for a few galaxies in the GASP sample, allowing to study other ISM components, such as atomic gas \citep{2019Ramatsoku,2020Ramatsoku,2020Deb,2021HealyDeb,2022Deb,2022Luber}, the molecular gas \citep{2018Moretti,2020Moretti,2020MorettiLetter}, and magnetic fields \citep{2021Muller}, and also young stellar populations \citep{2018George}. 
An unexpected result of the GASP project was the high fraction of active galactic nuclei (AGN) among ram pressure-stripped galaxies \citep{2017PoggiantiNat,2022Peluso,2022Poggianti}. 
This result was interpreted as an indication that ram pressure can drive gas flows towards the center and foster the AGN activity \citep[e.g.][]{2020Ricarte}. 
The galaxies analyzed in this work (hereafter referred as the GASP-ALMA sample) were studied by \cite{2017PoggiantiNat} and host indeed an AGN. 
This paper aims at answering the following open questions about the GASP-ALMA galaxies: 
What is the impact of ram pressure on the distribution and kinematics of the molecular gas?
Can we detect inflows of molecular gas that may feed the AGN? 

This paper is organized as follows. 
Section~\ref{sec:sample} presents the GASP-ALMA sample and summarizes the relevant pieces of information obtained by previous studies. 
We describe the data and methods used to carry out the analysis in Sects.~\ref{sec:data} and~\ref{sec:method}, respectively. 
For each galaxy, we present and discuss the results in Sect.~\ref{sec:results}. 
In Sect.~\ref{sec:discussion}, we compare our findings with other works in the literature. 
Section~\ref{sec:conclusions} summarizes this work and its conclusions. 

We adopt standard cosmological parameters ($h = 0.7$, $\Omega_M = 0.3$, and $\Omega_\lambda = 0.7$) and a \cite{2003Chabrier} initial mass function.

\section{The galaxy sample}\label{sec:sample}
The GASP-ALMA sample consists of four late-type galaxies, i.e. JO201, JO204, JO206, and JW100, located in different clusters at redshift $0.04 \lesssim z \lesssim 0.06$ and with relatively high stellar mass (see Table~\ref{tab:galaxyprops} and references therein). 
These galaxies are classified as  “jellyfish” because they show one-sided tails of ionized gas longer than the stellar disk diameter \citep[][]{2016Poggianti}. 
Thanks to the wealth of information provided by the GASP project and the availability of multi-wavelength observations, these galaxies have been extensively studied in the literature \citep[for a brief review, see][]{2022Poggianti}. 
Thus, we summarize some of the previous works that are relevant for our analysis. 

The galaxies in our sample are moving through the ICM with either super-sonic or transonic line-of-sight velocities and are located close to the cluster center \citep{2020Gullieuszik}. 
These properties indicate that the galaxies are in favorable conditions for strong ram pressure and move on very radial orbits, suggesting that they have recently entered into the cluster for the first time \citep{2017Yoon,2018Jaffe}. 
While JO204 and JO206 are relatively isolated for being cluster members \citep{2017Gullieuszik,2017Biviano}, JO201 and JW100 belong to a substructure of four and three galaxies, respectively \citep{2017Bellhouse,2019Poggianti}. 
Previous works show that, in all the GASP-ALMA galaxies, the stellar kinematics appears to be quite regular, while the ionized gas kinematics is very perturbed, as expected for galaxies undergoing ram pressure stripping \citep{2017Bellhouse,2017Gullieuszik,2017Poggianti,2018Jaffe,2019Poggianti}. 
Recent works showed that these galaxies host strongly asymmetric HI disks with long tails of atomic gas, and also have significantly reduced the HI content with respect to field galaxies \citep[$\gtrsim 50$~\%, ][]{2019Ramatsoku,2020Ramatsoku,2020Deb,2021HealyDeb,2022Deb}. 
This HI deficiency is however not coupled with a deficiency in the molecular gas reservoir, as these galaxies have H$_2$ masses that are 4-5 times higher than expected for galaxies with similar stellar mass \citep{2020Moretti,2020MorettiLetter}. 

\begin{deluxetable*}{lcccc}
	\tablecaption{Properties of the galaxy sample.}\label{tab:galaxyprops}
	\tablehead{
		\colhead{Property}										& \multicolumn4c{Galaxy}																\\
		\colhead{} 												& \colhead{JO201}	&\colhead{JO204}			& \colhead{JO206}			& \colhead{JW100}}
	\startdata
	Alternative names								& KAZ~364			& 2MASX~J10134686-0054514	& 2MASX J21134738+0228347	& IC~5337		\\	
	Morphological type								& Sab				& Sab						& Sb						& Sa			\\
	Cluster											& Abell~85			& Abell~957					& IIZW108					& Abell~2626	\\
	$z_\mathrm{clu}$								& 0.05568			& 0.04496					& 0.04889					& 0.05509		\\
	$z_\mathrm{gal}$								& 0.044631			& 0.042372					& 0.051089					& 0.061891		\\
	$V_\mathrm{gal}$ [\kms]							& -3138				& -743						& 629						& 1932			\\		$|V_\mathrm{gal}/\sigma_\mathrm{clu}|$			& 3.7				& 1.2						& 0.9						& 3.0			\\
	$R_\mathrm{clu}/R_\mathrm{200,cl}$				& 0.18				& 0.09						& 0.29						& 0.06			\\
	Distance [Mpc]									& 189.1				& 179.8						& 216.3						& 261.4			\\
	Physical scale [kpc/arcsec]						& 0.88				& 0.84						& 1.00						& 1.19			\\
	Center R.A. [J2000]								& 00:41:30.30		& 10:13:46.84				& 21:13:47.41				& 23:36:25.05	\\
	Center DEC [J2000]								& -09:15:45.9		& -00:54:51.27				& +02:28:35.50				& +21:09:02.64	\\
	$M_\star$ [$10^{10} \, M_\odot$]				& $6.2 \pm 0.8$		& $4.1 \pm 0.6$				& $9.1 \pm 0.9$				& $29 \pm 7$	\\
	$M_\mathrm{HI}$ [$10^9 \, M_\odot$]				& $1.7\pm0.5$		& $\gtrsim1.3 \pm 0.1$		& $3.2 \pm 0.9$				& $2.8 \pm 0.8$	\\
	M$_\mathrm{H_2}$ [$10^9 \, M_\odot$]			& $11.5\pm5.8$		& $5.7 \pm 2.9$				& $5.6 \pm 2.8$				& $16.5 \pm 8.3$\\
	\enddata
	\tablecomments{{\footnotesize 
			Galaxy names are in GASP convention; alternative names are also provided. 
			Morphological types are from \cite{2012Fasano}. 
			Cluster redshifts ($z_\mathrm{clu}$) are from \cite{2017Biviano}. 
			Galaxy redshifts ($z_\mathrm{gal}$) and distances from the cluster center ($R_\mathrm{clu}/R_\mathrm{200,cl}$) are from \cite{2017Bellhouse,2017Gullieuszik,2017Poggianti,2019Poggianti}. 
			Galaxy velocities relative to cluster are calculated as $V_\mathrm{gal}=c(z_\mathrm{gal}-z_\mathrm{clu})/(1+z_\mathrm{clu})$. 
			The optical center coordinates are from \cite{2017PoggiantiNat}. 
			Stellar masses ($M_\star$) and effective radii ($R_\mathrm{eff}$) are from \cite{2018Vulcani_b} and \cite{2020Franchetto}, respectively. 
			Atomic gas masses ($M_\mathrm{HI}$) are from \cite{2019Ramatsoku,2020Ramatsoku} and \cite{2020Deb,2022Deb}; a conservative uncertainty of 30\% is assumed. 
			Molecular gas masses (M$_\mathrm{H_2}$) are from \cite{2020Moretti}; a conservative uncertainty of 50\% is assumed. 
	}}
\end{deluxetable*}

As mentioned above, the galaxies in the GASP-ALMA sample host an AGN \citep[][]{2017PoggiantiNat,2019Radovich,2022Peluso}. 
It has been shown that the AGN is the main source of gas ionization in the central regions of these galaxies and, except for JO206, it also causes a low-velocity ($\approx 250 - 320$~\kms) wind of ionized gas \citep{2019Radovich}. 
\cite{2017PoggiantiNat} proposed that the AGN activity in these galaxies is triggered by the ram pressure, which can decrease the angular momentum of the gas and favor its inflow toward the center. 
In JO201, \cite{2019George} observed a cavity of about 8.6~kpc with reduced ultraviolet and CO flux around the AGN \citep[see also][]{2019Radovich}. 
By combining optical (MUSE) and sub-mm (ALMA) spectroscopic observations, these authors proposed that the cavity is due to AGN feedback that is either ionizing or sweeping away the gas, possibly reducing the star formation activity in the central regions. 
In JO204, \cite{2020Deb} found a redshifted absorption feature in the HI global profile, which could be ascribed to either a clumpy and fast rotating HI disc seen in front of the central radio continuum source or an inflow of atomic gas towards the central AGN.

\section{Data}\label{sec:data}
This section describes the multi-wavelength observations and data products that were used in this work, which primarily focuses on the molecular gas. 
Since the ram pressure can influence the kinematics and geometry of the molecular gas disk, we  analyze the stellar kinematics and use it as a reference. 
Sections~\ref{sec:data_alma} and~\ref{sec:data_muse} describe the data used to study the molecular gas and stellar component, respectively. 

\subsection{ALMA data}\label{sec:data_alma}
\begin{figure*} 
	\centering
	{\includegraphics[width=2.\columnwidth]{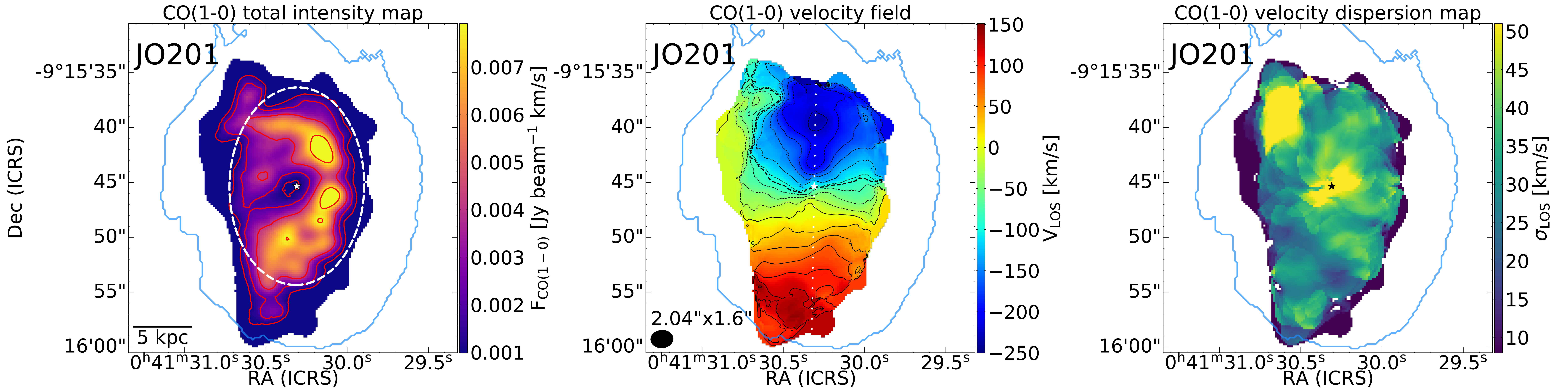}}
	{\includegraphics[width=2.\columnwidth]{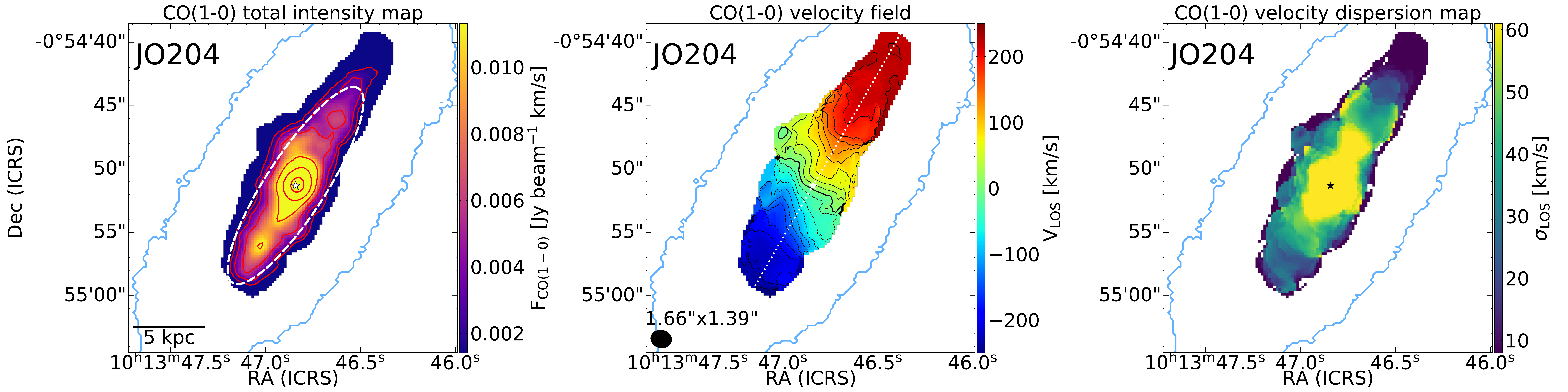}}
	{\includegraphics[width=2.\columnwidth]{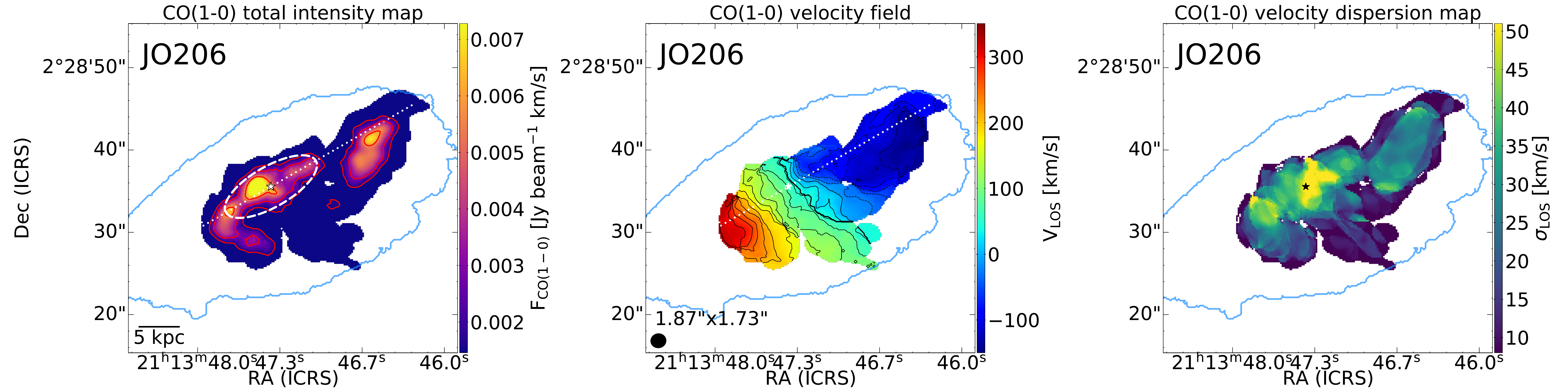}}
	{\includegraphics[width=2.\columnwidth]{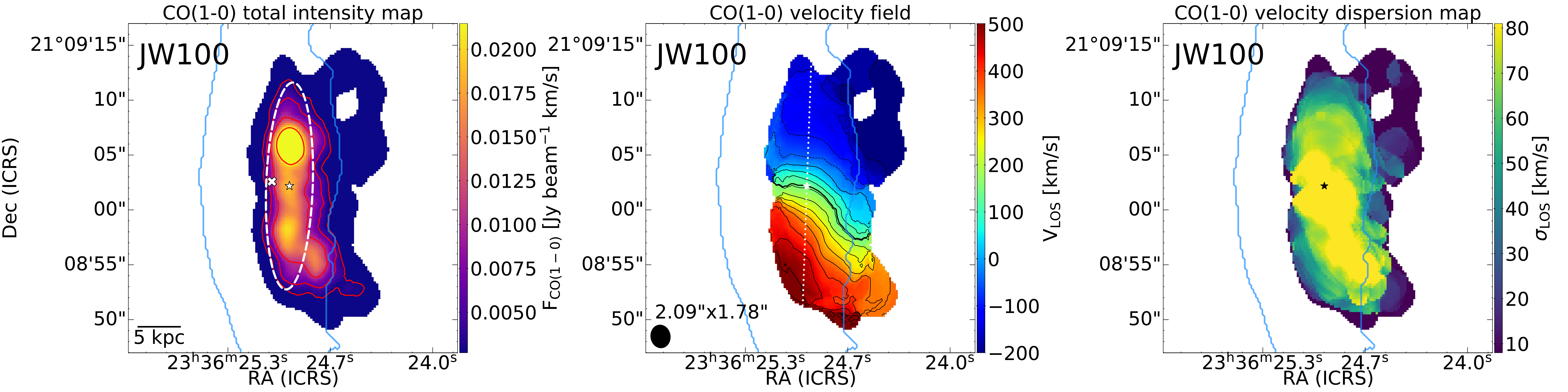}}
	\caption{Total intensity map (left column), velocity field (central column) and map of the observed velocity dispersion along the line of sight (right column) obtained from the CO(1-0) emission line datacubes for the four galaxies in the sample. 
	The stars and the white dashed ellipse indicate the kinematic center and the region used for modeling the gas kinematics, respectively (see Sect.~\ref{sec:method_molgas_kin}). 
	In JW100, the grey cross shows the optical center (see Sect.~\ref{sec:results_jw100}). 
	In the total intensity map, the red contours are at $2^n \sigma_\mathrm{tot}$, where $ \sigma_\mathrm{tot}$ is the noise in the total map \citep{2014Lelli_b,2017Iorio}, $n=1...10$, and $\sigma_\mathrm{tot}=0.9,1.4,1.4,2.7$~mJy/beam \kms for JO201, JO204, JO206, and JW100, respectively. 
	In the velocity field, the black curves are the iso-velocity contours, with the thick one being at the galaxy systemic velocity (see Sect.~\ref{sec:method_molgas_kin} for details). 
	The white dotted line in the velocity field is the kinematic major axis. 
	The bar and the ellipse in the bottom left corner respectively show the physical scale and beam of the observations. 
	The light-blue contour shows the most external isophote ($\approx 1.5\sigma$ above the background) encompassing the H$\alpha$ emission traced by MUSE, and it indicates the stellar disk defined by \cite{2020Gullieuszik}. 
	East is to the left and North to the top.}
	\label{fig:maps_co10}
\end{figure*}

We used CO(1--0) and CO(2--1) emission line observations obtained with ALMA during Cycle 5 (project 2017.1.00496.S; PI: Poggianti). 
These observations were already used by \cite{2020MorettiLetter} to study the molecular gas content of the GASP-ALMA galaxies. 
The datacubes used in this work are different from those used by \cite{2020MorettiLetter}, as the imaging procedure was re-performed to increase the spectral resolution (Mingozzi et al. to be submitted). 
Here, we use ALMA datacubes with angular resolution of $\approx 1-2$\arcsec (see Fig.~\ref{fig:maps_co10}) and velocity resolution of 10~\kms. 
Figure~\ref{fig:maps_co10} shows, for each galaxy, the moment maps of the CO(1--0) datacubes. 
These maps were obtained from the ALMA datacubes by applying a mask made by all pixels with signal-to-noise ratio (S/N) above 3 in a datacube smoothed by a factor 2 (i.e. in which each channel map is convolved with a beam 2 times larger than the original one). 

We note that \cite{2020Moretti,2020MorettiLetter} detected faint CO emission coming from the regions outside the stellar disk and coinciding with the ionized gas tails. 
The emission is not visible in the maps used in this work (Fig.~\ref{fig:maps_co10}), despite we used the same ALMA observations.  
This difference is due to the fact that our datacubes have better velocity resolution ($\Delta \upsilon = 10$~\kms) but lower S/N than those used by \cite{2020Moretti,2020MorettiLetter}, which have $\Delta \upsilon = 20$~\kms and to the different masking procedure adopted in the two works. 
For this work, we used the datacubes with $\Delta \upsilon = 10$~\kms, being the best compromise to have good velocity resolution and S/N in the regions within (or close to) the stellar disk. 
We note that we do not find significant differences in mass and size of the molecular gas disk with respect to the results obtained by \cite{2020Moretti,2020MorettiLetter}. 

\subsection{MUSE data}\label{sec:data_muse}
To analyze the stellar component, we used the MUSE observations obtained by the GASP survey \citep{2017Poggianti}. 
The data reduction and processing is detailed in \cite{2017Poggianti}. 
The wavelength coverage and spectral resolution of the final datacubes are $4800 \text{ \AA} < \lambda < 9300 \AA$ and $1770 < \mathrm{R}<3590$, respectively. 
The pixel size is 0.2~\arcsec$\times$0.2~\arcsec with a natural seeing of $\approx 1$~\arcsec. 
In this work, we use the $I$-band images and the stellar velocity fields extracted from the MUSE datacubes.  

\begin{figure*}
	\includegraphics[width=2.1\columnwidth]{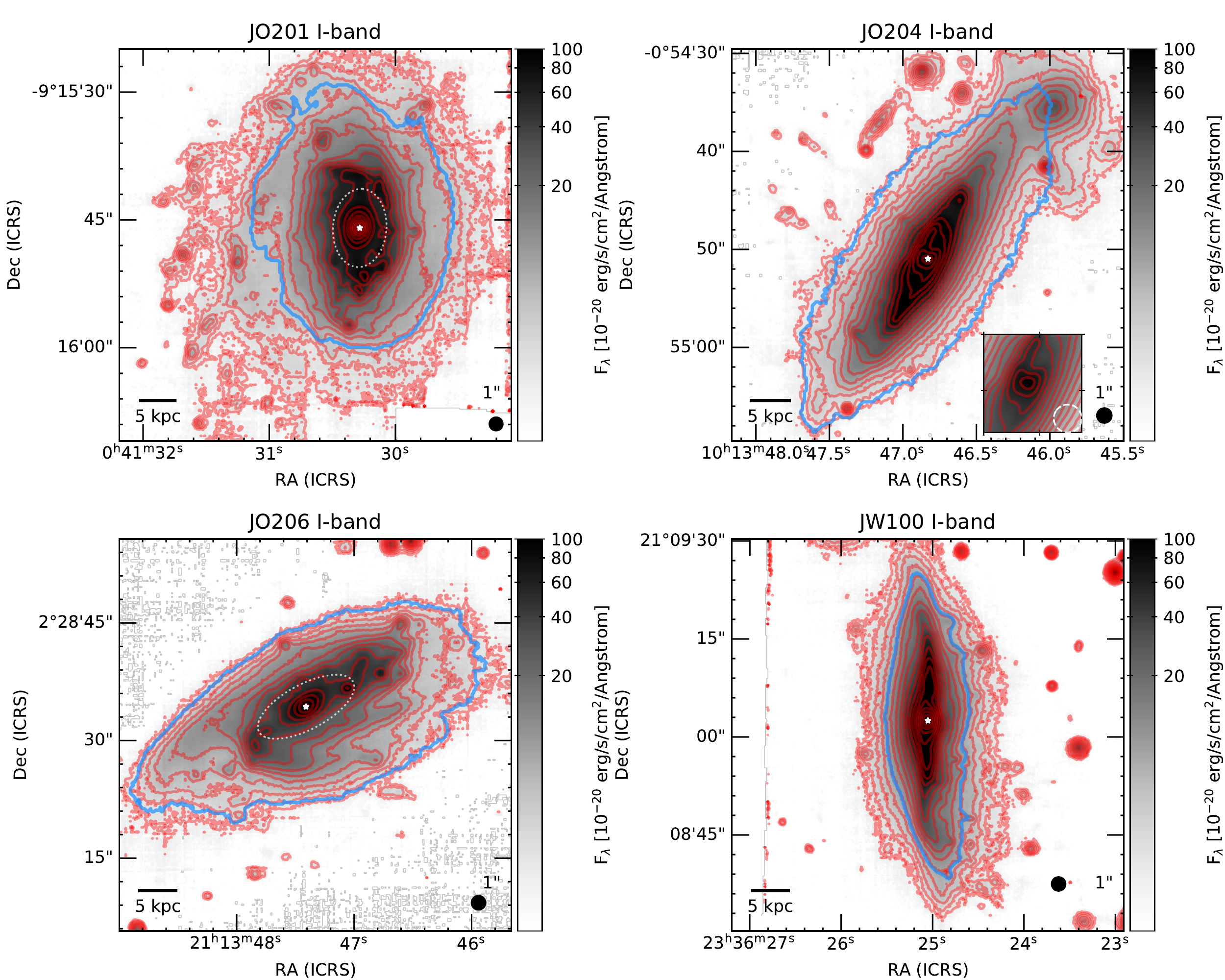}
	\caption{$I$-band images, extracted from the MUSE observations. 
		The red contours are at $2^n$ with $n$ going from 1 to 20 with steps of 0.5 (same units as colorbars). 
		The white stars show the galaxy center. 
		For JO201 and JO206, the white dotted ellipses indicate the regions influenced by the bar (see text).
		The light-blue contour indicates the stellar disk (same as Fig.~\ref{fig:maps_co10}). 
		The black dot in the bottom right corner shows the angular resolution of the MUSE observations. 
		The inset in the JO204 panel shows a zoom-in of the central regions of the galaxy, with the white circle showing the angular resolution of the observations. 
		East is to the left and north to the top. }
	\label{fig:iband_images}
\end{figure*}

The $I$-band images are very useful to identify stellar substructures, such as bulges and bars. 
These are typically dominated by intermediate-age ($> 1$~Gyr) stellar populations and hence generally brighter in $I$-band than at short wavelengths \citep[e.g.][]{2000Knapen}. 
Figure~\ref{fig:iband_images} shows, for each galaxy, the $I$-band images. 
These were obtained by \cite{2020Franchetto} from the integrated MUSE datacubes using the Cousins $I$-band filter response curve. 
\cite{2020Franchetto} also derived the center coordinates, the position angle, the inclination, and the $I$-band surface brightness profiles (see Appendix~\ref{ap:surface_brightness} for details) of these galaxies by fitting the galaxy isophotes with a series of concentric ellipses using the \textsc{iraf} task \texttt{ELLIPSE} \citep{1987Jedrzejewski}. 

We also use the azimuthally-averaged radial profiles of the stellar surface density obtained from the MUSE datacubes using the spectral synthesis code \textsc{Sinopsis} \citep{2011Fritz,2017Fritz}. 
This software performs a decomposition of galaxy spectra into a combination of stellar population models, fitting both the continuum and the main emission and absorption lines \citep[for a thorough description of the code features and outputs, see also][]{2017Poggianti,2022Werle}. 
\textsc{Sinopsis} provides the maps of several properties of the stellar populations, including their mass distribution. 
These maps were used to derive the radial profile of the stellar surface density for each galaxy \citep[see][]{2021Franchetto}. 

The stellar velocity fields are used to analyze the stellar disk kinematics, which is useful to interpret the kinematics of the molecular gas. 
Stellar velocity fields were extracted from the MUSE datacube using the Penalized Pixel-Fitting (\textsc{pPXF}) code \citep{2004CappellariEmsellem}. 
As a preliminary step, the observations were masked to remove spurious sources, such as stars and background galaxies. 
Spaxels in the MUSE data were binned through the Voronoi algorithm in order to reach S/N$>10$ in each bin. 
The observed spectra were fitted with the stellar population templates by \cite{2010Vazdekis} and using of single stellar populations. 
More details on the procedure can be found in \cite{2017Poggianti} and \cite{2018Moretti}. 
We just recall here that the fitting was performed including one kinematic component, that we attribute to the stellar disk. 
This is an oversimplification in the case of galaxies hosting a bulge and/or a bar, which are kinematically distinct component than the stellar disk \citep[see  e.g.][]{2017Tabor,2017Cappellari,2018Rizzo}. 
Since a multi-component fitting of the galaxy spectra is beyond the scope of this work, we discuss the caveats arising from this approach when necessary. 
We note though that the contamination from bulges and bars is mitigated by the fact that the fitting procedure to extract the stellar velocity field was performed on the part of the spectra at wavelengths shorter than the H$_\alpha$ emission line \citep{2017Poggianti}, which is more dominated by the young stellar population of the disk. 
In Appendix~\ref{ap:surface_brightness}, we show that the bulge luminosity is essentially negligible beyond the innermost 1-2~kpc from the center of our galaxies, even in the red part of the spectra.

\section{Method}\label{sec:method}
Our approach relies on the software $^\mathrm{3D}$\textsc{Barolo}\footnote{\url{https://editeodoro.github.io/Bbarolo/}} \citep[v.6.1][]{2015DiTeodoro,2021DiTeodoro}, which simulates galaxy observations assuming a tilted-ring model. 
This consists of a series of concentric annuli described by a set of geometric and kinematic parameters, which can all vary with the galactocentric distance $R$. 
The geometrical parameters are the coordinates of the center $x_0$ and $y_0$, the position angle $\phi$, and the disc inclination $i$. 
The kinematic parameters are the systemic velocity $V_\mathrm{sys}$, the rotation velocity $V_\mathrm{rot}$, the velocity dispersion $\sigma$, and the radial velocity in the disc plane $V_\mathrm{rad}$. 
The observed line-of-sight velocity is then \citep[e.g.][]{1987Begeman}
\begin{equation}\label{eq:vlos}
V_\mathrm{LOS,j} = V_\mathrm{sys,j} + \left(V_\mathrm{rot,j} \cos\theta_\mathrm{j} + V_\mathrm{rad,j} \sin\theta_\mathrm{j} \right) \sin i_\mathrm{j} \, ;
\end{equation} 
where $\theta$ is the azimuthal angle in the plane of the disc, $j=\star$ for the stellar disk, and $j=\mathrm{CO}$ for the molecular gas disk. 

$^\mathrm{3D}$\textsc{Barolo} (hereafter 3DB) was mainly designed to fit emission line observations working in 3D, meaning that the model is fitted to the datacube channel-by-channel. 
This approach allows us to use all the information in the datacube and to take into account both the angular resolution and the spectral resolution of the instrument. 
In a step prior to the fitting, 3DB convolves the model with the point spread function (PSF) or the beam of the instrument, while the instrumental spectral broadening is included in the model construction. 
The convolution with the PSF is required to correct for the so-called ``beam smearing" \citep[][]{1981Bosma,1987Begeman,2015DiTeodoro}. 
The finite size of the PSF smears the line emission on adjacent regions where the emitting material has different line-of-sight velocity, causing an artificial broadening of the profile. 
As a consequence, the rotation velocity and the velocity dispersion can be respectively underestimated and overestimated, if the beam smearing is not properly accounted for. 
This effect is particularly important if the angular resolution of the observations is low and where there are strong velocity gradients, as in the case of the inner regions of massive galaxies with steeply rising rotation curve. 
Moreover, the beam smearing effect is expected to become more and more relevant as the inclination angle of the galaxy increases. 
3DB normalizes the model using either the flux in each pixel of the total intensity map or the azimuthally-averaged flux in each ring. 
Finally, the model is fitted to the observations in order to find the set of free parameters that minimizes the residuals. 

With respect to 2D methods, which fit the velocity field, this 3D procedure not only corrects for the beam smearing effect, but also breaks the degeneracy between the rotation velocity and the velocity dispersion \citep[e.g.][]{1981Bosma,1987Begeman,2015DiTeodoro}. 
The 3DB task \texttt{3DFIT} is designed to model emission line datacubes working in 3D. 
The software also includes the task \texttt{2DFIT}, which can be used to model the 2D velocity fields. 
In this work, we use \texttt{3DFIT} and \texttt{2DFIT} to model the kinematics of the molecular gas disk and the stellar disk, respectively. 
For each component, we adopted an ad-hoc methodology, that is described in Sects.~\ref{sec:method_stellar_kins} and~\ref{sec:method_molgas_kin}. 

Before proceeding with the methodology presentation, a brief disclaimer is due. 
We stress that 3DB, like several other kinematic modeling software \citep[e.g.][]{1987Begeman,2015Kamphuis}, is specifically designed to model radially symmetric gas flows in discs. 
However, the galaxies studied in this work are subject to various local disturbances due to internal (bar, AGN feedback) and external (ram pressure) mechanisms, which are expected to produce deviations from this idealized kinematics. 
Our strategy here is to use 3DB to quantify the large-scale ordered motions (i.e., rotation and radial flows) in the molecular gas component, and to interpret possible deviations from such simple kinematics in terms of internal or external mechanisms.

\subsection{Modeling the stellar kinematics}\label{sec:method_stellar_kins}
We model the stellar kinematics using the task \texttt{2DFIT} on the velocity field (see Sect.~\ref{sec:data_muse}). 
We fixed the kinematic center at the optical center reported in \cite{2017PoggiantiNat} and $V_\mathrm{rad,\star}=0$~\kms. 
Since stars are not subject to the effect of ram pressure, we expect this to be a good approximation everywhere in the galaxy with the possible exception of the bar region (but see Sect.~\ref{sec:results}). 
We adopt the following three-step approach.
\begin{enumerate}
	\item We performed a preliminary run with $\phi_\star$, $i_\star$, $V_\mathrm{sys,\star}$, and $V_\mathrm{rot,\star}$ as free parameters. 
	The initial values of $\phi_\star$ and $i_\star$ were taken from \cite{2020Franchetto}. 
	\item We made a second run with $\phi_\star$, $i_\star$, and $V_\mathrm{rot,\star}$ as free parameters, fixing $V_\mathrm{sys,\star}$ at the median of the best-fit values from the first step.
	\item We run again 3DB with $V_\mathrm{rot,\star}$ as the free parameter, while $\phi_\star$ and $i_\star$ are regularized using a polynomial function with degree from zero to three, in order to avoid numerical oscillations.
\end{enumerate}
The ring spacing is fixed to 1\arcsec, which approximately corresponds to the angular resolution of the MUSE observations. 
This choice is also reasonable based on the size of the Voronoi bins. 
In all 3DB runs, we chose to give more weight to the regions close to the disc major axis (i.e. \texttt{wfunc=2}), in order to maximize the signal from the rotational motion. 

Before comparing the stellar and molecular gas kinematics, the rotation velocity of the stars ($V_\mathrm{rot,\star}$) must be corrected for the contribution of pressure support (i.e., asymmetric drift correction $V_\mathrm{AD,\star}$) to obtain the stellar circular velocity \citep{2008BinneyTremaine}
\begin{equation}
	V_\mathrm{circ,\star}^2=V_\mathrm{rot,\star}^2+V_\mathrm{AD,\star}^2 \, .
\end{equation}
We follow the same approach as \cite{2019Marasco} and calculate the asymmetric drift correction as
\begin{equation}\label{eq:asymmetric_drift_corr}
	V_\mathrm{AD,\star}^2(R) = - R \left( \frac{\sigma_{z,\star}^2(R)}{\beta} \right)  \frac{\partial \ln \left( \Sigma_\star(R) \sigma_{z,\star}^2(R) \right) }{\partial R} \, ,
\end{equation}
where $\beta \equiv \sigma_{z,\star}/\sigma_{R,\star}$ with $\sigma_{R,\star}$ and $\sigma_{z,\star}$ are the radial and vertical velocity dispersions of the stars, and $\Sigma_\star$ is the stellar surface density. 
We calculate Eq.~\ref{eq:asymmetric_drift_corr} using the radial profile of $\Sigma_\star$ described in Sect.~\ref{sec:data_muse}, and assuming $0.5<\beta<1$ based on the anisotropy measured in nearby spiral galaxies \citep{2010Bershady_b,2013Martisson_b}. 
For $\sigma_{z,\star}$, we adopt an exponential profile with e-folding length given by $2 R_\mathrm{d}$, with $R_\mathrm{d}$ being the exponential disk scale length (see Appendix~\ref{ap:surface_brightness}), and central velocity dispersion given by $\sigma_{z,\star} (R=0) = (0.248 \pm 0.038) \times V_\mathrm{rot,\star}(R=2.2 R_\mathrm{d})$ \citep{2013Martisson_b}. 
We also impose a floor of 15~\kms on $\sigma_{z,\star}$ to avoid unrealistically small values large radii \citep{2019Marasco}. 
In principle, \textsc{pPXF} delivers also the stellar velocity dispersion map. However, this quantity is not ideal to calculate $V_\mathrm{AD,\star}$, as it is affected by resolution effects due to the limited spatial and spectral resolution of the instrument, which artificially increase the stellar velocity dispersion and thus $V_\mathrm{AD,\star}$.

We note that the angular resolution of the MUSE observations is about 1~\arcsec, corresponding to about 1~kpc in our galaxy sample. 
Hence, we expect that the PSF smearing effect is relatively mild beyond the inner regions of the stellar disk. 
However, since the PSF smearing effect becomes more pronounced with increasing inclination angle of the galaxy, it may be important for JW100 due to its high inclination with respect to the line of sight. 

We also recall that the assumption of circular orbits might be inappropriate for the innermost regions of barred galaxies, as the stars move along elongated orbits \citep[e.g.][]{1993SellwoodWilkinson,2004KormendyKennicutt}. 
A more appropriate analysis would require a dynamical modeling of the stellar kinematics using, for instance, the full Gaussian-Hermite moments of the stellar velocity field \citep[e.g.][]{1994Emsellem,2002Cappellari,2008Cappellari,2020Cappellari}. 
Since this task is beyond the scope of this work, we just remind the reader that the circular velocity of the stellar component recovered by our methodology does not trace the dynamical mass of the galaxy in the bar region. 

Only galaxies for which the bar is inclined to both the projected major and minor axes show non-circular motions clearly \citep{1993SellwoodWilkinson}. 
Hence, we do not expect visible signatures of non-circular motions in JO201 and JO206. 
In Sanchez-Garcia et al. (in preparation), the stellar velocity field of the galaxies in the GASP sample is fitted using an ad-hoc approach to include large-scale non-circular motions induced by bars. 
Preliminary results show that, for the GASP-ALMA galaxies, the recovered stellar rotation velocity obtained by Sanchez-Garcia et al. is overall consistent with ours, suggesting that the non-circular motions are small compared to rotation. 

\subsection{Modeling the molecular gas kinematics}\label{sec:method_molgas_kin}
We model the molecular gas kinematics using the task \texttt{3DFIT} on the ALMA datacubes (see Sect.~\ref{sec:data_alma}). 
To reduce the free parameters in the model, we first fixed the kinematic center at the optical center reported in \cite{2017PoggiantiNat}. 
However, since the interaction with the ICM can displace the kinematic center of the gas from that of the stars \citep[e.g.][]{2008Kronberger_b,2022Boselli_rev}, we adjusted the kinematic center of the molecular gas when necessary. 
We also set $V_\mathrm{sys,CO}$ at the value obtained from the global profile of the emission line. 
When necessary, $V_\mathrm{sys,CO}$ was refined by a few \kms after inspecting the position-velocity diagrams (see Sect.~\ref{sec:results}). 
\begin{enumerate}
	\item We performed a first run with $V_\mathrm{rad,CO}=0$~\kms and leaving free the geometrical and kinematical parameters. 
	By setting the 3DB parameter \texttt{wfunc=2}, we chose to give more weight to the emission along the disc major axis, where most of the information on rotational motions lies ($\theta=0$° in Eq.~\ref{eq:vlos}). 
	\item We made a second run (i.e. \texttt{twostage=True}) in which the geometrical parameters are regularized using either a suitable function or the median value. 
	\item $V_\mathrm{rad,CO}$ is left free in the last run, while the other parameters are fixed to the best-fit values obtained previously. 
	By setting \texttt{wfunc=-2}, we give more weight to the emission along the disc minor axis, where the contribution of radial motions is the strongest ($\theta=90$° in Eq.~\ref{eq:vlos}).
\end{enumerate}
This procedure is substantially based on the approach developed by \cite{2021DiTeodoro}, who used 3DB to model the atomic gas kinematics in a sample of nearby galaxies in order to measure gas radial motions and mass flows. 
These authors used 21-cm observations with higher spatial resolution and better velocity resolution than our ALMA data. 
Radial motions are possibly stronger and easier to detect for galaxies affected by the ram pressure than in the case of \cite{2021DiTeodoro}'s galaxies, in which radial motions are of the order of a few \kms. 
We stress that the approach adopted in this work takes into account the radial motions within the galaxy disk, while motions perpendicular to the disk midplane are not considered \citep{2021DiTeodoro}. 
The direction (either inward or outward) of these radial motions cannot be determined unless the near/far sides of the galaxy are known. 
We infer this by assuming that spiral arms (when visible) trail the galaxy rotation, or by exploiting dust lanes crossing the disk. 
In 3DB's convention \citep{2021DiTeodoro}, radial motions with $V_\mathrm{rad,CO} < 0$~\kms point inward in a disk that rotates clockwise, while those with $V_\mathrm{rad,CO} > 0$~\kms point outward (vice versa for counterclockwise rotation). 

We used the 3DB task \texttt{ELLPROF} to derive the azimuthally-averaged radial profiles of the CO surface brightness. 
These profile were adopted for the normalization procedure of 3DB models and to derive the H$_2$ surface density $\Sigma_\mathrm{H_2}$. 

We also used the 3DB task \texttt{spacepar} to fully explore the parameter space for $V_\mathrm{rot,CO}$ and $\sigma_\mathrm{CO}$. 
This test is useful to check whether the model fitting converges to a good minimum of the parameter space. 
We anticipate that, while the best-fit $V_\mathrm{rot,CO}$ is generally well constrained, it is not always the case for $\sigma_\mathrm{CO}$. 
This is likely due to the complex shape of the emission line profiles. 

A possible caveat in our methodology is that the tilted-ring model is based on the assumption of concentric orbits, which might not be valid for the gas in galaxies affected by strong ram pressure or in an advanced stripping stage \citep[e.g.][]{2008Kronberger_b}. 
In these cases, the results of our analysis are very uncertain and should be taken with caution. 
However, if stripping is not too dramatic, modeling the gas kinematics using the tilted-ring approach may be possible for the disk regions where some or most of the gas has preserved its original motion. 
Stellar bars are also expected to induce non-circular motions due to the gas streaming along the bar \citep[e.g.][]{1993SellwoodWilkinson}. 
Indeed, the gas kinematics in barred galaxies is usually modeled using tools that are specifically designed to take into account non-axisymmetric distortions in the 2D velocity field \citep[e.g.][]{1999PhDThesis_Schoenmakers,2007Spekkens}. 
However, these methods fail when the bar is perpendicular to or parallel the disk major axis, being unable to break the degeneracy between the tangential and radial velocity components \citep[e.g.][]{2010Sellwood,2015Randriamampandry}. 
We thus decided to adopt the tilted-ring approach also in the case of JO201 and JO206, which host stellar bars aligned with the disk major axis.

\section{Results and discussion}\label{sec:results}
In this section, we present the best-fit models for the molecular gas kinematics and we then compare the stellar and molecular gas rotation curves. 
We discuss each galaxy individually in Sects.~\ref{sec:results_jo201}--\ref{sec:results_jw100} and summarize our findings in Sect.~\ref{sec:results_summary}. 
We analyzed both the CO(1--0) and the CO(2--1) datacubes, obtaining essentially the same results. 
Thus, we show here the best-fit models for the CO(1--0) data, which have a similar angular resolution but better S/N for the kinematic modeling compared to the CO(2--1) data. 
From here on, CO indicates CO(1--0) unless otherwise stated. 
Since the focus of this work is on the molecular gas, we show the best-fit model for the stellar kinematics only for JO201 in Fig~\ref{fig:2dmodel_jo201}, while the models for the rest of the sample can be found in Appendix~\ref{ap:stellar_kins}. 
\begin{figure*}
	\centering
	{\includegraphics[width=2.\columnwidth]{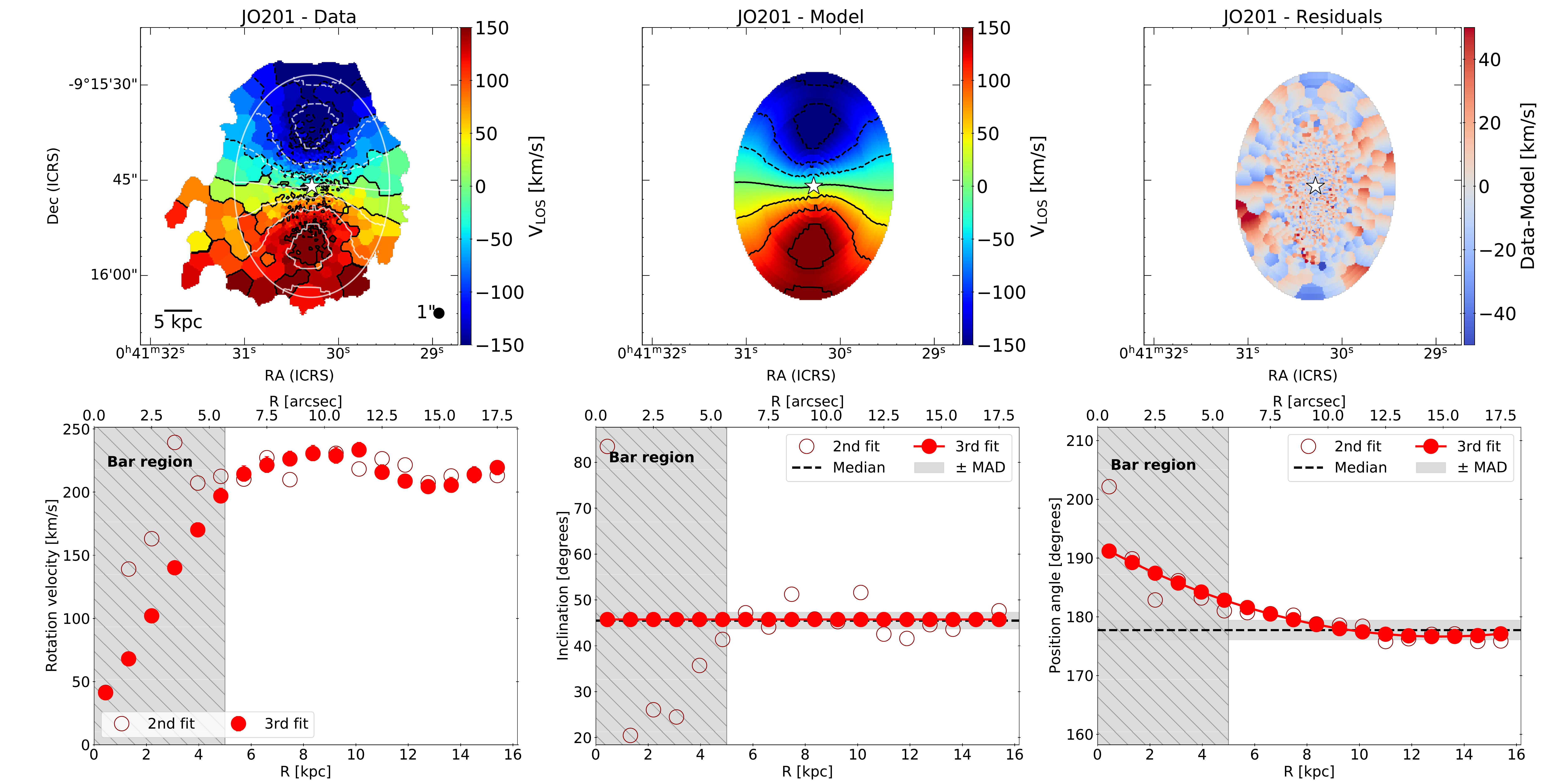}}
	\caption{\textit{Top row}: stellar velocity field (left), its best-fit model (center), and residual map (right) for JO201. 
		The white star indicates the disc center. 
		The black curves are the iso-velocity contours with steps of 50~\kms. 
		The thick black contour indicates $V_\mathrm{sys,\star}$. 
		The white contours in the left panel shows the best-fit model on the data.
		The bar and the circle in the bottom left and right corners respectively show the physical scale and the PSF of the observations. 
		\textit{Bottom row}: rotation velocity (left), inclination (center) , and position angle (right) as a function of the galactocentric distance for the best-fit models of the stellar velocity field. 
		The grey dashed area indicates the region influenced by the stellar bar. 
		The empty circles and the red points are for the 2nd and the 3rd steps of our procedure (see Sect.~\ref{sec:method_stellar_kins}), respectively. 
		The dashed black lines and the grey area indicate the median and the median absolute deviation, respectively.}
	\label{fig:2dmodel_jo201}
\end{figure*}

\subsection{JO201}\label{sec:results_jo201}
The $I$-band image in Fig.~\ref{fig:iband_images} shows that JO201 has a stellar bulge. 
Moreover, the elongated shape of the isophotes in the inner regions suggests that JO201 hosts a stellar bar, as reported by \cite{2019George}. 
\cite{2023SanchezGarcia} estimated that the bar length is $\approx 4.6$~kpc. 
We also note that the stellar disc of JO201 seems morphologically lopsided, being the east side slightly more extended than the west one. 

The top panels in Fig~\ref{fig:2dmodel_jo201} show, from left to right, the observed stellar velocity field, the best-fit model, and the map of the residuals between the data and the best-fit model. 
The bottom panels display the radial profile of the best-fit rotation velocity (left), inclination (center), and PA (right). 
The stellar velocity field is very well reproduced by the model. 
The residuals in the disk outskirts, where the Voronoi bins are the largest, tend to be higher than in the inner regions, but still within the velocity resolution of the MUSE observations, that is $\Delta v \approx 50$~\kms. 
We note that, for $R \lesssim 5$~kpc, the rotation velocity is much lower than expected for a galaxy with stellar mass $M_\star \approx 9 \times 10^{10}$~M$_\odot$ and hosting a stellar bulge. 
This feature can be explained by the fact that the stellar bar is aligned along the disk major axis.  
In a scenario where a large fraction of the stars in the bar move on elliptical orbits aligned parallel to the bar (the so-called $x_\mathrm{1}$ type; \citealt{2014Sellwood}), the velocity component along the line of sight has its minimum at the apocentre and then increases along the major axis.  
This can result in an underestimation of the rotation velocity in the regions influenced by the bar \citep[e.g.][]{2008Dicaire,2010Sellwood,2015Randriamampandry}. 
\begin{figure*}
	\centering
	\includegraphics[width=2.\columnwidth]{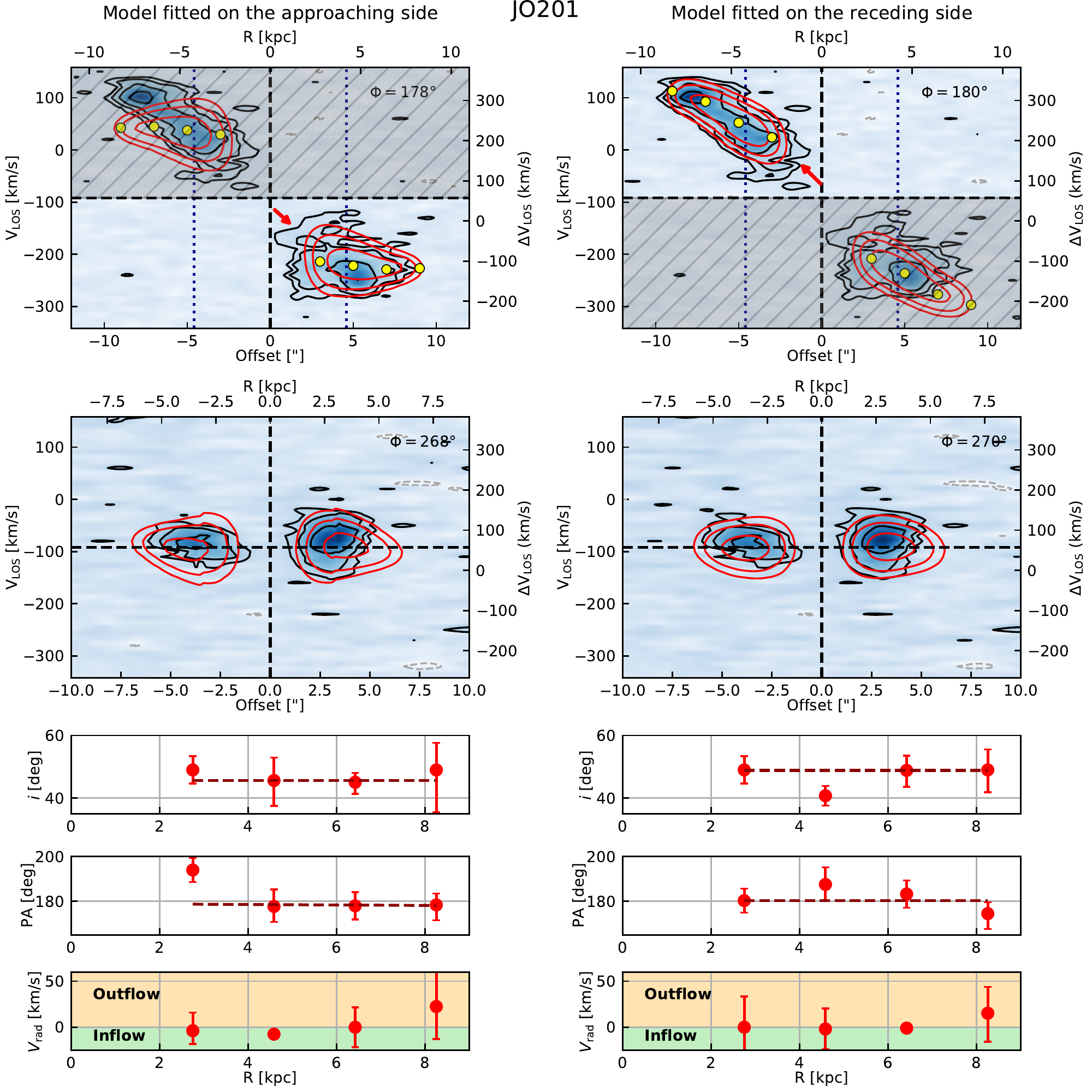}
	\caption{Best-fit models of the molecular gas kinematics for JO201 using CO(1--0) emission line observations. 
		The left and the right panels are for the approaching and the receding sides of the disc (the other side is shaded), respectively. 
		The first and the second rows show the PVD along the major and the minor axis, respectively. 
		The observed CO(1--0) emission is shown in blue with black and grey contours, while the red contours show the best-fit model. 
		All the contours are at $2^n \sigma_\mathrm{ch}$ with $n=1,...,10$ and $\sigma_\mathrm{ch}=0.7$~mJy/beam. 
		The yellow points indicate the projected rotation curves of the best-fit models. 
		The vertical blue dotted lines and the red arrows indicate the bar extent and the gas with anomalous kinematics (see text), respectively.
		In the last three rows, the panels show the profiles of inclination, PA, and radial velocity of the best-fit models. 
		The red points are the parameters from the 1st fitting step and the dark-red lines are the regularized profiles (see Sect.~\ref{sec:method_molgas_kin}). 
		The orange and green areas in the bottom panels indicate whether positive/negative values for $V_\mathrm{rad}$ mean radial gas outflow/inflow. }
	\label{fig:pvd_jo201_co10}
\end{figure*}

The total CO intensity map (top left panel in Fig.~\ref{fig:maps_co10}) gives useful indications about the effect and direction of ram pressure. 
In JO201, the west side of the disk shows compressed contours and regions with bright CO emission, possibly suggesting the ram pressure compressed this side of the disk \citep{2017Bellhouse}. 
The most evident feature in Fig.~\ref{fig:maps_co10} is arguably the presence of the ring-like structure surrounding the hole in the CO distribution in the innermost $\approx 3$~kpc \citep[see also][]{2018George,2019George}. 
The ring-like structure is also visible in the MUSE images shown by \cite{2017Bellhouse}. 
This feature can be explained by the presence of the bar driving the formation of a molecular gas ring around the co-rotation radius \citep[i.e. where the bar pattern equals the angular frequency of circular motions; see][]{1993SellwoodWilkinson,2004KormendyKennicutt}. 
At radii well inside co-rotation, gas is expected to fall toward the center. 
The molecular gas distribution in barred galaxies is typically very concentrated in the center \citep[e.g.][]{2004KormendyKennicutt}, while Figure~\ref{fig:maps_co10} clearly shows the lack of CO emission in the innermost regions of JO201. 
\cite{2019George} attribute this CO cavity to AGN feedback, which ionizes the molecular hydrogen (i.e. radiative feedback) and sweeps the gas from the center (i.e. mechanical feedback). 
The connection between nuclear activity and the gas distribution and kinematics is specifically tackled in the companion paper (Mingozzi et al. to be submitted). 

The CO velocity field of JO201 (2nd panel in the top row of Fig.~\ref{fig:maps_co10}) shows that the galaxy is kinematically lopsided, meaning that the velocity gradient in the receding and approaching sides of the disc are significantly different from each other \citep[e.g.][]{1994Richter,1999Swaters,1999PhDThesis_Schoenmakers,2015Shafi}. 
For this reason, we modeled the approaching side and receding side separately. 
We compare the observations with our best-fit models in Fig.~\ref{fig:pvd_jo201_co10}, where the left and the right panels are for the approaching and receding sides of the disc, respectively. 
The first and second rows in Fig.~\ref{fig:pvd_jo201_co10} are the position-velocity diagrams (PVDs) along the major and minor axis of the disc, respectively. 
Our rotating disc model can reproduce reasonably well the observations, indicating that the molecular gas in the disk preserved its original rotation, despite the interaction with the ICM. 
There is however some gas, which is indicated by the red arrow in Fig.~\ref{fig:pvd_jo201_co10}, moving with lower velocities than those predicted by the model. 
Since this gas is located at galactocentric distances smaller than the bar length, its anomalous kinematics is plausibly due to the bar influence.

By exploring the parameter space, we found that the best-fit value of the CO velocity dispersion is not well-constrained for the outermost ring, likely because of the low S/N. 
For $R \lesssim 5$~kpc, we obtain $\sigma_\mathrm{CO} \approx 25-40$~\kms, which can be explained by the non-circular motions due to the stellar bar. 
Outside the bar regions, we find $\sigma_\mathrm{CO} \approx 20$~\kms, which is about a factor 2 higher than the typical values of the molecular gas velocity dispersion in local isolated, unbarred galaxies \citep[e.g.][]{2020Bacchini_a}. 
This enhancement of $\sigma_\mathrm{CO}$ may be due to ram pressure increasing the molecular gas turbulence, either directly or by enhancing the star formation rate (SFR; see Sect.~\ref{sec:results_summary} for further discussion). 

We note that the best-fit values of radial velocity are consistent with zero, suggesting that the inclusion of radial motions does not significantly improve the fit. 
Hence, these values should be taken with caution. 
Based on the RGB image of JO201 shown by \cite{2017Bellhouse}, the spiral arm direction suggests clockwise rotation (i.e. $V_\mathrm{rad,CO} < 0$~\kms for inflows). 
Taken at face value, the inflow at $R \lesssim 5$~kpc with $V_\mathrm{rad,CO} \gtrsim -10$~\kms is comparable with the average values measured in the inner regions of nearby spiral galaxies \citep{2021DiTeodoro}. 
Beyond the bar region, the radial outflow with $V_\mathrm{rad,CO} \gtrsim 20$~\kms is consistent with being caused by ram pressure. 
However, since non-circular motions can be induced by any perturbation of the gravitational potential, we cannot exclude a different origin \citep[e.g.][]{2010Sellwood}.

\begin{figure}
	\centering
	{\includegraphics[width=1.\columnwidth]{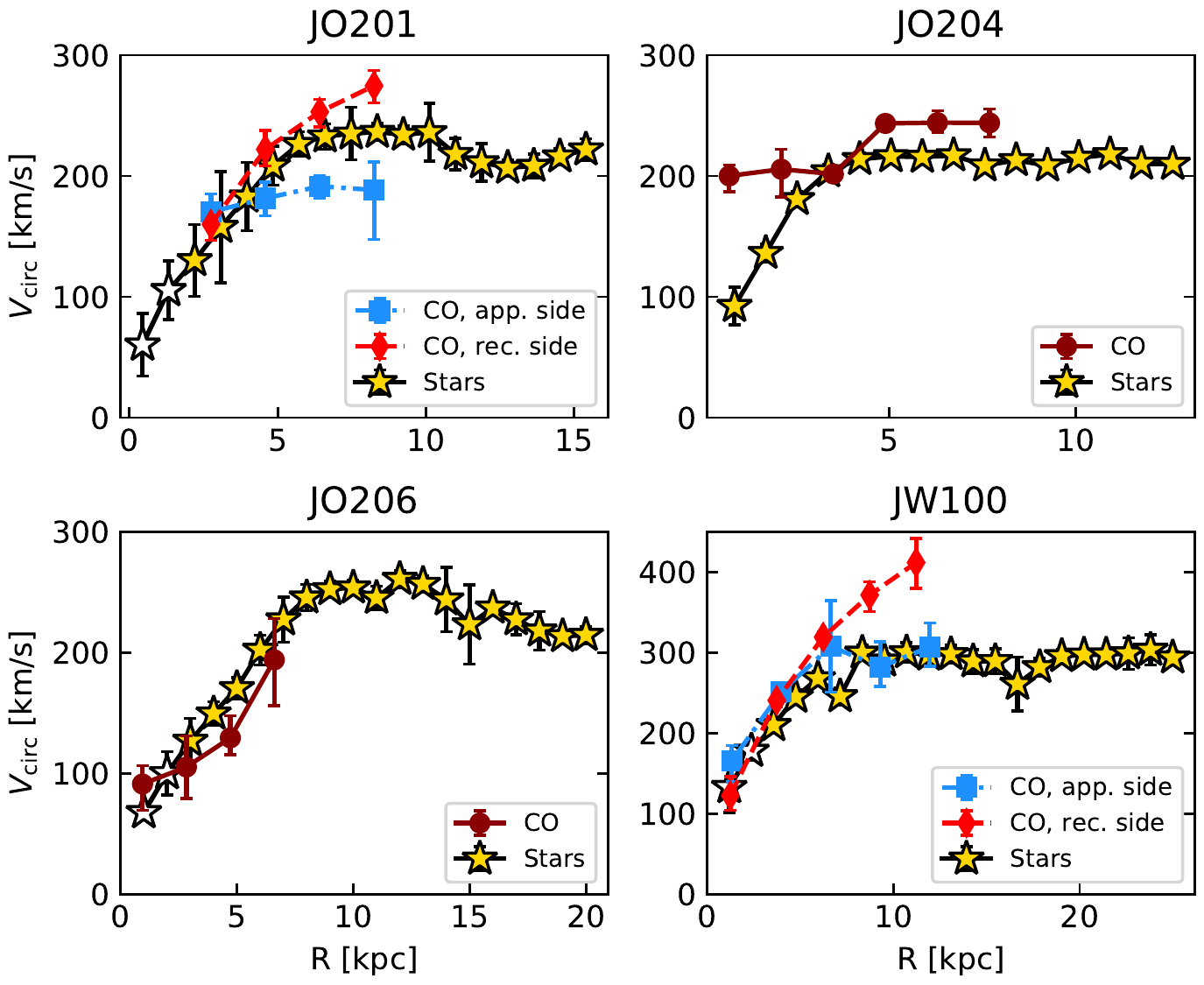}}
	\caption{Comparison of the stellar (yellow stars) and CO (darkred points) circular velocities for each galaxy in our sample. 
	The empty stars indicate the points that may be contaminated by the bulge kinematics (see Sect.~\ref{sec:data_muse}).	
	When the approaching side and the receding side of the disc are modeled separately, the resulting profiles are shown by the blue squares and the red diamonds, respectively.}
	\label{fig:vrot_CO}
\end{figure}

In Fig.~\ref{fig:vrot_CO} (top left), we compare the circular velocities inferred from the kinematics of the stellar and molecular gas disks. 
Within $R \approx 5$~kpc, the molecular gas and stellar circular velocities essentially coincide, but the velocity gradient is too shallow for a massive galaxy with a bulge such as JO201. 
As mentioned above, this is likely due to the stellar bar aligned along the major axis \citep[e.g.][]{2008Dicaire,2010Sellwood,2015Randriamampandry}. 
We cannot fully exclude that some contamination due to bulge component is also present for $R \lesssim 2$~kpc (see Appendix~\ref{ap:surface_brightness}). 
Indeed, the bulge kinematics are significantly pressure-supported, which may lead to an underestimated $V_\mathrm{rot,\star}$. 
Beyond the bar regions, the stellar velocity field of JO201 (Fig.~\ref{fig:2dmodel_jo201}) does not show any indication of the kinematic lopsidedness, contrary to the molecular gas. 
The receding side of the CO disc reaches slightly higher rotation velocities than the stellar disc, while the approaching side shows a lower velocity gradient. 
The kinematic lopsidedness in disc galaxies is typically ascribed to a triaxial potential, as in the presence of a stellar bar \citep[e.g.][]{1999Swaters,1999PhDThesis_Schoenmakers,2004Rhee}. 
However, the regular kinematics of the stellar disk seems to suggest that the molecular gas kinematics may be perturbed by some mechanisms that does not affect the stars, such as ram pressure. 
These distortions appear in the outer parts of the galaxy and in a symmetric way, as expected for face-on ram pressure \citep{2008Kronberger_b,2017Bellhouse,2019Bellhouse}. 
Indeed, JO201 is moving towards the observer at very high velocity (see Table~\ref{tab:galaxyprops}), implying that the approaching and receding sides of the disk move in the opposite and the same direction as the ram pressure, respectively. 
The ram pressure is thus expected to decelerate the approaching side of the molecular disk and accelerate the receding side \citep{2008Kronberger_b}, which is consistent with our results. 

We conclude that the molecular gas kinematics in the inner regions of JO201 is mainly dominated by the perturbations due to the stellar bar. 
In the outer parts of the molecular gas disk, the kinematic lopsidedness and radial motions (although rather uncertain) seem to suggest that the molecular gas in JO201 is affected by face-on ram pressure, despite other mechanisms cannot be ruled out. 

\subsection{JO204}\label{sec:results_jo204}
Before focusing on the molecular gas kinematics, it is worth noting two features of the stellar component. 
First, the innermost isophotes in the $I$-band image (Fig.~\ref{fig:iband_images}) show a boxy shape that might indicate the presence of a bar seen with high inclination with respect to the line-of-sight \citep[e.g.][]{1990Combes,1994Bettoni,1995Kuijken,1999Bureau,1999Merrifield}. 
Unfortunately, dust obscuration and projection effects hamper any attempt to estimate the bar length from the optical images. 
The second feature is visible in the stellar velocity field, which shows slightly distorted iso-velocity contours in the innermost regions (see Fig.~\ref{fig:2dmodel_jo204}). 
This S-shaped feature indicates the presence of non-circular motions and, possibly, of a stellar bar \citep[e.g.][Sanchez-Garcia et al. in preparation]{1989Bettoni,1997Vauterin,2004KormendyKennicutt,2015Cortes}. 
Indeed, the top right panel of Fig.~\ref{fig:2dmodel_jo204} shows residuals of a few tens of \kms in the regions close to the disc minor axis, indicating that a model based on circular orbits cannot fully reproduce the observations. 

In Fig.~\ref{fig:maps_co10}, the CO total intensity map shows that the molecular gas distribution is strongly concentrated in the center and two arm-like structures. 
Both features are typical of barred galaxies \citep[e.g.][]{1992Athanassoula_b,1999Bureau,1999Merrifield,2004KormendyKennicutt,2021Hogarth}. 
The iso-velocity contours in the CO velocity field (Fig.~\ref{fig:maps_co10}) are visibly distorted in the inner regions, which typically indicates the presence of non-circular motions. 
The CO velocity dispersion is also quite high in the inner regions of disk.

\begin{figure}
	{\includegraphics[width=1.\columnwidth]{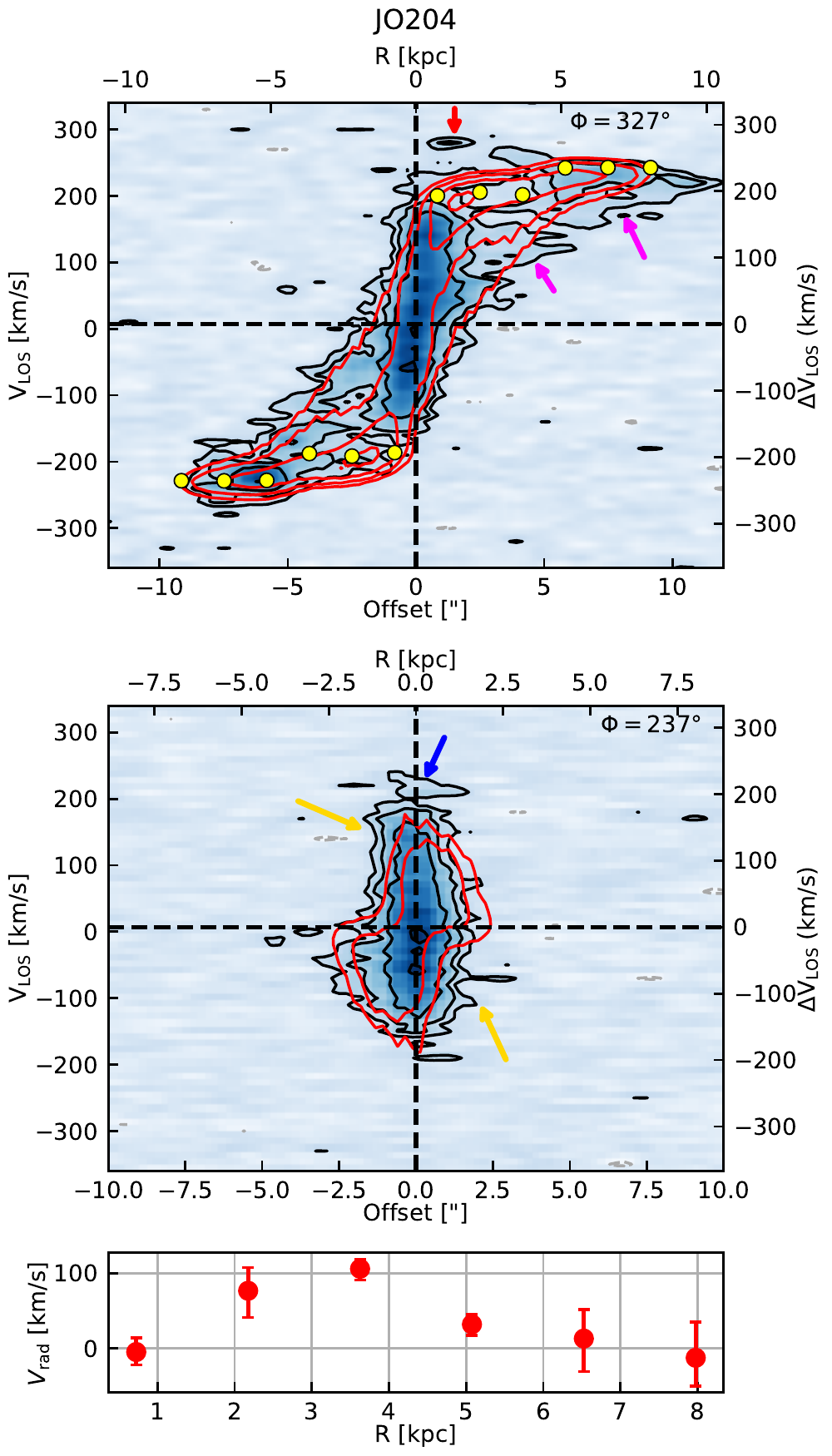}}
	\caption{Same as Fig.~\ref{fig:pvd_jo201_co10} but for JO204. 
		The model fitting is performed on the approaching and receding sides at the same time, and the inclination and PA are fixed to the values of 75° and 327°, respectively. 
		The arrows indicate the gas with anomalous kinematics (see text). 
		Here $\sigma_\mathrm{ch}=0.9$~mJy/beam.}
	\label{fig:pvd_jo204_co10}
\end{figure}

To model the CO kinematics, we modified the procedure described in Sec.~\ref{sec:method_molgas_kin}. 
After various tests, we decided to fit the data by fixing all the geometrical parameters and leaving free $V_\mathrm{rot,CO}$, $V_\mathrm{rad,CO}$, and $\sigma_\mathrm{CO}$, as this choice improved the comparison between the model and the data.  
We run 3DB using the \texttt{reverse} option, which performs the fit starting from the most external ring and then moving inward. 
This algorithm was designed to improve the fit for galaxies seen with high inclination ($i\gtrsim 70$°). 
Fig.~\ref{fig:pvd_jo204_co10} compares the best-fit model with the observations. 
In Fig.~\ref{fig:pvd_jo204_co10}, the PVD along the major axis shows that, overall, the model reproduces well the observations, except for two features. 
The first feature is indicated by the red arrow and consists in gas moving at $V_\mathrm{LOS,CO} \approx 270$~\kms at about 1~kpc from the center. 
One possibility is that this CO emission is probing the inner rise of the rotation curve of the molecular disk if the nuclear CO distribution is asymmetric between approaching and receding sides \citep[see for example][]{2022Lelli}. 
Alternatively, this central emission can be ascribed to complex non-circular motions caused by the stellar bar \citep[the so-called $x_\mathrm{2}$ orbits aligned perpendicular to the bar; see][]{1979Sancisi,2004KormendyKennicutt,2015Randriamampandry} or even feedback from stars or the AGN \citep[e.g.][]{2021Stuber}. 
Our finding is in agreement with the results obtained by \cite{2020Deb}, who revealed an absorption feature in the HI global profile that could be explained by high-velocity gas seen in front of the continuum emission from the AGN. 
The second feature that is not reproduced by the model is indicated by the magenta arrows.  
This emission comes from gas that moves at lower velocities than those predicted by our model. 
A possible explanation is that this gas is decelerated by the ram pressure component in the sky plane. 
Alternatively, this emission can be explained by gas in nearly circular orbits with a slightly lopsided distribution. 
We note indeed that the PVD along the major axis (top panel of Fig.~\ref{fig:pvd_jo204_co10}) resemble the characteristic X-shaped pattern, typical of barred galaxies seen at high inclination along the line of sight \citep[e.g.][]{1999Bureau,1999Merrifield,2004KormendyKennicutt,2013Alatalo,2021Hogarth}. 

In Fig.~\ref{fig:pvd_jo204_co10}, the PVD along the minor axis shows extended gas emission in the lower left quadrant, which can only be reproduced by a model with strong radial motions of $V_\mathrm{rad,CO} \gtrsim 50$~\kms. 
However, the same feature is not observed in the upper right quadrant, indicating that the CO distribution (or kinematics) is asymmetric. 
Unfortunately, JO204 does not have visible spiral arms and the dust lanes in the optical MUSE and Hubble Space Telescope (HST) images \citep[][]{2017Gullieuszik,2023Gullieuszik} do not allow to clearly identify the nearest side of the disk. 
Hence, we cannot infer the direction of rotation and radial motions. 
Since these non-circular motions are detected in the inner parts of the galaxy, one may speculate that they are inflows driven by a stellar bar. 
The magnitude of radial motions in JO204 is consistent with the values estimated in simulated barred galaxies \citep[see][]{2015Randriamampandry}, but about 2 times higher than those measured by \cite{2021DiTeodoro} in the atomic gas of real barred galaxies. 
The blue arrow indicates another feature that is not reproduced by our model. 
However, since this emission is very faint, it is unclear whether this is real emission from the galaxy. 
We find $\sigma_\mathrm{CO} \approx 10$~\kms, but this value is rather uncertain based on the inspection of the parameter space. 

Overall, our model can reproduce the molecular gas kinematics in JO204 reasonably well, despite the complexities due to the stellar bar and/or ram pressure. 
Figure~\ref{fig:vrot_CO} (top right panel) shows that the CO circular velocity is, on average, compatible with the stellar circular velocity. 
The difference in the innermost regions is most probably due to a combination of dust extinction and resolution effects, which may smooth the gradient of the stellar rotation curve (see Sect.~\ref{sec:method}). 
Although the uncertainties on $V_\mathrm{rot,CO}$ are likely underestimated, the CO and stellar circular velocity are slightly different at $R>3$~kpc. 
This discrepancy, if real, is difficult to explain given the complex gas kinematics and the limitations of our models. 
Similarly to the case of JO201, we conclude that the molecular gas kinematics is mainly rotation-dominated in JO204. 
Then, the stellar bar plays an important role in perturbing the gas kinematics and driving radial gas flows, while the ram pressure may play a minor role in the outskirts of the molecular gas disk of JO204. 

\subsection{JO206}\label{sec:results_jo206}
The $I$-band images in Fig.~\ref{fig:iband_images} shows that JO206 hosts a stellar bulge. 
In addition, the elongated shape of the isophotes in the inner regions indicate the presence of a stellar bar aligned with the disk major axis, as for JO201. 
The CO total intensity map in Fig.~\ref{fig:maps_co10} shows that the morphology of the molecular gas distribution is asymmetric, suggesting that the ram pressure is directed towards south-east. 
Hence, we expect the molecular gas kinematics to be particularly complex in JO206, as a consequence of the combined effects of bar perturbations and ram pressure stripping. 
Indeed, the iso-velocity contours in the CO velocity field (2nd panel in the third row in Fig.~\ref{fig:maps_co10}) are even more distorted than those of JO204, indicating stronger perturbations.

\begin{figure}
	{\includegraphics[width=1.\columnwidth]{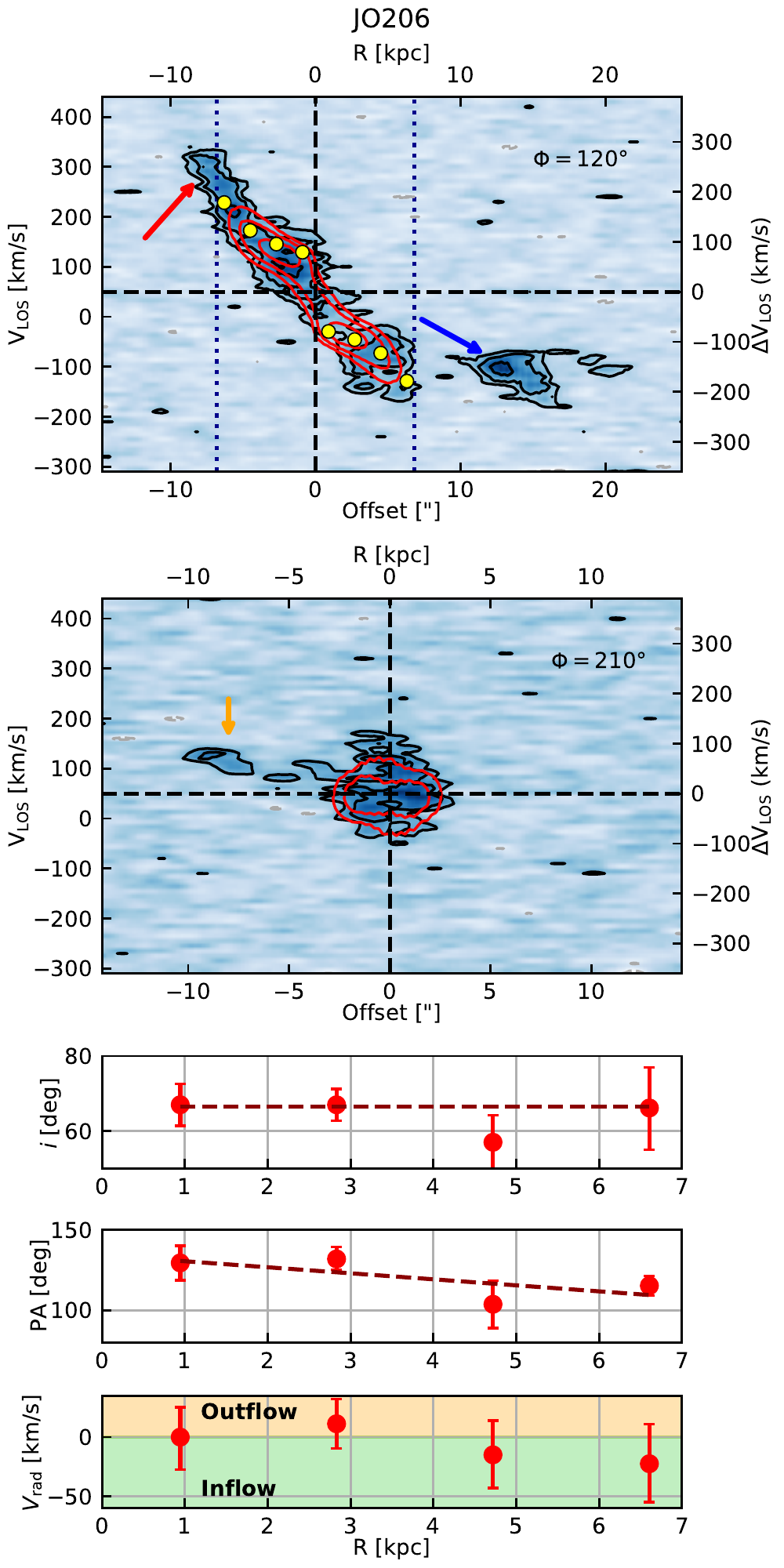}}
	\caption{Same as Fig.~\ref{fig:pvd_jo201_co10} but for JO206. 
		The model fitting is performed on both the approaching and receding sides, at the same time. 
		Here $\sigma_\mathrm{ch}=0.8$~mJy/beam.}
	\label{fig:pvd_jo206_co10}
\end{figure}

In the light of these considerations, we modeled only the regions of the molecular gas disk within $R \approx 6$~\arcsec, which essentially corresponds to the extent of its receding side (Fig.~\ref{fig:maps_co10}). 
We also adjusted the position of the kinematic center with respect to the optical center by applying a small shift of $\approx 0.23$~\arcsec. 
Figure~\ref{fig:pvd_jo206_co10} shows that our model is able to reproduce reasonably well the observations, except for the molecular gas with anomalous kinematics. 
The CO emission indicated by the blue arrow (offset $\approx 10-20$~\arcsec and -200~\kms$ \lesssim V_\mathrm{LOS,CO} \lesssim -100$~\kms) belongs to the tail of stripped gas (see also Fig.~\ref{fig:maps_co10}). 
Probably, this portion of gas disc was detached from the approaching side of the disc and decelerated by ram pressure. 
Another possibility is that the ram pressure displaced a portion of the disc at larger radii, thus its rotation velocity is lowered by conservation of angular momentum. 
The red arrow indicates the receding side of the molecular gas disk that has not been stripped. 
In the PVD along the minor axis (second panel in Fig.~\ref{fig:pvd_jo206_co10}), the CO emission indicated by the orange arrow belongs to the molecular gas in the stripped tail left behind by the galaxy. 

By exploring the parameter space, we found that the best-fit value of $\sigma_\mathrm{CO}$ is well constrained only for the 2nd and 3rd rings, where we obtained $\sigma_\mathrm{CO} \approx 30-40$~\kms (see also Fig.~\ref{fig:maps_co10}). 
These values are higher than those typically measured in nearby galaxies using CO observations with similar resolution \citep[e.g.][]{2020Bacchini_a}, which is not surprising given the complex kinematics of the molecular gas in JO206. 

We tentatively detect radial motions of $\approx -25$~\kms. 
Based on the MUSE optical images \citep{2017Poggianti,2019Bellhouse}, the spiral arm direction suggests clockwise rotation (i.e. $V_\mathrm{rad,CO} < 0$~\kms for inflows). 
However, including radial motions does not improve the best-fit model, as indicated by the fact that radial velocities are consistent with zero. 

The bottom left panel in Fig.~\ref{fig:vrot_CO} shows that the stellar and CO circular velocities are consistent within the uncertainties up to $R \approx 6.5$~kpc. 
As in the case of JO201 (see Sect.~\ref{sec:results_jo201}), the slow rise of the inner rotation curve is plausibly due to the fact that the stellar bar is aligned parallel to the disk major axis. 
This suggest that the kinematics of both the molecular gas and the stars are dominated by the stellar bar in these regions\footnote{This result is consistent with the preliminary estimate of the bar length obtain by Sanchez-Garcia et al. (in preparation), that is approximately 6.7~kpc.}.
We also note that the position and velocity of the molecular gas emission indicated by the red arrow (Fig.~\ref{fig:pvd_jo206_co10}, top panel) is perfectly compatible with the circular velocity profile of the stars (see Fig.~\ref{fig:vrot_CO}). 
On the contrary, the stripped tail indicated by the blue arrow  (Fig.~\ref{fig:pvd_jo206_co10}, top panel) is decelerated of about $70$\kms  with respect to the stars at the same galactocentric distance, suggesting that this material is decoupled from the disk rotation. 
The asymmetric perturbations on the molecular gas kinematics and the displaced kinematic center are signatures of edge-on ram pressure stripping \citep{2008Kronberger_b}. 

Overall, we conclude that the molecular gas kinematics is mainly perturbed by the stellar bar. 
Taken at face value, the radial motion in JO206 can be interpreted as a gas inflows driven by the bar, as they are within its region of influence. 
We also find clear indications of edge-on ram pressure stripping based on the presence of molecular gas emission detached from the galaxy and with kinematics decoupled from the main disk. 
This suggests that the ram pressure has a stronger effect on the molecular gas disk in JO206 than in JO201 and JO204. 

\subsection{JW100}\label{sec:results_jw100}
The $I$-band image in Fig.~\ref{fig:iband_images} seems to suggest that the nearly edge-on galaxy JW100 hosts a stellar bulge. 
Moreover, despite the fact that JW100 is strongly affected by projection effects and dust obscuration, we can tentatively identify the presence of a warp in the stellar disk based on the S-shape of the isophotes close to the galaxy major axis. 
Regarding the stellar kinematics, our model can successfully reproduce the observations and recover the stellar rotation curve (see Fig.~\ref{fig:2dmodel_jw100} in Appendinx~\ref{ap:stellar_kins}). 
However, we found quite high residuals in a ring at $R \approx 6$~\arcsec and in the disk outskirts. 
After various trails, we found no significant improvement in the residual map using different geometrical parameters and allowing for radial motions. 
This can be due to the combined effects of low S/N of the observations in the disk outskirts, strong projection effects due to the radial variation in disk inclination and PA, and asymmetric dust lanes \citep{2023Gullieuszik}.
Since JW100 belongs to a substructure of three galaxies in Abel~2626, we cannot rule out that the stellar kinematics is perturbed by tidal interaction. 

Figure~\ref{fig:maps_co10} clearly shows that the distribution and kinematics of the molecular gas in JW100 are strongly disturbed, suggesting that the ram pressure component in the sky plane is directed westward and contributes in pushing the gas outside the stellar disk. 
The case of JW100 may seem surprising, as the high mass of this galaxy is expected to produce a strong gravitational pull that can efficiently contrast the ram pressure stripping. 
However, the supersonic speed and the close proximity to the cluster center (see Table~\ref{tab:galaxyprops}) indicate that JW100 is in the most favorable conditions for experiencing strong ram pressure. 
Indeed, the iso-velocity contours in the CO velocity field (2nd panel in the last row in Fig.~\ref{fig:maps_co10}) are even more distorted than the rest of the GASP-ALMA sample. 

\begin{figure*}
	\centering
	{\includegraphics[width=2.\columnwidth]{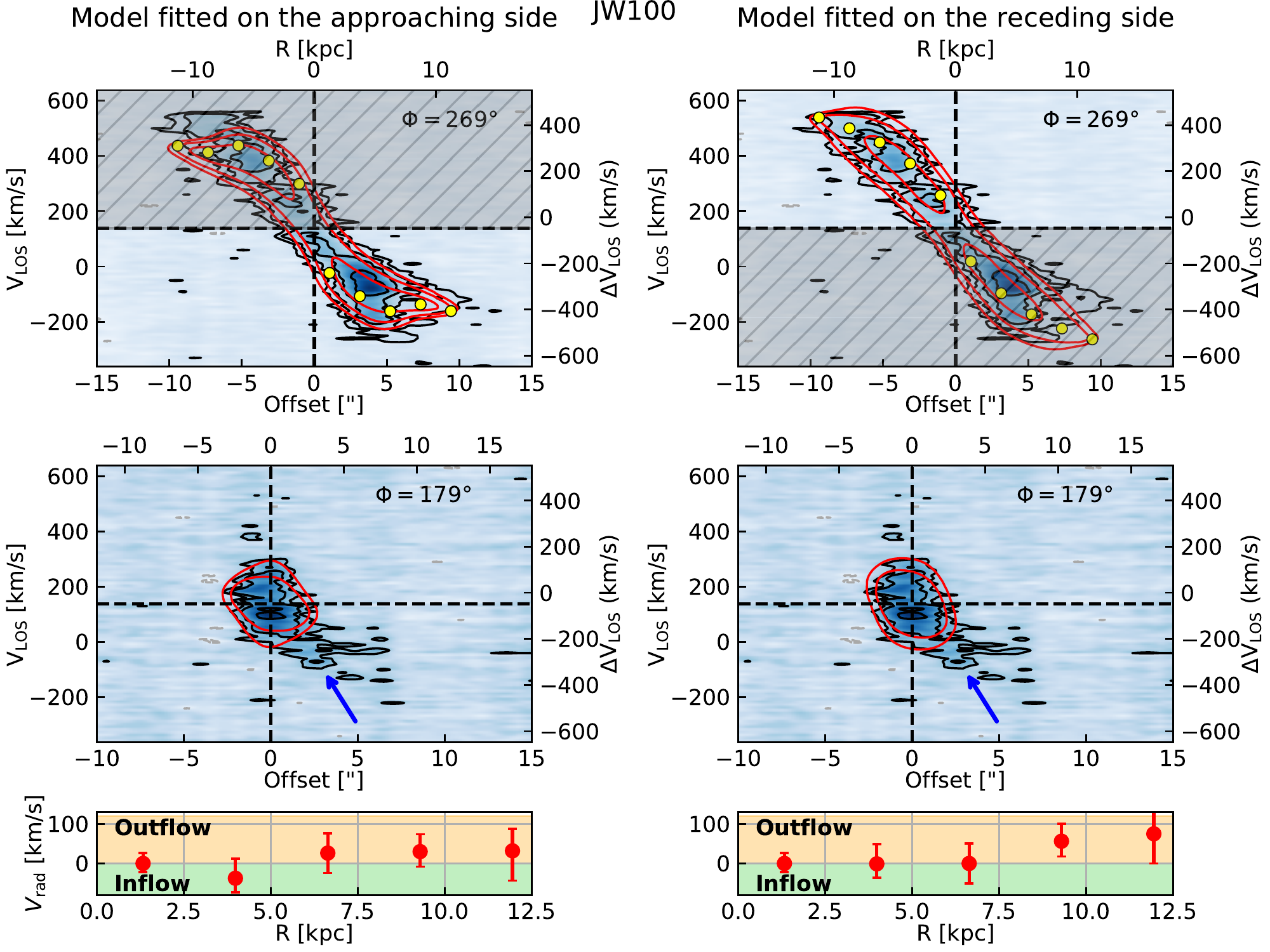}}
	\caption{Same as Fig.~\ref{fig:pvd_jo201_co10} but for JW100.  
		In the left and right panels, the model fitting is respectively done for the approaching and receding sides, and the kinematic center is $\approx 1.6$\arcsec westward from the optical center. 
		The inclination and PA are fixed to the values of 77° and 179°, respectively. 
		Here $\sigma_\mathrm{ch}=1.1$~mJy/beam.}
	\label{fig:pvd_jw100_co10}
\end{figure*}

We attempt to model the gas kinematics with the aim of understanding whether some gas has retained its original rotation. 
Hence, we run 3DB using the \texttt{reverse} option for highly inclined galaxies. 
We fixed the inclination and PA at the values obtained for the stellar disc and shifted the kinematic center $\approx 1.6$\arcsec westward from the optical center. 
We performed the fitting on the approaching and receding sides separately, as the PVD along the major axis is asymmetric with respect to $V_\mathrm{sys,CO}$. 
The resulting best-fit models are shown in the left and right panels of Fig.~\ref{fig:pvd_jw100_co10}, respectively. 
The models well reproduce the observations, except for the emission indicated by the blue arrow in the minor axis PVD. 
This emission comes from the molecular gas in the tail that is left behind by JW100 as it falls into the cluster receding from the observer. 
Indeed, the bottom panel in Fig.~\ref{fig:pvd_jw100_co10} shows the profiles of the radial velocity, which reaches $V_\mathrm{rad,CO} \approx 50-100$~\kms in the disk outskirts. 
Taken at face value, the radial velocities are larger than the $V_\mathrm{rad,CO}$ values of a few \kms that are typically measured in nearby galaxies \citep[e.g.][]{2021DiTeodoro}. 
Although our $V_\mathrm{rad,CO}$ measurements are uncertain, also the skewed shape of the CO emission in the PVD along the minor axis (blue arrows in Fig.~\ref{fig:pvd_jw100_co10}) suggests the presence of radial motions. 
We note that the emission from the stripped gas indicated by the blue arrow reaches even higher velocities ($\Delta V_\mathrm{LOS,CO} \approx -200$~\kms) than the model emission. 
The dust lanes in the HST images \citep{2023Gullieuszik} seem to suggest that the west side of JW100 is the nearest one and the galaxy is rotating clockwise. 
This implies that $V_\mathrm{rad,CO}>0$~\kms indicates an outward radial flow. 
This is in agreement with the morphology of the molecular gas disk, that clearly suggests an ongoing large-scale removal of molecular gas by ram pressure. 
We obtained $\sigma_\mathrm{CO} \approx 30-60$~\kms, possibly indicating that the molecular gas is highly turbulent (see also Fig.~\ref{fig:maps_co10}). 
This is not surprising given the strong perturbations affecting the molecular gas in JW100. 

The bottom right panel in Fig.~\ref{fig:vrot_CO} compares the circular velocity profile of the stellar disk and the molecular gas in JW100. 
We recall that different kinematic centers were used for the stellar and molecular gas components. 
Interestingly, the circular velocity of the approaching side of the molecular gas disk coincides with that of the stellar disk, flattening at $V_\mathrm{circ,CO} \approx 300$~\kms. 
On the contrary, the circular velocity of the receding side keeps on growing and reaches $V_\mathrm{circ,CO} \approx 400$~\kms. 
Similarly to JO206, the asymmetric perturbations on the molecular gas disk and the displaced kinematic center are signatures of edge-on ram pressure stripping \citep{2008Kronberger_b}. 
This is consistent with the fact that JW100 is falling into the cluster at very high velocity and its disk is seen at high inclination by the observer. 
We note that the circular velocity of JW100 rises less steeply than what is typically found in galaxies with similar stellar mass. 
Moreover, Figure~\ref{fig:iband_images} indicates that JW100 hosts a stellar bulge, which is expected to produce high circular velocities in the innermost regions of the galaxy. 
One may argue that the shallow and rather unusual gradient of $V_\mathrm{circ,\star}$ could be due to some contamination from the pressure-supported bulge kinematics, but this can explain only the regions at $R \lesssim 1-2$~kpc (see Appendix~\ref{ap:surface_brightness}). 
Since $V_\mathrm{circ,CO}$ and $V_\mathrm{circ,\star}$ coincide for $R \lesssim 7$~kpc, also the beam smearing does not seem a likely explanation. 
Other alternatives are the presence of a stellar bar aligned with the disk major axis or a dark matter halo with lower-than-average concentration \citep{2015Randriamampandry}.
Disentangling between these possibilities would require a dedicated mass modeling of the system which goes beyond the purpose of this study. 

In conclusion, our results confirm that the molecular gas disk of JW100 is dramatically affected by ram pressure \citep{2020Moretti}. 
The morphology and kinematics of the molecular gas disk indicate strong ram pressure both in the sky plane and along the line of sight. 
Gravitational interactions with other members in the same substructure may play a role, but we speculate that these effects are milder than ram pressure, as the stellar component does not seem to strongly perturb the molecular gas. 

\subsection{Summary}\label{sec:results_summary}
In JO201, the bar dominates the molecular gas kinematics for $R \lesssim 5$~kpc. 
At $R \gtrsim 5$~kpc, the rotation curve gradient is modified by some physical mechanisms, possibly face-on ram pressure. 
Since JO201 belongs to a substructure, a different origin (e.g. tidal interactions) is also possible. 
We tentatively measure radial gas outflows consistent with being due to ram pressure. 
The molecular gas velocity dispersion is about a factor 2 higher than the values typical of field galaxies, suggesting strong turbulent motions. 
Beyond the bar region, this can be either a direct or indirect consequence of ram pressure (or a combination of both). 
Indeed, the ram pressure can directly increase the gas turbulent energy and/or the SFR \citep[e.g.][]{2008Kronberger}, enhancing the turbulence driven by supernova feedback \citep{2020Bacchini_b}. 
This latter scenario seems plausible since the SFR of JO201 is about 2 times higher than field galaxies with similar stellar mass \citep{2018Vulcani_b}.

In JO204, clear bar signatures are found in the molecular gas distribution (central concentration, arm-like overdensities) and kinematics (PVD shape, strong radial motions), and the stellar kinematics (velocity field). 
Although the direction of radial motions remains unclear, a bar-driven inflow is a reasonable hypothesis. 
The molecular gas kinematics is dominated by rotation and, possibly, the bar influence, while the ram pressure is secondary. 
Indeed, the values of both $\sigma_\mathrm{CO}$ and the SFR are consistent with those measured in field galaxies \citep{2018Vulcani_b}.

The molecular gas kinematics in JO206 is dominated by the bar for $R \lesssim 6$~kpc. 
The tentative radial inflows of molecular gas are consistent with this scenario. 
We find clear signatures of edge-on ram pressure stripping for $R > 6$~kpc. 
The molecular gas velocity dispersion is significantly enhanced ($\sigma_\mathrm{CO} \approx$30-40~\kms), a likely consequence of the complex motions due to the combined influence of the bar and the ram pressure. 

In JW100, the molecular gas distribution and kinematics indicate ongoing ram pressure stripping. 
Since JW100 belongs to a substructure, mild tidal interactions may also play a role \citep{2019Poggianti}. 
We detect radial motions compatible with an outward gas flow. 
The velocity dispersion of the molecular gas is quite enhanced ($\sigma_\mathrm{CO} \approx$30-60~\kms), but the SFR is about 2.5 times lower than field galaxies with similar stellar mass \citep{2018Vulcani_b}, favoring the scenario in which the ram pressure directly enhances the gas turbulence.

\section{Comparison with previous works}\label{sec:discussion}

\subsection{The connection between stellar bars and AGN}
At least three out of four galaxies in the GASP-ALMA sample host a stellar bar. 
In JW100, the presence of the bar is difficult to confirm but arguably plausible, given that about 60\% of the disk galaxies with $10 \lesssim \log (M_\star/\mathrm{M}_\odot) \lesssim 11$ are bar hosts \citep{2009Aguerri,2012Masters,2016DiazGarcia}. 
The fraction of barred galaxies may be even higher in the central regions of clusters (e.g. \citealt{199Andersen,2009Barazza,2012MendezAbreu,2014Lansbury,2014Alonso}; but see also \citealt{2022Tawfeek} for a discussion).

The non-axisymmetric potential of a bar can trigger gas inflows by inducing torques and shocks \citep[e.g.][]{1992Athanassoula_a,1992Athanassoula_b,1993SellwoodWilkinson,2014Sellwood,2018Marasco}, often enhancing the molecular gas concentration in the galaxy center \citep[e.g.][]{2005Sheth,2006Regan,2022Yu}. 
Bar-driven inflows of gas may play an important role in fueling the central black hole and triggering the AGN activity \citep[e.g.][]{2013Alonso,2018Alonso,2020RosasGuevara,2022SilvaLima}. 
However, this topic is debated and the excess of AGN-hosts among barred galaxies seem to vanish when the dependence on the galaxy stellar mass and color are considered \citep[e.g.][]{1997Ho,2012Lee}.
Moreover, the bar alone is not always sufficient to efficiently feed the black hole, requiring the contribution of other mechanisms \citep{2008Combes,2014Sellwood,2015Fanali,2015Galloway}. 
Indeed, the molecular gas needs to lose almost all its angular momentum to fall into the black hole \citep[e.g.][]{1993SellwoodWilkinson,1999Krolik,2014Sellwood,2022Capelo}. 
\cite{2014Alonso} found that the location of spiral galaxies within their group or cluster influences both the AGN and bar fraction, suggesting that external mechanisms affecting galaxies in dense environments may contribute to triggering the AGN activity. 
For the specific case of jellyfish galaxies, the external mechanism might be the interaction with the ICM \citep{2017PoggiantiNat,2022Peluso}. 
Indeed, the ram pressure can make the gas lose its angular momentum and eventually move inward \citep[][]{2018RamosMertinez,2020Ricarte,2022Farber,2023Akerman}. 
Thus, the gas could easily reach the region influenced by the bar, which may drag it further inward, perhaps reaching the black hole. 
This picture is in agreement with the enhanced fraction of AGNs in ram pressure stripped galaxies such as those in our sample \citep[e.g.][but see \citealt{2019RomainOliveira} for a different conclusion]{2017PoggiantiNat,2022Peluso}. 
The relative importance of internal mechanisms, such as bars, and external processes, such as ram pressure, in fueling the AGN activity is a compelling and debated topic \citep{2018Alonso,2020Kim,2022Boselli}, which would require higher statistics than the four galaxies studied here. 
This task goes beyond the scope of this paper and we leave it to future studies. 

\subsection{Comparison with Virgo galaxies}\label{sec:discussion_vertico}
There is growing evidence that the ram pressure can affect the molecular gas in cluster galaxies. 
In this section, we compare the GASP-ALMA sample with the galaxies in Virgo cluster, in order to increase the statistics. 
Other cases of ram pressure affecting the molecular disk have been found in Coma \citep{2017Jachym}, Norma \citep{2014Jachym}, and Fornax \citep{2019Zabel}, just to mention some examples. 

\cite{2017Lee} studied the molecular gas kinematics in three disk galaxies. 
These author did not find clear signs of molecular gas stripping, but they showed that the morphological and kinematical disturbances in the molecular and atomic gas disks are closely related to each other, suggesting that the molecular gas can be also affected by strong ram pressure even if it is not globally stripped. 
They also ascribed the perturbation in the innermost regions of their galaxies to the presence of a stellar bar, rather than to ram pressure. 
As discussed in Sects.~\ref{sec:results_jo201} and~\ref{sec:results_jo204}, our results for JO201 and JO204 are consistent with \cite{2017Lee}'s findings. 
Interestingly, all the molecular gas disks in \cite{2017Lee} sample are kinematically lopsided, at least to some degree, possibly indicating that the molecular gas was either accelerated or decelerated by ram pressure \citep[see also][]{2020Cramer}. 
They also found CO clumps that are kinematically decoupled from the molecular gas disk, suggesting that this gas was displaced by the ram pressure, as in the case of JO206 (see Sect.~\ref{sec:results_jo206}). 

Recently, \cite{2021Brown} presented the first results of the Virgo Environment Traced in CO (VERTICO) survey, which maps CO emission in 51 galaxies in Virgo cluster using ALMA. 
This authors derived the mass-size relation for the molecular gas disk for VERTICO galaxies. 
They showed that the scatter in the relation is minimized if the disk size is defined as the radius where the azimuthally-averaged H$_2$ surface density reaches $\Sigma_\mathrm{H_2}=5~M_\odot \mathrm{pc}^{-2}$ ($R_5$). 
As a control sample, \cite{2021Brown} used the field galaxies in the Heterodyne Receiver Array CO Line Extragalactic Survey \citep[HERACLES,][]{2009Leroy}. 
\cite{2021Brown} found that the best-fit relations for the galaxies in Virgo and in the field are consistent. 
They concluded that $R_{5}$-M$_\mathrm{H_2}$ relation does not significantly depend on the environment, in agreement with the studies on the HI size–mass relation \citep{2016Wang,2019Stevens}, and that galaxies affected by environmental processes move along the size-mass relation rather than deviating from it.

\begin{figure}
	\centering
	\includegraphics[width=1\columnwidth]{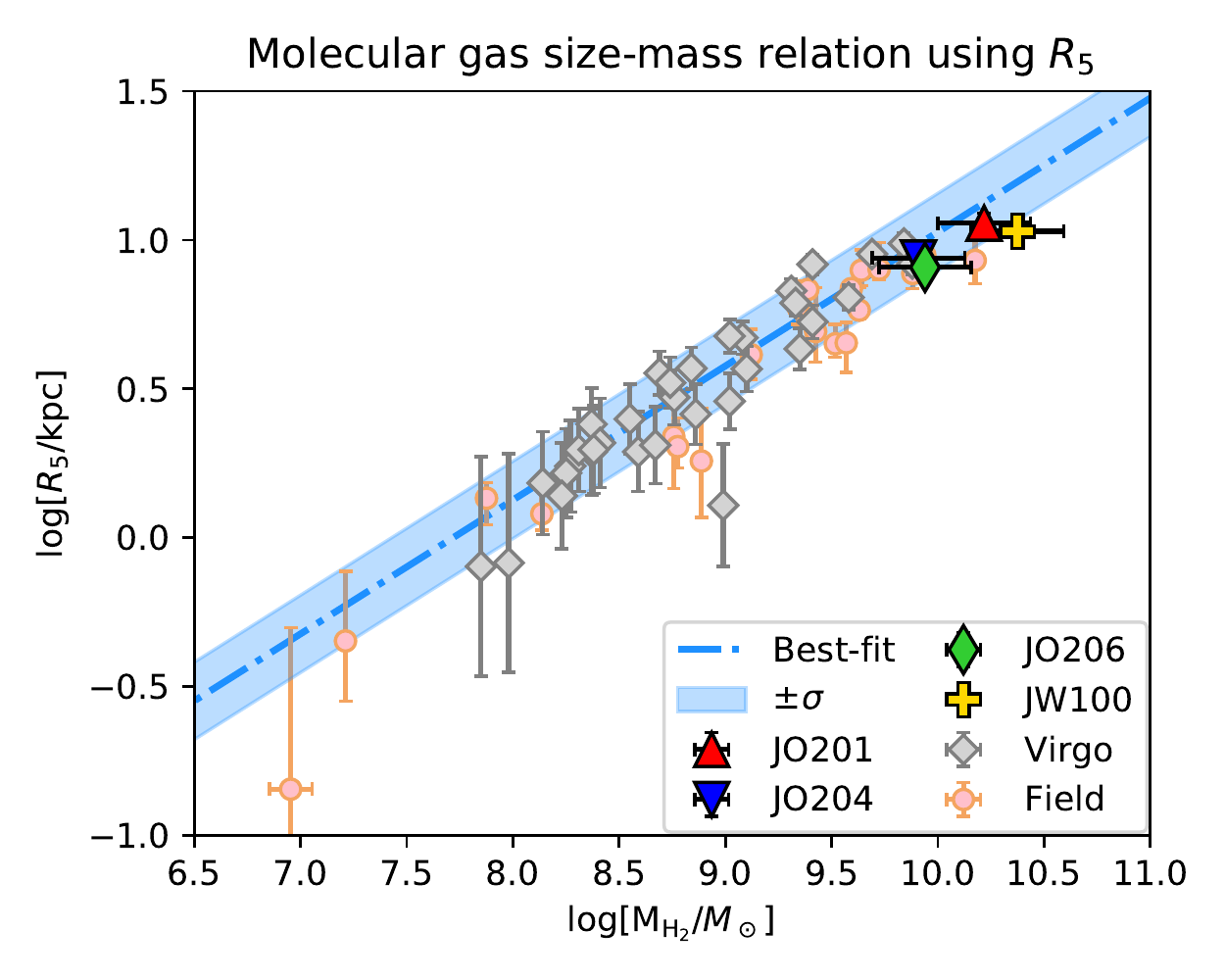}
	\caption{
		Molecular gas mass-size relation based on $R_\mathrm{5}$ (see Sect.~\ref{sec:discussion_vertico}). 
		Each galaxy in the GASP-ALMA sample is indicated by a colored symbol.  
		Grey diamonds and pink points show galaxies in the Virgo cluster and nearby field galaxies (see text), respectively. 
		The best-fit relation obtained by \cite{2021Brown} for the VERTICO and HERACLES samples is shown by the dash-dotted line, while its scatter is represented by the shaded area. 
		}
	\label{fig:vertico_mass_size}
\end{figure}

In Fig.~\ref{fig:vertico_mass_size}, we compare our galaxies with the $R_{5}$-M$_\mathrm{H_2}$ relation from \cite{2021Brown}. 
We assumed the Milky Way CO-to-H$_2$ conversion factor for consistency with \cite{2021Brown}. 
Our galaxies are within the scatter of the $R_{5}$-M$_\mathrm{H_2}$ relation derived by \cite{2021Brown}, confirming that this scaling relation does not show any clear dependence on environment, even in extreme ram pressure cases as the galaxies of our sample. 
We note that the GASP-ALMA sample tend to be slightly below the relation, suggesting that the molecular gas distribution is more centrally concentrated than the average for these samples. 
This can be due to the combined effect of stellar bars, which tend to increase the gas density in the inner regions of the disk \citep[e.g.][]{2004KormendyKennicutt}, and ram pressure, which compresses the molecular disk. 
The GASP-ALMA galaxies stand out against the other two samples because of their high M$_\mathrm{H_2}$, being up to $\approx 0.5$~dex more massive than the Virgo and control samples \citep[see also][]{2020Moretti,2020MorettiLetter}. 
On the other hand, it has been shown that our galaxies are up to 50\% deficient in HI with respect to field galaxies with similar mass and size  \citep[][]{2019Ramatsoku,2020Ramatsoku,2020Deb,2021HealyDeb,2022Deb}. 
Taken together, these results suggest an unusually efficient conversion of HI to H$_2$ \citep{2020MorettiLetter}.
These properties are in agreement with the recent results by \cite{2022Zabel} for Virgo galaxies. 
They found that the galaxies showing clear signs of ongoing ram pressure stripping affecting the HI disk are from H$_2$-normal to H$_2$-rich. 
This was interpreted as an indication that ram pressure stripping is not effective at reducing global molecular gas fractions on the timescales in which
such features are still clearly visible. 
This is likely because the stripping is less severe on H$_2$ than on HI, as the molecular gas is denser and more gravitationally bound to the galaxy than the atomic gas \citep{2017Lee,2022Boselli}. 
The atomic gas disk in our galaxies shows indeed signs of truncation and the ram pressure stripping is much more dramatic than for the molecular gas disk \citep[][]{2019Ramatsoku,2020Ramatsoku,2020Deb,2022Deb}

\subsection{Baryonic Tully-Fisher relation}
\begin{figure}
	\centering
	{\includegraphics[width=1.\columnwidth]{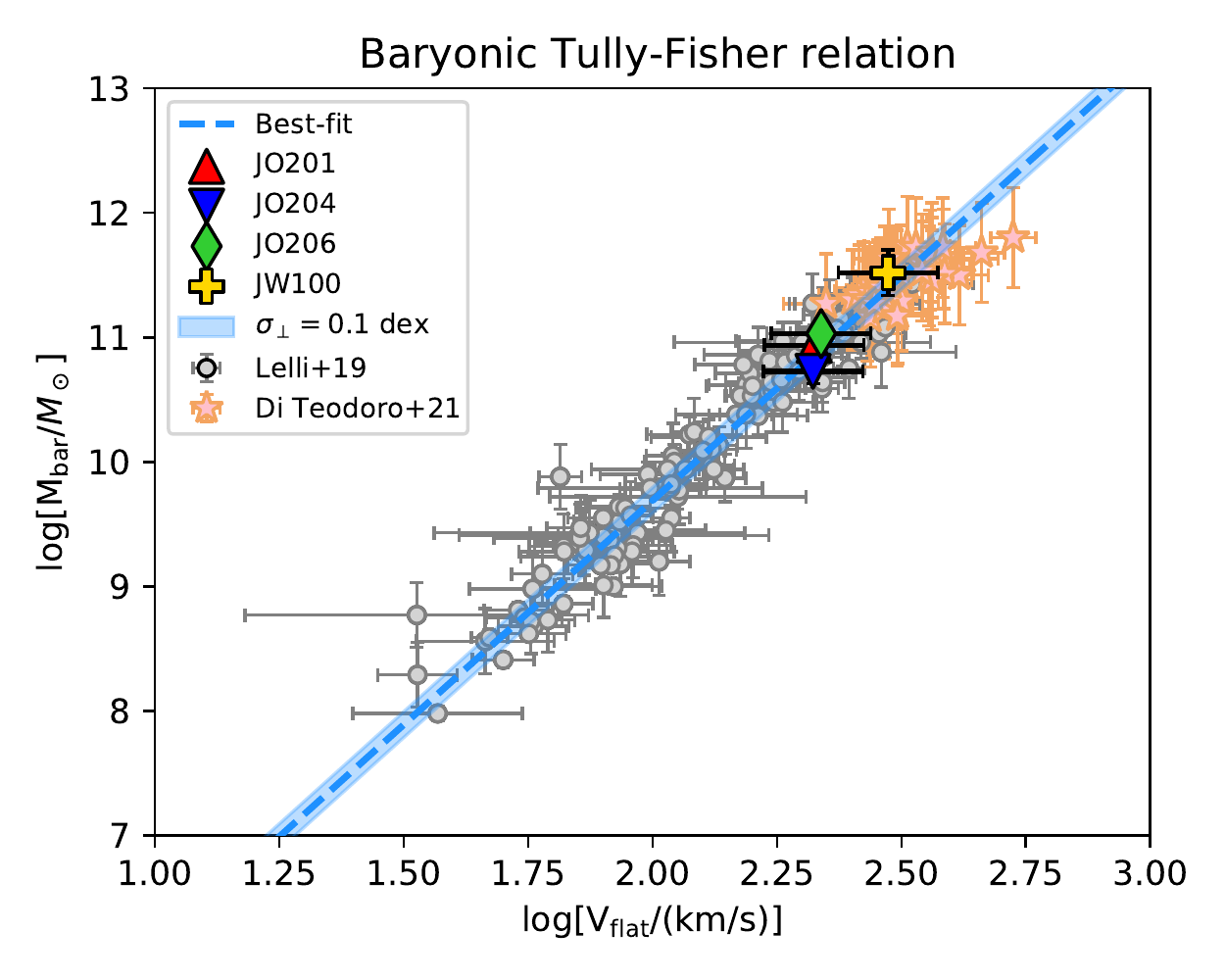}}
	\caption{Baryonic Tully-Fisher relation for the four galaxies in the GASP-ALMA sample (triangles, diamond, and cross). 
		The grey points show the spiral and dwarf galaxies from \cite{2019Lelli}, while the pink stars are for the massive disks from \cite{2021bDiTeodoro}. 
		The dashed line is their best-fit relation with the shaded area showing the orthogonal intrinsic scatter.}
	\label{fig:btfr}
\end{figure}

Rotation curves of disk galaxies are typically used to derive fundamental scaling relations \citep[e.g.][]{2001Verheijen,2016Lelli,2017Ponomareva,2017Iorio,2018Posti,2021aManceraPina,2021bManceraPina,2021DiTeodoro}. 
In particular, the baryonic Tully-Fisher relation (hereafter BTFR) is a very tight correlation between the mass of baryons and the rotation velocity of galaxies, being a useful test-case to check the robustness of the stellar rotation derived in this work. 
The BTFR is usually derived using HI rotation curves, as the atomic gas disk is the most extended baryonic component, allowing to probe the flat part of the galaxy rotation curve. 
In jellyfish galaxies, the atomic gas disk is stripped or truncated by the ram pressure and the HI kinematics is strongly perturbed, hampering the usage of HI observations to study scaling relations. 
The ionized gas is not a good alternative to HI, as not only it is less spatially extended but also more diffuse and thus easier to strip. 
The results of this work suggest that the molecular gas is more resilient to ram pressure, but its spatial extent is still very limited. 
Therefore, the stellar component is likely the best way to derive scaling relations in the case of jellyfish galaxies, provided that observations with high spatial resolution and sensitivity are available. 
The GASP sample is ideal to perform this exercise, thanks to the high spatial resolution and sensitivity of the MUSE observations. 
In Fig.~\ref{fig:btfr}, we show that the galaxies in the GASP-ALMA sample closely follow the BTFR derived by \cite{2021bDiTeodoro} using a sample of about 200 galaxies from high-mass disks to dwarf galaxies \citep{2019Lelli}. 
We calculated the velocity in the flat part of the rotation curve ($V_\mathrm{flat}$) as the average of the outermost 5 measurements of the stellar rotation velocity (see Fig.~\ref{fig:vrot_CO}). 
The baryonic mass was calculated as $M_\mathrm{bar} = M_\star + 1.33 \left( M_\mathrm{HI} + M_\mathrm{H_2} \right)$, where the masses of atomic gas ($M_\mathrm{HI}$) and molecular gas ($M_\mathrm{H_2}$) are taken from Table~\ref{tab:galaxyprops} and the multiplicative factor 1.33 accounts for the Helium content. 
We checked that considering only the gas mass within the stellar disk or the total gas mass (including the gas in the stripped tail) does not change the results, as the gas mass is largely dominated by molecular gas component which is mostly concentrated within the galaxy disk. 
We also checked that our galaxies fall on the stellar Tully-Fisher relation (not shown here), which is not surprising given that the baryonic mass is largely dominated by the stellar component. 
These simple tests indicate that the GASP sample can be used to study important scaling relations of baryons and, potentially, dark matter \citep[e.g.][]{2016Lelli,2018Posti,2021aManceraPina,2022DiTeodoro}. 
This will be addressed in future work by fully exploiting the richness and quality of the MUSE observations obtained with the GASP survey (Bacchini et al., in preparation). 

\section{Summary and conclusions}\label{sec:conclusions}
Galaxies in dense environments, such as clusters, can be affected by the ram pressure due to the interaction with the ICM. 
This process leaves the stellar disk essentially unperturbed, but it can have a strong impact on the morphology, kinematics and overall gas content, with important consequences on the evolution of galaxies \citep{2021Cortese}. 
In this context, jellyfish galaxies are ideal cases to study the impact of ram pressure on the gas components. 
In this work, we have studied the distribution and kinematics of the molecular gas in a sample of four jellyfish galaxies in the GASP sample \citep[][]{2017Poggianti}. 
These galaxies were observed with ALMA to detect the CO(1--0) and CO(2--1) emission \citep{2020Moretti,2020MorettiLetter}. 
Thanks to the wealth of information obtained from MUSE and ALMA observations provided by the GASP survey, we could analyze the stellar and CO distribution and kinematics. 
We used the software 3DB based on the tilted-ring approach to model the stellar velocity field and the CO emission line datacubes. 
We identified the gas with anomalous velocity that cannot be explained by a rotation disk and used the information on the stellar distribution and kinematics to understand the origin of this anomalous gas. 
We reached the following conclusions. 
\begin{enumerate}	
	\item At least three (JO201, JO204, and JO206) out of four galaxies in the GASP-ALMA sample are barred. 
	In JO201 and JO206, the bars aligned with the disk major axis are visible in the $I$-band images and explain the shallow gradient of the circular velocity in the inner regions of these galaxies. 
	In JO204, various bar signatures are found in the distribution of the molecular gas and the kinematics of both the molecular gas and the stars. 
	In JW100, the disk inclination and dust obscuration do not allow us to unambiguously identify a bar. 
			
	\item The molecular gas kinematics in JO201 and JO206 are mainly dominated by non-circular motions in the region influenced by the bar, while the ram pressure becomes important at larger galactocentric distance. 
	The ram pressure plays a secondary role for the molecular gas kinematics of JO204, which is mainly rotation-dominated. 
	Clear indications of molecular gas stripping are found in two galaxies, JO206 and JW100. 
	In JO206, some molecular gas is detached and kinematically decoupled from the main disk. 
	In JW100, the molecular gas disk is displaced with respect to the stellar disk and its kinematics is strongly perturbed. 
	Since JO201 and JW100 belong to cluster substructures, other mechanisms than ram pressure might be also at play. 
	
	\item Radial flows of molecular gas are manifestly present in two galaxies (JO204 and JW100), but this is less clear in the other two objects (JO201 and JO206). 
	These gas flows are consistent with being bar-driven inflows in JO206 and ram pressure-driven outflows in JO201 and JW100. 
	The direction of radial motions remains unclear for JO204. 
	
	\item The molecular gas velocity dispersion in JO201, JO206, and JW100 tends to be enhanced with respect to field galaxies, suggesting that the gas is very turbulent. 
	In the case of JO201 and JO206, this can be explained by the complex motions induced by the bar within its region of influence or, beyond the bar region, by the the ram pressure, which can enhance the gas turbulence directly and/or by increasing the SFR. 
	In the case of JW100, the most likely scenario is that the gas turbulence is directly enhanced by the ram pressure. 
	
	\item Our galaxies fall within the scatter of the molecular gas mass-size relation derived for field and Virgo galaxies by \citep{2021Brown}, confirming that the relation is essentially independent of environment. 
\end{enumerate}
Overall, our results are consistent with a scenario in which the molecular gas is affected by ram pressure on different timescales and less severely than the atomic and ionized gas, likely because the molecular gas is denser and more gravitationally bound to the galaxy than the other gas phases. 
The galaxies in the GASP-ALMA sample host an AGN \citep{2017PoggiantiNat,2022Peluso}. 
Both stellar bars and ram pressure can contribute to efficiently drive molecular gas towards the galaxy center, possibly feeding the central black hole and triggering the nuclear activity. 
Since the relative importance of bars and ram pressure in fueling the AGN has not been fully understood yet, we hope that our work may foster future studies. 
In this work, we have shown that high-resolution observations of the molecular gas emission can be very useful in identifying stellar bars and radial flows. 
Future effort will be devoted to further study the bar-AGN connection by expanding the GASP-ALMA sample. 
Moreover, we have shown that the GASP sample is potentially very useful to investigate the impact of environment on scaling relations of galaxies. 
In future work, we plan to address this topic by fully exploiting the richness and quality of the MUSE observations obtained with the GASP survey. 

\begin{acknowledgments}
We thank the referee for carefully reading the paper and providing constructive comments. 
This paper makes use of the following ALMA data: ADS/JAO.ALMA\#2017.1.00496.S.
ALMA is a partnership of ESO (representing its member states), NSF (USA) and NINS (Japan), together with NRC (Canada) and NSC and ASIAA (Taiwan), in cooperation with the Republic of Chile. The Joint ALMA Observatory is operated by ESO, AUI/NRAO and NAOJ. 
CB acknowledges financial support from the European Research Council (ERC) under the European Union's Horizon 2020 research and innovation programme (grant agreement No. 833824).
CB would like to thank E. Di Teodoro, F. Rizzo, and F. Fraternali, for useful discussions and the help with the kinematic modeling. 

\facility{ALMA, MUSE@VLT}
\software{$^\mathrm{3D}$\textsc{Barolo} \citep{2015DiTeodoro}, APLpy \citep{2012Robitaille}, Astropy \citep{2013Astropy,2018Astropy}.}
\end{acknowledgments}

\appendix

\section{Modeling of the I-band surface brightness profiles}\label{ap:surface_brightness}
We modeled the $I$-band surface brightness profile in order to estimate i) the size of the region where the bulge contribution may contaminate the velocity field of the stellar disk, and ii) the exponential disk scale length. 
The $I$-band surface brightness profile is fitted using a two-component model including the disk and the bulge. 
The disk is modeled using an exponential profile
\begin{equation}\label{eq:exp_disc}
I_\mathrm{d}(R) = I_\mathrm{d,0} \exp{\left(-\frac{R}{R_\mathrm{d}} \right)} ,
\end{equation}
where $I_\mathrm{d,0}$ and $R_\mathrm{d}$ are respectively the central surface brightness and the exponential disk scale length. 
The bulge component is modeled using a S\'{e}rsic profile
\begin{equation}
I_\mathrm{b}(R) = I_\mathrm{b,e} \exp{\left[-k \left(\frac{R}{R_\mathrm{b,e}}-1 \right)^\frac{1}{n} \right]} ,
\end{equation}
where $k=2n-1/3+4/(405n)$, $n$ is the Sersic index, and $R_\mathrm{b,e}$ is the effective radius. 
We excluded from the fit the innermost 1\arcsec, which grossly corresponds to the PSF size. 
The $I$-band surface brightness profile in the innermost regions of JO204 shows no feature indicating the presence of a central component. 
We therefore considered only the exponential disk in the modeling of JO204. 
For the sake of simplicity, we do not attempt to model complex features such as bars or rings. 

\begin{figure*}
	\centering
	{\includegraphics[width=1.\textwidth]{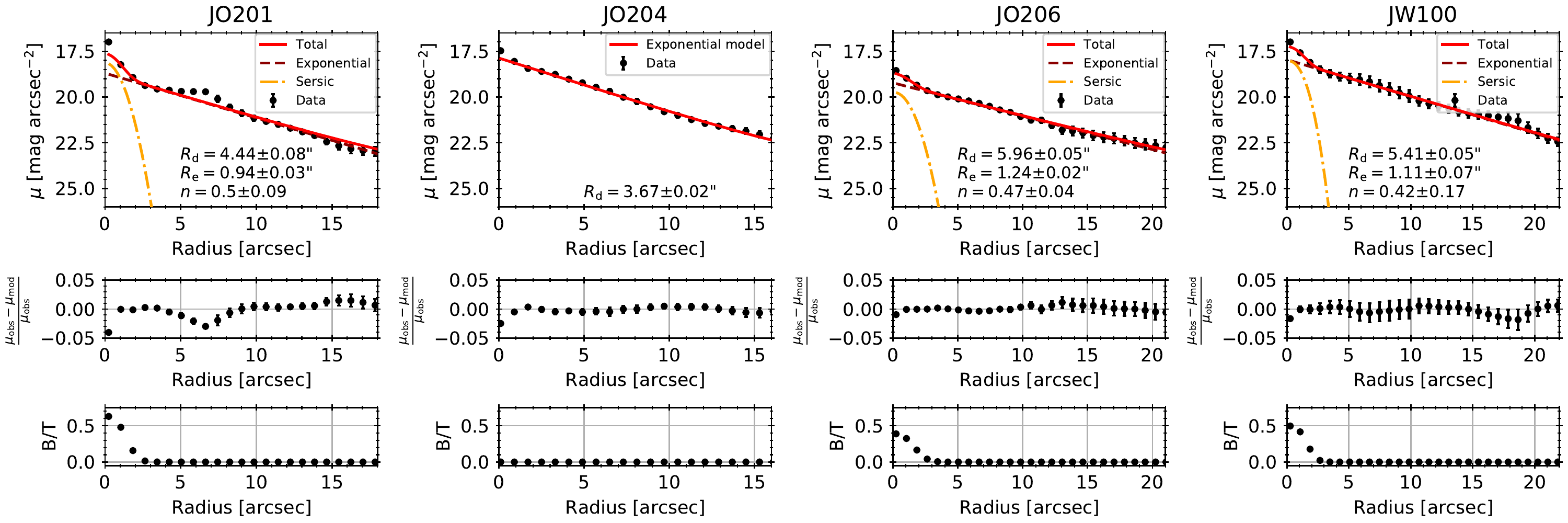}}
	\caption{Top: Radial profiles of the observed $I$-band surface brightness (black points) and the best-fit models (red curve) for each galaxy in our sample. 
	The contribution of an exponential disc model is shown by the dark red dashed lines, while the Sersic model for the bulge component is indicated by the dash-dotted orange curves. 
	The best-fit values of $R_\mathrm{d}$, $R_\mathrm{b,e}$ and $n$ are reported in each panel. 
	Middle: normalized residuals between the observed $I$-band surface brightness and the best-fit model. 
	Bottom: bulge-to-total luminosity ratio.}
	\label{fig:surf_bright_dens_fits}
\end{figure*}

The top panels in Fig.~\ref{fig:surf_bright_dens_fits} shows, for each galaxy, the observed radial profile of the surface brightness and the best-fit models with the corresponding. 
In each panel, we also report the best-fit values of $R_\mathrm{d}$, $n$, and $R_\mathrm{b,e}$. 
The normalized residuals between the observed profiles and the models are shown by the middle panels in Fig.~\ref{fig:surf_bright_dens_fits}. 
The bottom panels of Fig.~\ref{fig:surf_bright_dens_fits} provide the radial profiles bulge-to-total luminosity ratio, showing that the bulge contribution is less than 20\% beyond $R\approx 1-2$~kpc, even in the $I$-band. 
The bump at $5\mathrm{\arcsec} \lesssim R \lesssim 8\mathrm{\arcsec}$ in the surface brightness profile and residuals of JO201 is due to the bar emission \citep{2000Knapen,2004KormendyKennicutt}, which we neglected for simplicity (but see \citealt{2019George} for an analysis of JO201 including the bar). 

\section{Best-fit models of the stellar velocity field}\label{ap:stellar_kins}
\begin{figure*}
	\centering
	{\includegraphics[width=1.\columnwidth]{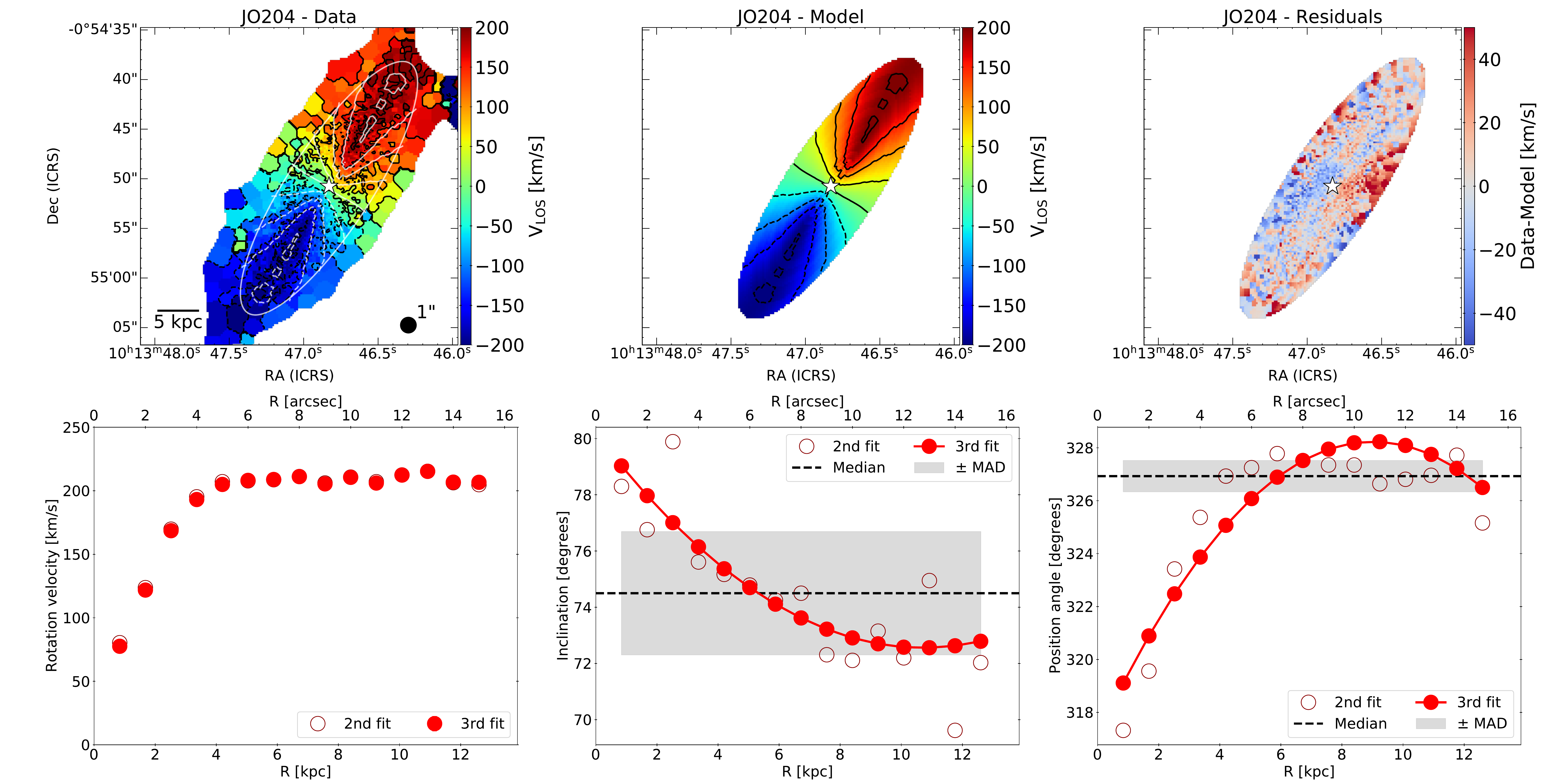}}
	\caption{Same as Fig.~\ref{fig:2dmodel_jo201} but for JO204.}
	\label{fig:2dmodel_jo204}
\end{figure*}

\begin{figure*}
	\centering
	{\includegraphics[width=1.\columnwidth]{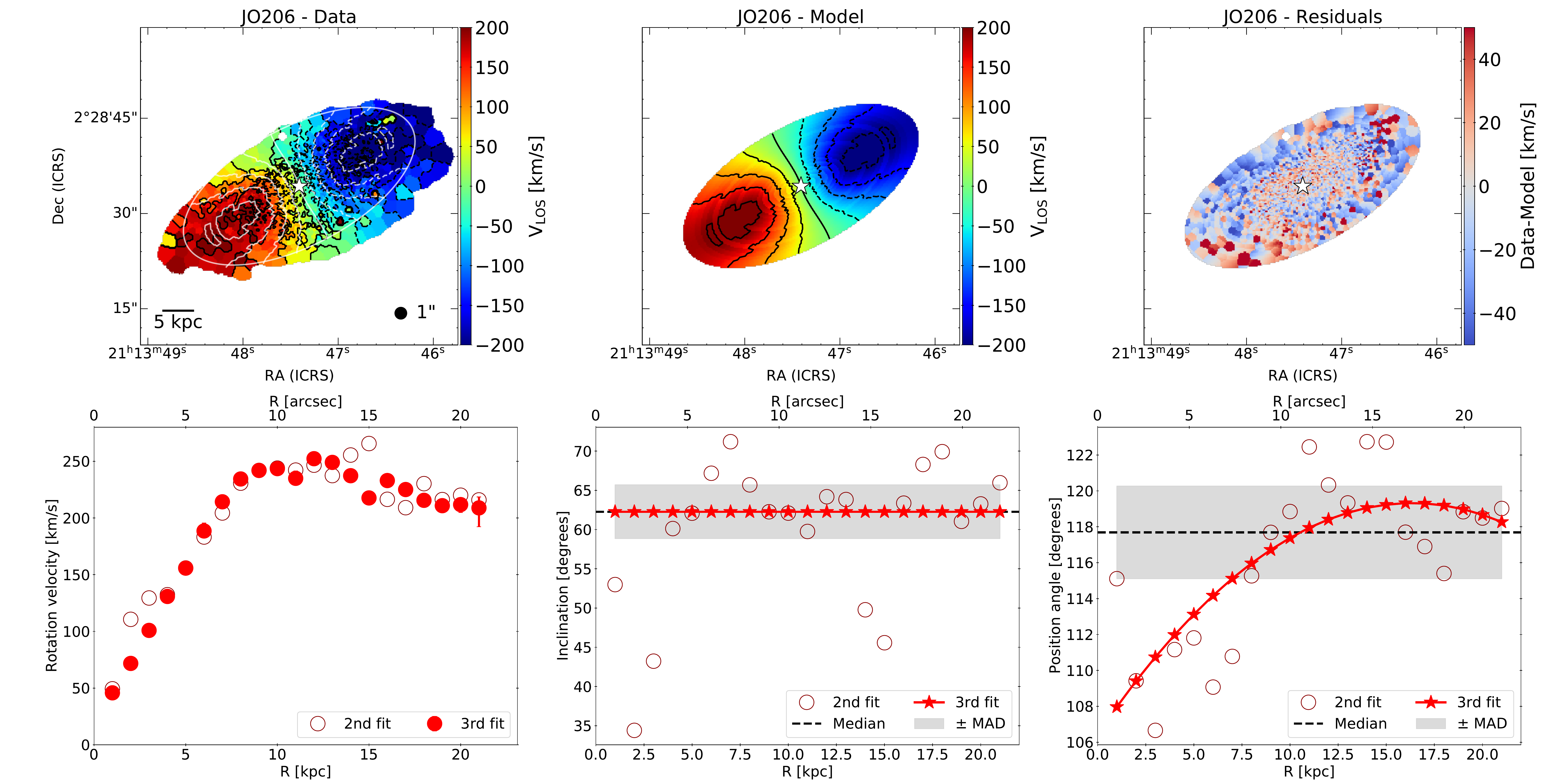}}
	\caption{Same as Fig.~\ref{fig:2dmodel_jo201} but for JO206.}
	\label{fig:2dmodel_jo206}
\end{figure*}

\begin{figure*}
	\centering
	{\includegraphics[width=1.\columnwidth]{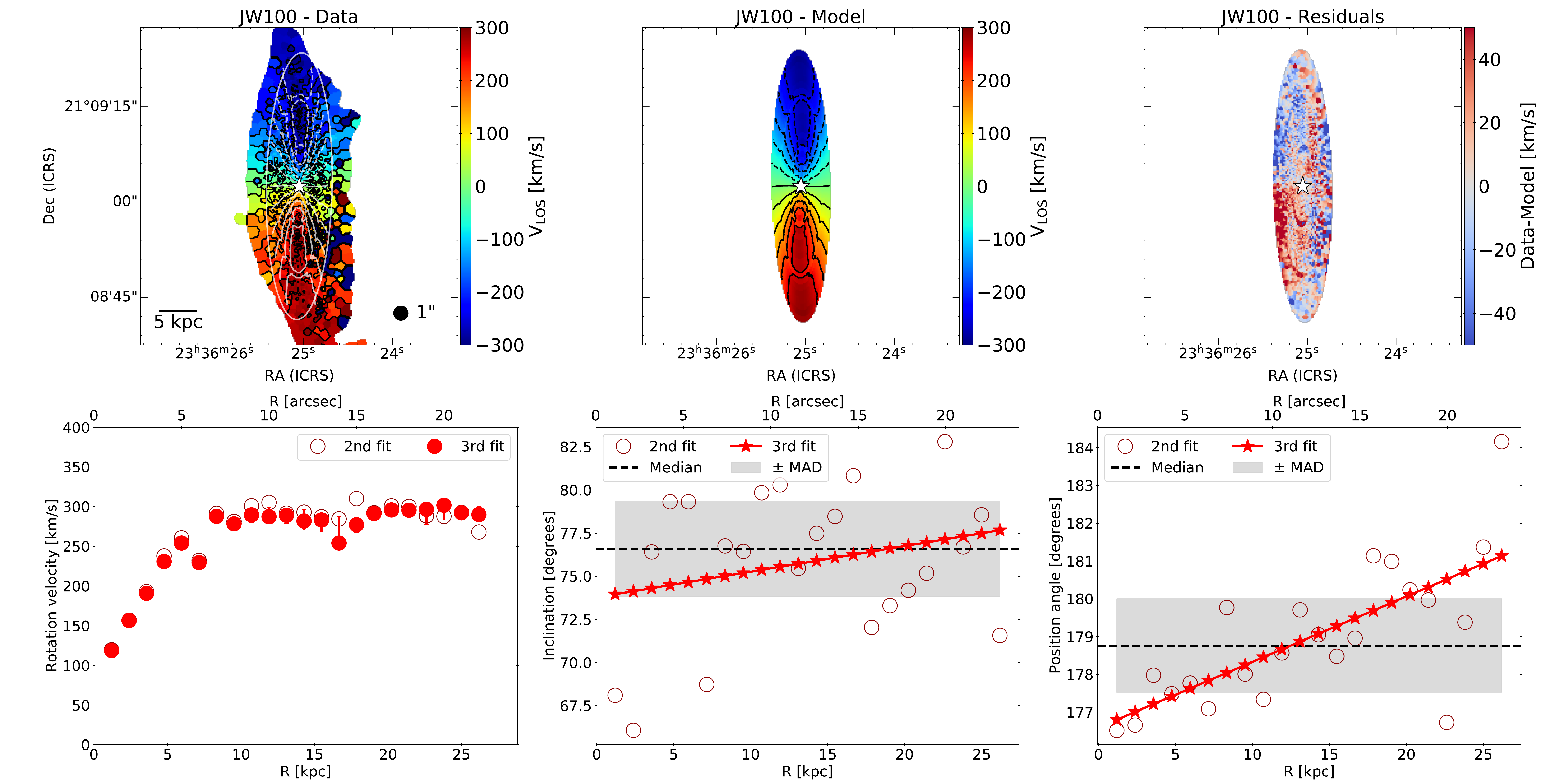}}	
	\caption{Same as Fig.~\ref{fig:2dmodel_jo201} but for JW100.}
	\label{fig:2dmodel_jw100}
\end{figure*}

This section provides the best-fit model of the stellar disk for JO204, JO206, and JW100. 
The top panels in Figs.~\ref{fig:2dmodel_jo204},~\ref{fig:2dmodel_jo206}, and~\ref{fig:2dmodel_jw100} show the observed stellar velocity field, the best-fit model, and the map of the residuals. 
The bottom panels display the stellar rotation curve and the radial profiles of the PA and inclination. 
Overall, the stellar kinematics is well reproduced by our models. 
However, the residual map of JO204 clearly shows a pattern in the central regions of the disk, which likely indicates the presence of a stellar bar (see also Sect.~\ref{sec:results_jo204}). 
The residual map of JW100 highlights some regions where the model does not fully reproduces the observations. 
The origin of these differences is tricky to understand and may be due to asymmetric dust observation or the warp along the line of sight, or both (see also Sect.~\ref{sec:results_jw100}). 

%
\bibliography{sample631}{}

\begin{thebibliography}{}
\expandafter\ifx\csname natexlab\endcsname\relax\def\natexlab#1{#1}\fi
\providecommand{\url}[1]{\href{#1}{#1}}
\providecommand{\dodoi}[1]{doi:~\href{http://doi.org/#1}{\nolinkurl{#1}}}
\providecommand{\doeprint}[1]{\href{http://ascl.net/#1}{\nolinkurl{http://ascl.net/#1}}}
\providecommand{\doarXiv}[1]{\href{https://arxiv.org/abs/#1}{\nolinkurl{https://arxiv.org/abs/#1}}}

\bibitem[{{Aguerri} {et~al.}(2009){Aguerri}, {M{\'e}ndez-Abreu}, \&
  {Corsini}}]{2009Aguerri}
{Aguerri}, J.~A.~L., {M{\'e}ndez-Abreu}, J., \& {Corsini}, E.~M. 2009, \aap,
  495, 491, \dodoi{10.1051/0004-6361:200810931}

\bibitem[{{Akerman} {et~al.}(2023){Akerman}, {Tonnesen}, {Poggianti}, {Smith},
  \& {Marasco}}]{2023Akerman}
{Akerman}, N., {Tonnesen}, S., {Poggianti}, B.~M., {Smith}, R., \& {Marasco},
  A. 2023, arXiv e-prints, arXiv:2301.09652, \dodoi{10.48550/arXiv.2301.09652}

\bibitem[{{Alatalo} {et~al.}(2013){Alatalo}, {Davis}, {Bureau}, {Young},
  {Blitz}, {Crocker}, {Bayet}, {Bois}, {Bournaud}, {Cappellari}, {Davies}, {de
  Zeeuw}, {Duc}, {Emsellem}, {Khochfar}, {Krajnovi{\'c}}, {Kuntschner},
  {Lablanche}, {Morganti}, {McDermid}, {Naab}, {Oosterloo}, {Sarzi}, {Scott},
  {Serra}, \& {Weijmans}}]{2013Alatalo}
{Alatalo}, K., {Davis}, T.~A., {Bureau}, M., {et~al.} 2013, \mnras, 432, 1796,
  \dodoi{10.1093/mnras/sts299}

\bibitem[{{Alonso} {et~al.}(2013){Alonso}, {Coldwell}, \&
  {Lambas}}]{2013Alonso}
{Alonso}, M.~S., {Coldwell}, G., \& {Lambas}, D.~G. 2013, \aap, 549, A141,
  \dodoi{10.1051/0004-6361/201220117}

\bibitem[{{Alonso} {et~al.}(2018){Alonso}, {Coldwell}, {Duplancic}, {Mesa}, \&
  {Lambas}}]{2018Alonso}
{Alonso}, S., {Coldwell}, G., {Duplancic}, F., {Mesa}, V., \& {Lambas}, D.~G.
  2018, \aap, 618, A149, \dodoi{10.1051/0004-6361/201832796}

\bibitem[{{Alonso} {et~al.}(2014){Alonso}, {Coldwell}, \&
  {Lambas}}]{2014Alonso}
{Alonso}, S., {Coldwell}, G., \& {Lambas}, D.~G. 2014, \aap, 572, A86,
  \dodoi{10.1051/0004-6361/201424523}

\bibitem[{{Andersen}(1996)}]{199Andersen}
{Andersen}, V. 1996, \aj, 111, 1805, \dodoi{10.1086/117918}

\bibitem[{{Astropy Collaboration} {et~al.}(2013){Astropy Collaboration},
  {Robitaille}, {Tollerud}, {Greenfield}, {Droettboom}, {Bray}, {Aldcroft},
  {Davis}, {Ginsburg}, {Price-Whelan}, {Kerzendorf}, {Conley}, {Crighton},
  {Barbary}, {Muna}, {Ferguson}, {Grollier}, {Parikh}, {Nair}, {Unther},
  {Deil}, {Woillez}, {Conseil}, {Kramer}, {Turner}, {Singer}, {Fox}, {Weaver},
  {Zabalza}, {Edwards}, {Azalee Bostroem}, {Burke}, {Casey}, {Crawford},
  {Dencheva}, {Ely}, {Jenness}, {Labrie}, {Lim}, {Pierfederici}, {Pontzen},
  {Ptak}, {Refsdal}, {Servillat}, \& {Streicher}}]{2013Astropy}
{Astropy Collaboration}, {Robitaille}, T.~P., {Tollerud}, E.~J., {et~al.} 2013,
  \aap, 558, A33, \dodoi{10.1051/0004-6361/201322068}

\bibitem[{{Astropy Collaboration} {et~al.}(2018){Astropy Collaboration},
  {Price-Whelan}, {Sip{\H{o}}cz}, {G{\"u}nther}, {Lim}, {Crawford}, {Conseil},
  {Shupe}, {Craig}, {Dencheva}, {Ginsburg}, {VanderPlas}, {Bradley},
  {P{\'e}rez-Su{\'a}rez}, {de Val-Borro}, {Aldcroft}, {Cruz}, {Robitaille},
  {Tollerud}, {Ardelean}, {Babej}, {Bach}, {Bachetti}, {Bakanov}, {Bamford},
  {Barentsen}, {Barmby}, {Baumbach}, {Berry}, {Biscani}, {Boquien}, {Bostroem},
  {Bouma}, {Brammer}, {Bray}, {Breytenbach}, {Buddelmeijer}, {Burke},
  {Calderone}, {Cano Rodr{\'\i}guez}, {Cara}, {Cardoso}, {Cheedella}, {Copin},
  {Corrales}, {Crichton}, {D'Avella}, {Deil}, {Depagne}, {Dietrich}, {Donath},
  {Droettboom}, {Earl}, {Erben}, {Fabbro}, {Ferreira}, {Finethy}, {Fox},
  {Garrison}, {Gibbons}, {Goldstein}, {Gommers}, {Greco}, {Greenfield},
  {Groener}, {Grollier}, {Hagen}, {Hirst}, {Homeier}, {Horton}, {Hosseinzadeh},
  {Hu}, {Hunkeler}, {Ivezi{\'c}}, {Jain}, {Jenness}, {Kanarek}, {Kendrew},
  {Kern}, {Kerzendorf}, {Khvalko}, {King}, {Kirkby}, {Kulkarni}, {Kumar},
  {Lee}, {Lenz}, {Littlefair}, {Ma}, {Macleod}, {Mastropietro}, {McCully},
  {Montagnac}, {Morris}, {Mueller}, {Mumford}, {Muna}, {Murphy}, {Nelson},
  {Nguyen}, {Ninan}, {N{\"o}the}, {Ogaz}, {Oh}, {Parejko}, {Parley}, {Pascual},
  {Patil}, {Patil}, {Plunkett}, {Prochaska}, {Rastogi}, {Reddy Janga},
  {Sabater}, {Sakurikar}, {Seifert}, {Sherbert}, {Sherwood-Taylor}, {Shih},
  {Sick}, {Silbiger}, {Singanamalla}, {Singer}, {Sladen}, {Sooley},
  {Sornarajah}, {Streicher}, {Teuben}, {Thomas}, {Tremblay}, {Turner},
  {Terr{\'o}n}, {van Kerkwijk}, {de la Vega}, {Watkins}, {Weaver}, {Whitmore},
  {Woillez}, {Zabalza}, \& {Astropy Contributors}}]{2018Astropy}
{Astropy Collaboration}, {Price-Whelan}, A.~M., {Sip{\H{o}}cz}, B.~M., {et~al.}
  2018, \aj, 156, 123, \dodoi{10.3847/1538-3881/aabc4f}

\bibitem[{{Athanassoula}(1992{\natexlab{a}})}]{1992Athanassoula_b}
{Athanassoula}, E. 1992{\natexlab{a}}, \mnras, 259, 345,
  \dodoi{10.1093/mnras/259.2.345}

\bibitem[{{Athanassoula}(1992{\natexlab{b}})}]{1992Athanassoula_a}
---. 1992{\natexlab{b}}, \mnras, 259, 328, \dodoi{10.1093/mnras/259.2.328}

\bibitem[{{Bacchini} {et~al.}(2019{\natexlab{a}}){Bacchini}, {Fraternali},
  {Iorio}, \& {Pezzulli}}]{2019Bacchini_a}
{Bacchini}, C., {Fraternali}, F., {Iorio}, G., \& {Pezzulli}, G.
  2019{\natexlab{a}}, \aap, 622, A64, \dodoi{10.1051/0004-6361/201834382}

\bibitem[{{Bacchini} {et~al.}(2020{\natexlab{a}}){Bacchini}, {Fraternali},
  {Iorio}, {Pezzulli}, {Marasco}, \& {Nipoti}}]{2020Bacchini_a}
{Bacchini}, C., {Fraternali}, F., {Iorio}, G., {et~al.} 2020{\natexlab{a}},
  \aap, 641, A70, \dodoi{10.1051/0004-6361/202038223}

\bibitem[{{Bacchini} {et~al.}(2020{\natexlab{b}}){Bacchini}, {Fraternali},
  {Pezzulli}, \& {Marasco}}]{2020Bacchini_b}
{Bacchini}, C., {Fraternali}, F., {Pezzulli}, G., \& {Marasco}, A.
  2020{\natexlab{b}}, \aap, 644, A125, \dodoi{10.1051/0004-6361/202038962}

\bibitem[{{Bacchini} {et~al.}(2019{\natexlab{b}}){Bacchini}, {Fraternali},
  {Pezzulli}, {Marasco}, {Iorio}, \& {Nipoti}}]{2019Bacchini_b}
{Bacchini}, C., {Fraternali}, F., {Pezzulli}, G., {et~al.} 2019{\natexlab{b}},
  \aap, 632, A127, \dodoi{10.1051/0004-6361/201936559}

\bibitem[{{Barazza} {et~al.}(2009){Barazza}, {Jablonka}, {Desai}, {Jogee},
  {Arag{\'o}n-Salamanca}, {De Lucia}, {Saglia}, {Halliday}, {Poggianti},
  {Dalcanton}, {Rudnick}, {Milvang-Jensen}, {Noll}, {Simard}, {Clowe},
  {Pell{\'o}}, {White}, \& {Zaritsky}}]{2009Barazza}
{Barazza}, F.~D., {Jablonka}, P., {Desai}, V., {et~al.} 2009, \aap, 497, 713,
  \dodoi{10.1051/0004-6361/200810352}

\bibitem[{{Barnes} \& {Hernquist}(1992)}]{1992BarnesHernquist}
{Barnes}, J.~E., \& {Hernquist}, L. 1992, \araa, 30, 705,
  \dodoi{10.1146/annurev.aa.30.090192.003421}

\bibitem[{{Begeman}(1987)}]{1987Begeman}
{Begeman}, K.~G. 1987, PhD thesis, -

\bibitem[{{Bellhouse} {et~al.}(2017){Bellhouse}, {Jaff{\'e}}, {Hau}, {McGee},
  {Poggianti}, {Moretti}, {Gullieuszik}, {Bettoni}, {Fasano}, {D'Onofrio},
  {Fritz}, {Omizzolo}, {Sheen}, \& {Vulcani}}]{2017Bellhouse}
{Bellhouse}, C., {Jaff{\'e}}, Y.~L., {Hau}, G.~K.~T., {et~al.} 2017, \apj, 844,
  49, \dodoi{10.3847/1538-4357/aa7875}

\bibitem[{{Bellhouse} {et~al.}(2019){Bellhouse}, {Jaff{\'e}}, {McGee},
  {Poggianti}, {Smith}, {Tonnesen}, {Fritz}, {Hau}, {Gullieuszik}, {Vulcani},
  {Fasano}, {Moretti}, {George}, {Bettoni}, {D'Onofrio}, {Omizzolo}, \&
  {Sheen}}]{2019Bellhouse}
{Bellhouse}, C., {Jaff{\'e}}, Y.~L., {McGee}, S.~L., {et~al.} 2019, \mnras,
  485, 1157, \dodoi{10.1093/mnras/stz460}

\bibitem[{{Bershady} {et~al.}(2010){Bershady}, {Verheijen}, {Westfall},
  {Andersen}, {Swaters}, \& {Martinsson}}]{2010Bershady_b}
{Bershady}, M.~A., {Verheijen}, M. A.~W., {Westfall}, K.~B., {et~al.} 2010,
  \apj, 716, 234, \dodoi{10.1088/0004-637X/716/1/234}

\bibitem[{{Bettoni}(1989)}]{1989Bettoni}
{Bettoni}, D. 1989, \aj, 97, 79, \dodoi{10.1086/114958}

\bibitem[{{Bettoni} \& {Galletta}(1994)}]{1994Bettoni}
{Bettoni}, D., \& {Galletta}, G. 1994, \aap, 281, 1

\bibitem[{{Binney} \& {Tremaine}(2008)}]{2008BinneyTremaine}
{Binney}, J., \& {Tremaine}, S. 2008, {Galactic Dynamics: Second Edition}

\bibitem[{{Biviano} {et~al.}(2017){Biviano}, {Moretti}, {Paccagnella},
  {Poggianti}, {Bettoni}, {Gullieuszik}, {Vulcani}, {Fasano}, {D'Onofrio},
  {Fritz}, \& {Cava}}]{2017Biviano}
{Biviano}, A., {Moretti}, A., {Paccagnella}, A., {et~al.} 2017, \aap, 607, A81,
  \dodoi{10.1051/0004-6361/201731289}

\bibitem[{{Boselli} {et~al.}(2014){Boselli}, {Cortese}, {Boquien}, {Boissier},
  {Catinella}, {Gavazzi}, {Lagos}, \& {Saintonge}}]{2014Boselli_c}
{Boselli}, A., {Cortese}, L., {Boquien}, M., {et~al.} 2014, \aap, 564, A67,
  \dodoi{10.1051/0004-6361/201322313}

\bibitem[{{Boselli} {et~al.}(2022{\natexlab{a}}){Boselli}, {Fossati}, \&
  {Sun}}]{2022Boselli}
{Boselli}, A., {Fossati}, M., \& {Sun}, M. 2022{\natexlab{a}}, \aapr, 30, 3,
  \dodoi{10.1007/s00159-022-00140-3}

\bibitem[{{Boselli} {et~al.}(2022{\natexlab{b}}){Boselli}, {Fossati}, \&
  {Sun}}]{2022Boselli_rev}
---. 2022{\natexlab{b}}, \aapr, 30, 3, \dodoi{10.1007/s00159-022-00140-3}

\bibitem[{{Boselli} \& {Gavazzi}(2006)}]{2006BoselliGavazzi}
{Boselli}, A., \& {Gavazzi}, G. 2006, \pasp, 118, 517, \dodoi{10.1086/500691}

\bibitem[{{Boselli} {et~al.}(2019){Boselli}, {Epinat}, {Contini},
  {Abril-Melgarejo}, {Boogaard}, {Pointecouteau}, {Ventou}, {Brinchmann},
  {Carton}, {Finley}, {Michel-Dansac}, {Soucail}, \&
  {Weilbacher}}]{2019Boselli}
{Boselli}, A., {Epinat}, B., {Contini}, T., {et~al.} 2019, \aap, 631, A114,
  \dodoi{10.1051/0004-6361/201936133}

\bibitem[{{Bosma}(1981)}]{1981Bosma}
{Bosma}, A. 1981, \aj, 86, 1825, \dodoi{10.1086/113063}

\bibitem[{{Bravo-Alfaro} {et~al.}(2000){Bravo-Alfaro}, {Cayatte}, {van Gorkom},
  \& {Balkowski}}]{2000BravoAlfaro}
{Bravo-Alfaro}, H., {Cayatte}, V., {van Gorkom}, J.~H., \& {Balkowski}, C.
  2000, \aj, 119, 580, \dodoi{10.108.f6/301194}

\bibitem[{{Brown} {et~al.}(2021){Brown}, {Wilson}, {Zabel}, {Davis}, {Boselli},
  {Chung}, {Ellison}, {Lagos}, {Stevens}, {Cortese}, {Bah{\'e}}, {Bisaria},
  {Bolatto}, {Cashmore}, {Catinella}, {Chown}, {Diemer}, {Elahi}, {Hani},
  {Jim{\'e}nez-Donaire}, {Lee}, {Leidig}, {Mok}, {Olsen}, {Parker}, {Roberts},
  {Smith}, {Spekkens}, {Thorp}, {Tonnesen}, {Vienneau}, {Villanueva}, {Vogel},
  {Wadsley}, {Welker}, \& {Yoon}}]{2021Brown}
{Brown}, T., {Wilson}, C.~D., {Zabel}, N., {et~al.} 2021, \apjs, 257, 21,
  \dodoi{10.3847/1538-4365/ac28f5}

\bibitem[{{Bureau} \& {Freeman}(1999)}]{1999Bureau}
{Bureau}, M., \& {Freeman}, K.~C. 1999, \aj, 118, 126, \dodoi{10.1086/300922}

\bibitem[{{Byrd} \& {Valtonen}(1990)}]{1990Byrd}
{Byrd}, G., \& {Valtonen}, M. 1990, \apj, 350, 89, \dodoi{10.1086/168362}

\bibitem[{{Capelo} {et~al.}(2022){Capelo}, {Feruglio}, {Hickox}, \&
  {Tombesi}}]{2022Capelo}
{Capelo}, P.~R., {Feruglio}, C., {Hickox}, R.~C., \& {Tombesi}, F. 2022, arXiv
  e-prints, arXiv:2211.00765.
\newblock \doarXiv{2211.00765}

\bibitem[{{Cappellari}(2002)}]{2002Cappellari}
{Cappellari}, M. 2002, \mnras, 333, 400,
  \dodoi{10.1046/j.1365-8711.2002.05412.x}

\bibitem[{{Cappellari}(2008)}]{2008Cappellari}
---. 2008, \mnras, 390, 71, \dodoi{10.1111/j.1365-2966.2008.13754.x}

\bibitem[{{Cappellari}(2017)}]{2017Cappellari}
---. 2017, \mnras, 466, 798, \dodoi{10.1093/mnras/stw3020}

\bibitem[{{Cappellari}(2020)}]{2020Cappellari}
---. 2020, \mnras, 494, 4819, \dodoi{10.1093/mnras/staa959}

\bibitem[{{Cappellari} \& {Emsellem}(2004)}]{2004CappellariEmsellem}
{Cappellari}, M., \& {Emsellem}, E. 2004, \pasp, 116, 138,
  \dodoi{10.1086/381875}

\bibitem[{{Cayatte} {et~al.}(1990){Cayatte}, {van Gorkom}, {Balkowski}, \&
  {Kotanyi}}]{1990Cayatte}
{Cayatte}, V., {van Gorkom}, J.~H., {Balkowski}, C., \& {Kotanyi}, C. 1990,
  \aj, 100, 604, \dodoi{10.1086/115545}

\bibitem[{{Chabrier}(2003)}]{2003Chabrier}
{Chabrier}, G. 2003, \pasp, 115, 763, \dodoi{10.1086/376392}

\bibitem[{{Chamaraux} {et~al.}(1980){Chamaraux}, {Balkowski}, \&
  {Gerard}}]{1980Chamaraux}
{Chamaraux}, P., {Balkowski}, C., \& {Gerard}, E. 1980, \aap, 83, 38

\bibitem[{{Chung} {et~al.}(2009){Chung}, {van Gorkom}, {Kenney}, {Crowl}, \&
  {Vollmer}}]{2009Chung}
{Chung}, A., {van Gorkom}, J.~H., {Kenney}, J. D.~P., {Crowl}, H., \&
  {Vollmer}, B. 2009, \aj, 138, 1741, \dodoi{10.1088/0004-6256/138/6/1741}

\bibitem[{{Chung} {et~al.}(2007){Chung}, {van Gorkom}, {Kenney}, \&
  {Vollmer}}]{2007Chung}
{Chung}, A., {van Gorkom}, J.~H., {Kenney}, J. D.~P., \& {Vollmer}, B. 2007,
  \apjl, 659, L115, \dodoi{10.1086/518034}

\bibitem[{{Combes}(2008)}]{2008Combes}
{Combes}, F. 2008, in Formation and Evolution of Galaxy Bulges, ed.
  M.~{Bureau}, E.~{Athanassoula}, \& B.~{Barbuy}, Vol. 245, 151--160,
  \dodoi{10.1017/S1743921308017535}

\bibitem[{{Combes} {et~al.}(1990){Combes}, {Debbasch}, {Friedli}, \&
  {Pfenniger}}]{1990Combes}
{Combes}, F., {Debbasch}, F., {Friedli}, D., \& {Pfenniger}, D. 1990, \aap,
  233, 82

\bibitem[{{Cort{\'e}s} {et~al.}(2015){Cort{\'e}s}, {Kenney}, \&
  {Hardy}}]{2015Cortes}
{Cort{\'e}s}, J.~R., {Kenney}, J. D.~P., \& {Hardy}, E. 2015, \apjs, 216, 9,
  \dodoi{10.1088/0067-0049/216/1/9}

\bibitem[{{Cortese} {et~al.}(2021){Cortese}, {Catinella}, \&
  {Smith}}]{2021Cortese}
{Cortese}, L., {Catinella}, B., \& {Smith}, R. 2021, \pasa, 38, e035,
  \dodoi{10.1017/pasa.2021.18}

\bibitem[{{Cortese} {et~al.}(2007){Cortese}, {Marcillac}, {Richard},
  {Bravo-Alfaro}, {Kneib}, {Rieke}, {Covone}, {Egami}, {Rigby}, {Czoske}, \&
  {Davies}}]{2007Cortese}
{Cortese}, L., {Marcillac}, D., {Richard}, J., {et~al.} 2007, \mnras, 376, 157,
  \dodoi{10.1111/j.1365-2966.2006.11369.x}

\bibitem[{{Cowie} \& {Songaila}(1977)}]{1977Cowie}
{Cowie}, L.~L., \& {Songaila}, A. 1977, \nat, 266, 501,
  \dodoi{10.1038/266501a0}

\bibitem[{{Cramer} {et~al.}(2020){Cramer}, {Kenney}, {Cortes}, {Cortes P.~C.},
  {Vlahakis}, {J{\'a}chym}, {Pompei}, \& {Rubio}}]{2020Cramer}
{Cramer}, W.~J., {Kenney}, J.~D.~P., {Cortes}, J.~R., {et~al.} 2020, \apj, 901,
  95, \dodoi{10.3847/1538-4357/abaf54}

\bibitem[{{Davis} {et~al.}(2013){Davis}, {Alatalo}, {Bureau}, {Cappellari},
  {Scott}, {Young}, {Blitz}, {Crocker}, {Bayet}, {Bois}, {Bournaud}, {Davies},
  {de Zeeuw}, {Duc}, {Emsellem}, {Khochfar}, {Krajnovi{\'c}}, {Kuntschner},
  {Lablanche}, {McDermid}, {Morganti}, {Naab}, {Oosterloo}, {Sarzi}, {Serra},
  \& {Weijmans}}]{2013Davis}
{Davis}, T.~A., {Alatalo}, K., {Bureau}, M., {et~al.} 2013, \mnras, 429, 534,
  \dodoi{10.1093/mnras/sts353}

\bibitem[{{Deb} {et~al.}(2020){Deb}, {Verheijen}, {Gullieuszik}, {Poggianti},
  {van Gorkom}, {Ramatsoku}, {Serra}, {Moretti}, {Vulcani}, {Bettoni},
  {Jaff{\'e}}, {Tonnesen}, \& {Fritz}}]{2020Deb}
{Deb}, T., {Verheijen}, M. A.~W., {Gullieuszik}, M., {et~al.} 2020, \mnras,
  494, 5029, \dodoi{10.1093/mnras/staa968}

\bibitem[{{Deb} {et~al.}(2022){Deb}, {Verheijen}, {Poggianti}, {Moretti}, {van
  der Hulst}, {Vulcani}, {Ramatsoku}, {Serra}, {Healy}, {Gullieuszik},
  {Bacchini}, {Ignesti}, {M{\"u}ller}, {Zabel}, {Luber}, {Jaff{\"e}}, \&
  {Gitti}}]{2022Deb}
{Deb}, T., {Verheijen}, M. A.~W., {Poggianti}, B.~M., {et~al.} 2022, \mnras,
  516, 2683, \dodoi{10.1093/mnras/stac2441}

\bibitem[{{Di Teodoro} \& {Fraternali}(2015)}]{2015DiTeodoro}
{Di Teodoro}, E.~M., \& {Fraternali}, F. 2015, \mnras, 451, 3021,
  \dodoi{10.1093/mnras/stv1213}

\bibitem[{{Di Teodoro} \& {Peek}(2021)}]{2021DiTeodoro}
{Di Teodoro}, E.~M., \& {Peek}, J. E.~G. 2021, arXiv e-prints,
  arXiv:2110.01618.
\newblock \doarXiv{2110.01618}

\bibitem[{{Di Teodoro} {et~al.}(2021){Di Teodoro}, {Posti}, {Ogle}, {Fall}, \&
  {Jarrett}}]{2021bDiTeodoro}
{Di Teodoro}, E.~M., {Posti}, L., {Ogle}, P.~M., {Fall}, S.~M., \& {Jarrett},
  T. 2021, \mnras, 507, 5820, \dodoi{10.1093/mnras/stab2549}

\bibitem[{{Di Teodoro} {et~al.}(2022){Di Teodoro}, {Posti}, {Fall}, {Ogle},
  {Jarrett}, {Appleton}, {Cluver}, {Haynes}, \& {Lisenfeld}}]{2022DiTeodoro}
{Di Teodoro}, E.~M., {Posti}, L., {Fall}, S.~M., {et~al.} 2022, arXiv e-prints,
  arXiv:2207.02906.
\newblock \doarXiv{2207.02906}

\bibitem[{{D{\'\i}az-Garc{\'\i}a} {et~al.}(2016){D{\'\i}az-Garc{\'\i}a},
  {Salo}, {Laurikainen}, \& {Herrera-Endoqui}}]{2016DiazGarcia}
{D{\'\i}az-Garc{\'\i}a}, S., {Salo}, H., {Laurikainen}, E., \&
  {Herrera-Endoqui}, M. 2016, \aap, 587, A160,
  \dodoi{10.1051/0004-6361/201526161}

\bibitem[{{Dicaire} {et~al.}(2008){Dicaire}, {Carignan}, {Amram}, {Hernandez},
  {Chemin}, {Daigle}, {de Denus-Baillargeon}, {Balkowski}, {Boselli}, {Fathi},
  \& {Kennicutt}}]{2008Dicaire}
{Dicaire}, I., {Carignan}, C., {Amram}, P., {et~al.} 2008, \mnras, 385, 553,
  \dodoi{10.1111/j.1365-2966.2008.12868.x}

\bibitem[{{Emsellem} {et~al.}(1994){Emsellem}, {Monnet}, \&
  {Bacon}}]{1994Emsellem}
{Emsellem}, E., {Monnet}, G., \& {Bacon}, R. 1994, \aap, 285, 723

\bibitem[{{Fanali} {et~al.}(2015){Fanali}, {Dotti}, {Fiacconi}, \&
  {Haardt}}]{2015Fanali}
{Fanali}, R., {Dotti}, M., {Fiacconi}, D., \& {Haardt}, F. 2015, \mnras, 454,
  3641, \dodoi{10.1093/mnras/stv2247}

\bibitem[{{Farber} {et~al.}(2022){Farber}, {Ruszkowski}, {Tonnesen}, \&
  {Holguin}}]{2022Farber}
{Farber}, R.~J., {Ruszkowski}, M., {Tonnesen}, S., \& {Holguin}, F. 2022,
  \mnras, 512, 5927, \dodoi{10.1093/mnras/stac794}

\bibitem[{{Fasano} {et~al.}(2012){Fasano}, {Vanzella}, {Dressler}, {Poggianti},
  {Moles}, {Bettoni}, {Valentinuzzi}, {Moretti}, {D'Onofrio}, {Varela},
  {Couch}, {Kj{\ae}rgaard}, {Fritz}, {Omizzolo}, \& {Cava}}]{2012Fasano}
{Fasano}, G., {Vanzella}, E., {Dressler}, A., {et~al.} 2012, \mnras, 420, 926,
  \dodoi{10.1111/j.1365-2966.2011.19798.x}

\bibitem[{{Franchetto} {et~al.}(2020){Franchetto}, {Vulcani}, {Poggianti},
  {Gullieuszik}, {Mingozzi}, {Moretti}, {Tomi{\v{c}}i{\'c}}, {Fritz},
  {Bettoni}, \& {Jaff{\'e}}}]{2020Franchetto}
{Franchetto}, A., {Vulcani}, B., {Poggianti}, B.~M., {et~al.} 2020, \apj, 895,
  106, \dodoi{10.3847/1538-4357/ab8db9}

\bibitem[{{Franchetto} {et~al.}(2021){Franchetto}, {Mingozzi}, {Poggianti},
  {Vulcani}, {Bacchini}, {Gullieuszik}, {Moretti}, {Tomi{\v{c}}i{\'c}}, \&
  {Fritz}}]{2021Franchetto}
{Franchetto}, A., {Mingozzi}, M., {Poggianti}, B.~M., {et~al.} 2021, \apj, 923,
  28, \dodoi{10.3847/1538-4357/ac2510}

\bibitem[{{Fritz} {et~al.}(2011){Fritz}, {Poggianti}, {Cava}, {Valentinuzzi},
  {Moretti}, {Bettoni}, {Bressan}, {Couch}, {D'Onofrio}, {Dressler}, {Fasano},
  {Kj{\ae}rgaard}, {Moles}, {Omizzolo}, \& {Varela}}]{2011Fritz}
{Fritz}, J., {Poggianti}, B.~M., {Cava}, A., {et~al.} 2011, \aap, 526, A45,
  \dodoi{10.1051/0004-6361/201015214}

\bibitem[{{Fritz} {et~al.}(2017){Fritz}, {Moretti}, {Gullieuszik}, {Poggianti},
  {Bruzual}, {Vulcani}, {Nicastro}, {Jaff{\'e}}, {Cervantes Sodi}, {Bettoni},
  {Biviano}, {Fasano}, {Charlot}, {Bellhouse}, \& {Hau}}]{2017Fritz}
{Fritz}, J., {Moretti}, A., {Gullieuszik}, M., {et~al.} 2017, \apj, 848, 132,
  \dodoi{10.3847/1538-4357/aa8f51}

\bibitem[{{Fumagalli} {et~al.}(2009){Fumagalli}, {Krumholz}, {Prochaska},
  {Gavazzi}, \& {Boselli}}]{2009Fumagalli}
{Fumagalli}, M., {Krumholz}, M.~R., {Prochaska}, J.~X., {Gavazzi}, G., \&
  {Boselli}, A. 2009, \apj, 697, 1811, \dodoi{10.1088/0004-637X/697/2/1811}

\bibitem[{{Galloway} {et~al.}(2015){Galloway}, {Willett}, {Fortson},
  {Cardamone}, {Schawinski}, {Cheung}, {Lintott}, {Masters}, {Melvin}, \&
  {Simmons}}]{2015Galloway}
{Galloway}, M.~A., {Willett}, K.~W., {Fortson}, L.~F., {et~al.} 2015, \mnras,
  448, 3442, \dodoi{10.1093/mnras/stv235}

\bibitem[{{Gavazzi} {et~al.}(2018){Gavazzi}, {Consolandi}, {Gutierrez},
  {Boselli}, \& {Yoshida}}]{2018Gavazzi}
{Gavazzi}, G., {Consolandi}, G., {Gutierrez}, M.~L., {Boselli}, A., \&
  {Yoshida}, M. 2018, \aap, 618, A130, \dodoi{10.1051/0004-6361/201833427}

\bibitem[{{George} {et~al.}(2018){George}, {Poggianti}, {Gullieuszik},
  {Fasano}, {Bellhouse}, {Postma}, {Moretti}, {Jaff{\'e}}, {Vulcani},
  {Bettoni}, {Fritz}, {C{\^o}t{\'e}}, {Ghosh}, {Hutchings}, {Mohan},
  {Sreekumar}, {Stalin}, {Subramaniam}, \& {Tandon}}]{2018George}
{George}, K., {Poggianti}, B.~M., {Gullieuszik}, M., {et~al.} 2018, \mnras,
  479, 4126, \dodoi{10.1093/mnras/sty1452}

\bibitem[{{George} {et~al.}(2019){George}, {Poggianti}, {Bellhouse},
  {Radovich}, {Fritz}, {Paladino}, {Bettoni}, {Jaff{\'e}}, {Moretti},
  {Gullieuszik}, {Vulcani}, {Fasano}, {Stalin}, {Subramaniam}, \&
  {Tandon}}]{2019George}
{George}, K., {Poggianti}, B.~M., {Bellhouse}, C., {et~al.} 2019, \mnras, 487,
  3102, \dodoi{10.1093/mnras/stz1443}

\bibitem[{{Giovanelli} \& {Haynes}(1985)}]{1985Giovanelli}
{Giovanelli}, R., \& {Haynes}, M.~P. 1985, \apj, 292, 404,
  \dodoi{10.1086/163170}

\bibitem[{{Gullieuszik} {et~al.}(2017){Gullieuszik}, {Poggianti}, {Moretti},
  {Fritz}, {Jaff{\'e}}, {Hau}, {Bischko}, {Bellhouse}, {Bettoni}, {Fasano},
  {Vulcani}, {D'Onofrio}, \& {Biviano}}]{2017Gullieuszik}
{Gullieuszik}, M., {Poggianti}, B.~M., {Moretti}, A., {et~al.} 2017, \apj, 846,
  27, \dodoi{10.3847/1538-4357/aa8322}

\bibitem[{{Gullieuszik} {et~al.}(2020){Gullieuszik}, {Poggianti}, {McGee},
  {Moretti}, {Vulcani}, {Tonnesen}, {Roediger}, {Jaff{\'e}}, {Fritz},
  {Franchetto}, {Omizzolo}, {Bettoni}, {Radovich}, \&
  {Wolter}}]{2020Gullieuszik}
{Gullieuszik}, M., {Poggianti}, B.~M., {McGee}, S.~L., {et~al.} 2020, \apj,
  899, 13, \dodoi{10.3847/1538-4357/aba3cb}

\bibitem[{{Gullieuszik} {et~al.}(2023){Gullieuszik}, {Giunchi}, {Poggianti},
  {Moretti}, {Scarlata}, {Calzetti}, {Werle}, {Zanella}, {Radovich},
  {Bellhouse}, {Bettoni}, {Franchetto}, {Fritz}, {Jaff{\'e}}, {McGee},
  {Mingozzi}, {Omizzolo}, {Tonnesen}, {Verheijen}, \&
  {Vulcani}}]{2023Gullieuszik}
{Gullieuszik}, M., {Giunchi}, E., {Poggianti}, B.~M., {et~al.} 2023, arXiv
  e-prints, arXiv:2301.08279, \dodoi{10.48550/arXiv.2301.08279}

\bibitem[{{Gunn} \& {Gott}(1972)}]{1972GunnGott}
{Gunn}, J.~E., \& {Gott}, J.~Richard, I. 1972, \apj, 176, 1,
  \dodoi{10.1086/151605}

\bibitem[{{Haynes} {et~al.}(1984){Haynes}, {Giovanelli}, \&
  {Chincarini}}]{1984Haynes}
{Haynes}, M.~P., {Giovanelli}, R., \& {Chincarini}, G.~L. 1984, \araa, 22, 445,
  \dodoi{10.1146/annurev.aa.22.090184.002305}

\bibitem[{{Healy} {et~al.}(2021){Healy}, {Deb}, {Verheijen}, {Blyth}, {Serra},
  {Ramatsoku}, \& {Vulcani}}]{2021HealyDeb}
{Healy}, J., {Deb}, T., {Verheijen}, M.~A.~W., {et~al.} 2021, \aap, 654, A173,
  \dodoi{10.1051/0004-6361/202141377}

\bibitem[{{Hess} {et~al.}(2022){Hess}, {Kotulla}, {Chen}, {Carignan},
  {Gallagher}, {Jarrett}, \& {Kraan-Korteweg}}]{2022Hess}
{Hess}, K.~M., {Kotulla}, R., {Chen}, H., {et~al.} 2022, \aap, 668, A184,
  \dodoi{10.1051/0004-6361/202243412}

\bibitem[{{Ho} {et~al.}(1997){Ho}, {Filippenko}, \& {Sargent}}]{1997Ho}
{Ho}, L.~C., {Filippenko}, A.~V., \& {Sargent}, W. L.~W. 1997, \apj, 487, 591,
  \dodoi{10.1086/304643}

\bibitem[{{Hogarth} {et~al.}(2021){Hogarth}, {Saintonge}, {Cortese}, {Davis},
  {Croom}, {Bland-Hawthorn}, {Brough}, {Bryant}, {Catinella}, {Fletcher},
  {Groves}, {Lawrence}, {L{\'o}pez-S{\'a}nchez}, {Owers}, {Richards},
  {Roberts-Borsani}, {Taylor}, {van de Sande}, \& {Scott}}]{2021Hogarth}
{Hogarth}, L.~M., {Saintonge}, A., {Cortese}, L., {et~al.} 2021, \mnras, 500,
  3802, \dodoi{10.1093/mnras/staa3512}

\bibitem[{{Iorio} {et~al.}(2017){Iorio}, {Fraternali}, {Nipoti}, {Di Teodoro},
  {Read}, \& {Battaglia}}]{2017Iorio}
{Iorio}, G., {Fraternali}, F., {Nipoti}, C., {et~al.} 2017, \mnras, 466, 4159,
  \dodoi{10.1093/mnras/stw3285}

\bibitem[{{J{\'a}chym} {et~al.}(2014){J{\'a}chym}, {Combes}, {Cortese}, {Sun},
  \& {Kenney}}]{2014Jachym}
{J{\'a}chym}, P., {Combes}, F., {Cortese}, L., {Sun}, M., \& {Kenney}, J. D.~P.
  2014, \apj, 792, 11, \dodoi{10.1088/0004-637X/792/1/11}

\bibitem[{{J{\'a}chym} {et~al.}(2017){J{\'a}chym}, {Sun}, {Kenney}, {Cortese},
  {Combes}, {Yagi}, {Yoshida}, {Palou{\v{s}}}, \& {Roediger}}]{2017Jachym}
{J{\'a}chym}, P., {Sun}, M., {Kenney}, J. D.~P., {et~al.} 2017, \apj, 839, 114,
  \dodoi{10.3847/1538-4357/aa6af5}

\bibitem[{{Jaff{\'e}} {et~al.}(2018){Jaff{\'e}}, {Poggianti}, {Moretti},
  {Gullieuszik}, {Smith}, {Vulcani}, {Fasano}, {Fritz}, {Tonnesen}, {Bettoni},
  {Hau}, {Biviano}, {Bellhouse}, \& {McGee}}]{2018Jaffe}
{Jaff{\'e}}, Y.~L., {Poggianti}, B.~M., {Moretti}, A., {et~al.} 2018, \mnras,
  476, 4753, \dodoi{10.1093/mnras/sty500}

\bibitem[{{Jedrzejewski}(1987)}]{1987Jedrzejewski}
{Jedrzejewski}, R.~I. 1987, \mnras, 226, 747, \dodoi{10.1093/mnras/226.4.747}

\bibitem[{{Kamphuis} {et~al.}(2015){Kamphuis}, {J{\'o}zsa}, {Oh}, {Spekkens},
  {Urbancic}, {Serra}, {Koribalski}, \& {Dettmar}}]{2015Kamphuis}
{Kamphuis}, P., {J{\'o}zsa}, G.~I.~G., {Oh}, S. .~H., {et~al.} 2015, \mnras,
  452, 3139, \dodoi{10.1093/mnras/stv1480}

\bibitem[{{Kenney} {et~al.}(2004){Kenney}, {van Gorkom}, \&
  {Vollmer}}]{2004Kenney}
{Kenney}, J. D.~P., {van Gorkom}, J.~H., \& {Vollmer}, B. 2004, \aj, 127, 3361,
  \dodoi{10.1086/420805}

\bibitem[{{Kim} \& {Choi}(2020)}]{2020Kim}
{Kim}, M., \& {Choi}, Y.-Y. 2020, \apjl, 901, L38,
  \dodoi{10.3847/2041-8213/abb66f}

\bibitem[{{Knapen} {et~al.}(2000){Knapen}, {Shlosman}, \&
  {Peletier}}]{2000Knapen}
{Knapen}, J.~H., {Shlosman}, I., \& {Peletier}, R.~F. 2000, \apj, 529, 93,
  \dodoi{10.1086/308266}

\bibitem[{{K{\"o}ppen} {et~al.}(2018){K{\"o}ppen}, {J{\'a}chym}, {Taylor}, \&
  {Palou{\v{s}}}}]{2018Koppen}
{K{\"o}ppen}, J., {J{\'a}chym}, P., {Taylor}, R., \& {Palou{\v{s}}}, J. 2018,
  \mnras, 479, 4367, \dodoi{10.1093/mnras/sty1610}

\bibitem[{{Kormendy} \& {Kennicutt}(2004)}]{2004KormendyKennicutt}
{Kormendy}, J., \& {Kennicutt}, Robert~C., J. 2004, \araa, 42, 603,
  \dodoi{10.1146/annurev.astro.42.053102.134024}

\bibitem[{{Krolik}(1999)}]{1999Krolik}
{Krolik}, J.~H. 1999, {Active galactic nuclei : from the central black hole to
  the galactic environment}

\bibitem[{{Kronberger} {et~al.}(2008{\natexlab{a}}){Kronberger}, {Kapferer},
  {Ferrari}, {Unterguggenberger}, \& {Schindler}}]{2008Kronberger}
{Kronberger}, T., {Kapferer}, W., {Ferrari}, C., {Unterguggenberger}, S., \&
  {Schindler}, S. 2008{\natexlab{a}}, \aap, 481, 337,
  \dodoi{10.1051/0004-6361:20078904}

\bibitem[{{Kronberger} {et~al.}(2006){Kronberger}, {Kapferer}, {Schindler},
  {B{\"o}hm}, {Kutdemir}, \& {Ziegler}}]{2006Kronberger}
{Kronberger}, T., {Kapferer}, W., {Schindler}, S., {et~al.} 2006, \aap, 458,
  69, \dodoi{10.1051/0004-6361:20064976}

\bibitem[{{Kronberger} {et~al.}(2008{\natexlab{b}}){Kronberger}, {Kapferer},
  {Unterguggenberger}, {Schindler}, \& {Ziegler}}]{2008Kronberger_b}
{Kronberger}, T., {Kapferer}, W., {Unterguggenberger}, S., {Schindler}, S., \&
  {Ziegler}, B.~L. 2008{\natexlab{b}}, \aap, 483, 783,
  \dodoi{10.1051/0004-6361:200809387}

\bibitem[{{Kuijken} \& {Merrifield}(1995)}]{1995Kuijken}
{Kuijken}, K., \& {Merrifield}, M.~R. 1995, \apjl, 443, L13,
  \dodoi{10.1086/187824}

\bibitem[{{Lansbury} {et~al.}(2014){Lansbury}, {Lucey}, \&
  {Smith}}]{2014Lansbury}
{Lansbury}, G.~B., {Lucey}, J.~R., \& {Smith}, R.~J. 2014, \mnras, 439, 1749,
  \dodoi{10.1093/mnras/stu049}

\bibitem[{{Larson} {et~al.}(1980){Larson}, {Tinsley}, \&
  {Caldwell}}]{1980Larson}
{Larson}, R.~B., {Tinsley}, B.~M., \& {Caldwell}, C.~N. 1980, \apj, 237, 692,
  \dodoi{10.1086/157917}

\bibitem[{{Lee} {et~al.}(2017){Lee}, {Chung}, {Tonnesen}, {Kenney}, {Wong},
  {Vollmer}, {Petitpas}, {Crowl}, \& {van Gorkom}}]{2017Lee}
{Lee}, B., {Chung}, A., {Tonnesen}, S., {et~al.} 2017, \mnras, 466, 1382,
  \dodoi{10.1093/mnras/stw3162}

\bibitem[{{Lee} {et~al.}(2012){Lee}, {Woo}, {Lee}, {Hwang}, {Lee}, {Sohn}, \&
  {Lee}}]{2012Lee}
{Lee}, G.-H., {Woo}, J.-H., {Lee}, M.~G., {et~al.} 2012, \apj, 750, 141,
  \dodoi{10.1088/0004-637X/750/2/141}

\bibitem[{{Lee} {et~al.}(2022){Lee}, {Sheen}, {Yoon}, {Jaff{\'e}}, \&
  {Chung}}]{2022Lee}
{Lee}, S., {Sheen}, Y.-K., {Yoon}, H., {Jaff{\'e}}, Y., \& {Chung}, A. 2022,
  \mnras, 517, 2912, \dodoi{10.1093/mnras/stac2821}

\bibitem[{{Lelli} {et~al.}(2022){Lelli}, {Davis}, {Bureau}, {Cappellari},
  {Liu}, {Ruffa}, {Smith}, \& {Williams}}]{2022Lelli}
{Lelli}, F., {Davis}, T.~A., {Bureau}, M., {et~al.} 2022, \mnras, 516, 4066,
  \dodoi{10.1093/mnras/stac2493}

\bibitem[{{Lelli} {et~al.}(2016{\natexlab{a}}){Lelli}, {McGaugh}, \&
  {Schombert}}]{2016Lelli_c}
{Lelli}, F., {McGaugh}, S.~S., \& {Schombert}, J.~M. 2016{\natexlab{a}}, \aj,
  152, 157, \dodoi{10.3847/0004-6256/152/6/157}

\bibitem[{{Lelli} {et~al.}(2016{\natexlab{b}}){Lelli}, {McGaugh}, \&
  {Schombert}}]{2016Lelli}
---. 2016{\natexlab{b}}, \apjl, 816, L14, \dodoi{10.3847/2041-8205/816/1/L14}

\bibitem[{{Lelli} {et~al.}(2019){Lelli}, {McGaugh}, {Schombert}, {Desmond}, \&
  {Katz}}]{2019Lelli}
{Lelli}, F., {McGaugh}, S.~S., {Schombert}, J.~M., {Desmond}, H., \& {Katz}, H.
  2019, \mnras, 484, 3267, \dodoi{10.1093/mnras/stz205}

\bibitem[{{Lelli} {et~al.}(2014){Lelli}, {Verheijen}, \&
  {Fraternali}}]{2014Lelli_b}
{Lelli}, F., {Verheijen}, M., \& {Fraternali}, F. 2014, \aap, 566, A71,
  \dodoi{10.1051/0004-6361/201322657}

\bibitem[{{Leroy} {et~al.}(2008){Leroy}, {Walter}, {Brinks}, {Bigiel}, {de
  Blok}, {Madore}, \& {Thornley}}]{2008Leroy}
{Leroy}, A.~K., {Walter}, F., {Brinks}, E., {et~al.} 2008, \aj, 136, 2782,
  \dodoi{10.1088/0004-6256/136/6/2782}

\bibitem[{{Leroy} {et~al.}(2009){Leroy}, {Walter}, {Bigiel}, {Usero}, {Weiss},
  {Brinks}, {de Blok}, {Kennicutt}, {Schuster}, {Kramer}, {Wiesemeyer}, \&
  {Roussel}}]{2009Leroy}
{Leroy}, A.~K., {Walter}, F., {Bigiel}, F., {et~al.} 2009, \aj, 137, 4670,
  \dodoi{10.1088/0004-6256/137/6/4670}

\bibitem[{{Loni} {et~al.}(2021){Loni}, {Serra}, {Kleiner}, {Cortese},
  {Catinella}, {Koribalski}, {Jarrett}, {Molnar}, {Davis}, {Iodice},
  {Lee-Waddell}, {Loi}, {Maccagni}, {Peletier}, {Popping}, {Ramatsoku},
  {Smith}, \& {Zabel}}]{2021Loni}
{Loni}, A., {Serra}, P., {Kleiner}, D., {et~al.} 2021, \aap, 648, A31,
  \dodoi{10.1051/0004-6361/202039803}

\bibitem[{{Luber} {et~al.}(2022){Luber}, {M{\"u}ller}, {van Gorkom},
  {Poggianti}, {Vulcani}, {Franchetto}, {Bacchini}, {Bettoni}, {Deb}, {Fritz},
  {Gullieuszik}, {Ignesti}, {Jaffe}, {Moretti}, {Paladino}, {Ramatsoku},
  {Serra}, {Smith}, {Tomicic}, {Tonnesen}, {Verheijen}, \&
  {Wolter}}]{2022Luber}
{Luber}, N., {M{\"u}ller}, A., {van Gorkom}, J.~H., {et~al.} 2022, \apj, 927,
  39, \dodoi{10.3847/1538-4357/ac469a}

\bibitem[{{Mancera Pi{\~n}a} {et~al.}(2021{\natexlab{a}}){Mancera Pi{\~n}a},
  {Posti}, {Fraternali}, {Adams}, \& {Oosterloo}}]{2021aManceraPina}
{Mancera Pi{\~n}a}, P.~E., {Posti}, L., {Fraternali}, F., {Adams}, E. A.~K., \&
  {Oosterloo}, T. 2021{\natexlab{a}}, \aap, 647, A76,
  \dodoi{10.1051/0004-6361/202039340}

\bibitem[{{Mancera Pi{\~n}a} {et~al.}(2021{\natexlab{b}}){Mancera Pi{\~n}a},
  {Posti}, {Pezzulli}, {Fraternali}, {Fall}, {Oosterloo}, \&
  {Adams}}]{2021bManceraPina}
{Mancera Pi{\~n}a}, P.~E., {Posti}, L., {Pezzulli}, G., {et~al.}
  2021{\natexlab{b}}, \aap, 651, L15, \dodoi{10.1051/0004-6361/202141574}

\bibitem[{{Mapelli} {et~al.}(2008){Mapelli}, {Moore}, \&
  {Bland-Hawthorn}}]{2008Mapelli}
{Mapelli}, M., {Moore}, B., \& {Bland-Hawthorn}, J. 2008, \mnras, 388, 697,
  \dodoi{10.1111/j.1365-2966.2008.13421.x}

\bibitem[{{Marasco} {et~al.}(2019){Marasco}, {Fraternali}, {Posti}, {Ijtsma},
  {Di Teodoro}, \& {Oosterloo}}]{2019Marasco}
{Marasco}, A., {Fraternali}, F., {Posti}, L., {et~al.} 2019, \aap, 621, L6,
  \dodoi{10.1051/0004-6361/201834456}

\bibitem[{{Marasco} {et~al.}(2017){Marasco}, {Fraternali}, {van der Hulst}, \&
  {Oosterloo}}]{2017Marasco}
{Marasco}, A., {Fraternali}, F., {van der Hulst}, J.~M., \& {Oosterloo}, T.
  2017, \aap, 607, A106, \dodoi{10.1051/0004-6361/201731054}

\bibitem[{{Marasco} {et~al.}(2018){Marasco}, {Oman}, {Navarro}, {Frenk}, \&
  {Oosterloo}}]{2018Marasco}
{Marasco}, A., {Oman}, K.~A., {Navarro}, J.~F., {Frenk}, C.~S., \& {Oosterloo},
  T. 2018, \mnras, 476, 2168, \dodoi{10.1093/mnras/sty354}

\bibitem[{{Martinsson} {et~al.}(2013){Martinsson}, {Verheijen}, {Westfall},
  {Bershady}, {Schechtman-Rook}, {Andersen}, \& {Swaters}}]{2013Martisson_b}
{Martinsson}, T. P.~K., {Verheijen}, M. A.~W., {Westfall}, K.~B., {et~al.}
  2013, \aap, 557, A130, \dodoi{10.1051/0004-6361/201220515}

\bibitem[{{Masters} {et~al.}(2012){Masters}, {Nichol}, {Haynes}, {Keel},
  {Lintott}, {Simmons}, {Skibba}, {Bamford}, {Giovanelli}, \&
  {Schawinski}}]{2012Masters}
{Masters}, K.~L., {Nichol}, R.~C., {Haynes}, M.~P., {et~al.} 2012, \mnras, 424,
  2180, \dodoi{10.1111/j.1365-2966.2012.21377.x}

\bibitem[{{M{\'e}ndez-Abreu} {et~al.}(2012){M{\'e}ndez-Abreu},
  {S{\'a}nchez-Janssen}, {Aguerri}, {Corsini}, \&
  {Zarattini}}]{2012MendezAbreu}
{M{\'e}ndez-Abreu}, J., {S{\'a}nchez-Janssen}, R., {Aguerri}, J.~A.~L.,
  {Corsini}, E.~M., \& {Zarattini}, S. 2012, \apjl, 761, L6,
  \dodoi{10.1088/2041-8205/761/1/L6}

\bibitem[{{Merluzzi} {et~al.}(2013){Merluzzi}, {Busarello}, {Dopita}, {Haines},
  {Steinhauser}, {Mercurio}, {Rifatto}, {Smith}, \& {Schindler}}]{2013Merluzzi}
{Merluzzi}, P., {Busarello}, G., {Dopita}, M.~A., {et~al.} 2013, \mnras, 429,
  1747, \dodoi{10.1093/mnras/sts466}

\bibitem[{{Merrifield} \& {Kuijken}(1999)}]{1999Merrifield}
{Merrifield}, M.~R., \& {Kuijken}, K. 1999, \aap, 345, L47.
\newblock \doarXiv{astro-ph/9904158}

\bibitem[{{Merritt}(1983)}]{1983Merritt}
{Merritt}, D. 1983, \apj, 264, 24, \dodoi{10.1086/160571}

\bibitem[{{Moretti} {et~al.}(2018){Moretti}, {Paladino}, {Poggianti},
  {D'Onofrio}, {Bettoni}, {Gullieuszik}, {Jaff{\'e}}, {Vulcani}, {Fasano},
  {Fritz}, \& {Torstensson}}]{2018Moretti}
{Moretti}, A., {Paladino}, R., {Poggianti}, B.~M., {et~al.} 2018, \mnras, 480,
  2508, \dodoi{10.1093/mnras/sty2021}

\bibitem[{{Moretti} {et~al.}(2020{\natexlab{a}}){Moretti}, {Paladino},
  {Poggianti}, {Serra}, {Roediger}, {Gullieuszik}, {Tomi{\v{c}}i{\'c}},
  {Radovich}, {Vulcani}, {Jaff{\'e}}, {Fritz}, {Bettoni}, {Ramatsoku}, \&
  {Wolter}}]{2020Moretti}
---. 2020{\natexlab{a}}, \apj, 889, 9, \dodoi{10.3847/1538-4357/ab616a}

\bibitem[{{Moretti} {et~al.}(2020{\natexlab{b}}){Moretti}, {Paladino},
  {Poggianti}, {Serra}, {Ramatsoku}, {Franchetto}, {Deb}, {Gullieuszik},
  {Tomi{\v{c}}i{\'c}}, {Mingozzi}, {Vulcani}, {Radovich}, {Bettoni}, \&
  {Fritz}}]{2020MorettiLetter}
---. 2020{\natexlab{b}}, \apjl, 897, L30, \dodoi{10.3847/2041-8213/ab9f3b}

\bibitem[{{Moretti} {et~al.}(2022){Moretti}, {Radovich}, {Poggianti},
  {Vulcani}, {Gullieuszik}, {Werle}, {Bellhouse}, {Bacchini}, {Fritz},
  {Soucail}, {Richard}, {Franchetto}, {Tomi{\v{c}}i{\'c}}, \&
  {Omizzolo}}]{2022Moretti}
{Moretti}, A., {Radovich}, M., {Poggianti}, B.~M., {et~al.} 2022, \apj, 925, 4,
  \dodoi{10.3847/1538-4357/ac36c7}

\bibitem[{{M{\"u}ller} {et~al.}(2021){M{\"u}ller}, {Poggianti}, {Pfrommer},
  {Adebahr}, {Serra}, {Ignesti}, {Sparre}, {Gitti}, {Dettmar}, {Vulcani}, \&
  {Moretti}}]{2021Muller}
{M{\"u}ller}, A., {Poggianti}, B.~M., {Pfrommer}, C., {et~al.} 2021, Nature
  Astronomy, 5, 159, \dodoi{10.1038/s41550-020-01234-7}

\bibitem[{{Nulsen}(1982)}]{1982Nulsen}
{Nulsen}, P.~E.~J. 1982, \mnras, 198, 1007, \dodoi{10.1093/mnras/198.4.1007}

\bibitem[{{Olling}(1996)}]{1996Olling}
{Olling}, R.~P. 1996, \aj, 112, 457, \dodoi{10.1086/118028}

\bibitem[{{Peluso} {et~al.}(2022){Peluso}, {Vulcani}, {Poggianti}, {Moretti},
  {Radovich}, {Smith}, {Jaff{\'e}}, {Crossett}, {Gullieuszik}, {Fritz}, \&
  {Ignesti}}]{2022Peluso}
{Peluso}, G., {Vulcani}, B., {Poggianti}, B.~M., {et~al.} 2022, \apj, 927, 130,
  \dodoi{10.3847/1538-4357/ac4225}

\bibitem[{{Poggianti} \& {the GASP team}(2022)}]{2022Poggianti}
{Poggianti}, B.~M., \& {the GASP team}. 2022, arXiv e-prints, arXiv:2211.12297.
\newblock \doarXiv{2211.12297}

\bibitem[{{Poggianti} {et~al.}(2016){Poggianti}, {Fasano}, {Omizzolo},
  {Gullieuszik}, {Bettoni}, {Moretti}, {Paccagnella}, {Jaff{\'e}}, {Vulcani},
  {Fritz}, {Couch}, \& {D'Onofrio}}]{2016Poggianti}
{Poggianti}, B.~M., {Fasano}, G., {Omizzolo}, A., {et~al.} 2016, \aj, 151, 78,
  \dodoi{10.3847/0004-6256/151/3/78}

\bibitem[{{Poggianti} {et~al.}(2017{\natexlab{a}}){Poggianti}, {Jaff{\'e}},
  {Moretti}, {Gullieuszik}, {Radovich}, {Tonnesen}, {Fritz}, {Bettoni},
  {Vulcani}, {Fasano}, {Bellhouse}, {Hau}, \& {Omizzolo}}]{2017PoggiantiNat}
{Poggianti}, B.~M., {Jaff{\'e}}, Y.~L., {Moretti}, A., {et~al.}
  2017{\natexlab{a}}, \nat, 548, 304, \dodoi{10.1038/nature23462}

\bibitem[{{Poggianti} {et~al.}(2017{\natexlab{b}}){Poggianti}, {Moretti},
  {Gullieuszik}, {Fritz}, {Jaff{\'e}}, {Bettoni}, {Fasano}, {Bellhouse}, {Hau},
  {Vulcani}, {Biviano}, {Omizzolo}, {Paccagnella}, {D'Onofrio}, {Cava},
  {Sheen}, {Couch}, \& {Owers}}]{2017Poggianti}
{Poggianti}, B.~M., {Moretti}, A., {Gullieuszik}, M., {et~al.}
  2017{\natexlab{b}}, \apj, 844, 48, \dodoi{10.3847/1538-4357/aa78ed}

\bibitem[{{Poggianti} {et~al.}(2019){Poggianti}, {Gullieuszik}, {Tonnesen},
  {Moretti}, {Vulcani}, {Radovich}, {Jaff{\'e}}, {Fritz}, {Bettoni},
  {Franchetto}, {Fasano}, {Bellhouse}, \& {Omizzolo}}]{2019Poggianti}
{Poggianti}, B.~M., {Gullieuszik}, M., {Tonnesen}, S., {et~al.} 2019, \mnras,
  482, 4466, \dodoi{10.1093/mnras/sty2999}

\bibitem[{{Ponomareva} {et~al.}(2017){Ponomareva}, {Verheijen}, {Peletier}, \&
  {Bosma}}]{2017Ponomareva}
{Ponomareva}, A.~A., {Verheijen}, M. A.~W., {Peletier}, R.~F., \& {Bosma}, A.
  2017, \mnras, 469, 2387, \dodoi{10.1093/mnras/stx1018}

\bibitem[{{Posti} {et~al.}(2018){Posti}, {Fraternali}, {Di Teodoro}, \&
  {Pezzulli}}]{2018Posti}
{Posti}, L., {Fraternali}, F., {Di Teodoro}, E.~M., \& {Pezzulli}, G. 2018,
  \aap, 612, L6, \dodoi{10.1051/0004-6361/201833091}

\bibitem[{{Radovich} {et~al.}(2019){Radovich}, {Poggianti}, {Jaff{\'e}},
  {Moretti}, {Bettoni}, {Gullieuszik}, {Vulcani}, \& {Fritz}}]{2019Radovich}
{Radovich}, M., {Poggianti}, B., {Jaff{\'e}}, Y.~L., {et~al.} 2019, \mnras,
  486, 486, \dodoi{10.1093/mnras/stz809}

\bibitem[{{Ramatsoku} {et~al.}(2019){Ramatsoku}, {Serra}, {Poggianti},
  {Moretti}, {Gullieuszik}, {Bettoni}, {Deb}, {Fritz}, {van Gorkom},
  {Jaff{\'e}}, {Tonnesen}, {Verheijen}, {Vulcani}, {Hugo}, {J{\'o}zsa},
  {Maccagni}, {Makhathini}, {Ramaila}, {Smirnov}, \& {Thorat}}]{2019Ramatsoku}
{Ramatsoku}, M., {Serra}, P., {Poggianti}, B.~M., {et~al.} 2019, \mnras, 487,
  4580, \dodoi{10.1093/mnras/stz1609}

\bibitem[{{Ramatsoku} {et~al.}(2020){Ramatsoku}, {Serra}, {Poggianti},
  {Moretti}, {Gullieuszik}, {Bettoni}, {Deb}, {Franchetto}, {van Gorkom},
  {Jaff{\'e}}, {Tonnesen}, {Verheijen}, {Vulcani}, {Andati}, {de Blok},
  {J{\'o}zsa}, {Kamphuis}, {Kleiner}, {Maccagni}, {Makhathini}, {Moln{\'a}r},
  {Ramaila}, {Smirnov}, \& {Thorat}}]{2020Ramatsoku}
---. 2020, \aap, 640, A22, \dodoi{10.1051/0004-6361/202037759}

\bibitem[{{Ramos-Mart{\'\i}nez} {et~al.}(2018){Ramos-Mart{\'\i}nez},
  {G{\'o}mez}, \& {P{\'e}rez-Villegas}}]{2018RamosMertinez}
{Ramos-Mart{\'\i}nez}, M., {G{\'o}mez}, G.~C., \& {P{\'e}rez-Villegas}, {\'A}.
  2018, \mnras, 476, 3781, \dodoi{10.1093/mnras/sty393}

\bibitem[{{Randriamampandry} {et~al.}(2015){Randriamampandry}, {Combes},
  {Carignan}, \& {Deg}}]{2015Randriamampandry}
{Randriamampandry}, T.~H., {Combes}, F., {Carignan}, C., \& {Deg}, N. 2015,
  \mnras, 454, 3743, \dodoi{10.1093/mnras/stv2147}

\bibitem[{{Regan} {et~al.}(2006){Regan}, {Thornley}, {Vogel}, {Sheth},
  {Draine}, {Hollenbach}, {Meyer}, {Dale}, {Engelbracht}, {Kennicutt}, {Armus},
  {Buckalew}, {Calzetti}, {Gordon}, {Helou}, {Leitherer}, {Malhotra}, {Murphy},
  {Rieke}, {Rieke}, \& {Smith}}]{2006Regan}
{Regan}, M.~W., {Thornley}, M.~D., {Vogel}, S.~N., {et~al.} 2006, \apj, 652,
  1112, \dodoi{10.1086/505382}

\bibitem[{{Rhee} {et~al.}(2004){Rhee}, {Valenzuela}, {Klypin}, {Holtzman}, \&
  {Moorthy}}]{2004Rhee}
{Rhee}, G., {Valenzuela}, O., {Klypin}, A., {Holtzman}, J., \& {Moorthy}, B.
  2004, \apj, 617, 1059, \dodoi{10.1086/425565}

\bibitem[{{Ricarte} {et~al.}(2020){Ricarte}, {Tremmel}, {Natarajan}, \&
  {Quinn}}]{2020Ricarte}
{Ricarte}, A., {Tremmel}, M., {Natarajan}, P., \& {Quinn}, T. 2020, \apjl, 895,
  L8, \dodoi{10.3847/2041-8213/ab9022}

\bibitem[{{Richter} \& {Sancisi}(1994)}]{1994Richter}
{Richter}, O.~G., \& {Sancisi}, R. 1994, \aap, 290, L9

\bibitem[{{Rizzo} {et~al.}(2018){Rizzo}, {Fraternali}, \& {Iorio}}]{2018Rizzo}
{Rizzo}, F., {Fraternali}, F., \& {Iorio}, G. 2018, \mnras, 476, 2137,
  \dodoi{10.1093/mnras/sty347}

\bibitem[{{Robitaille} \& {Bressert}(2012)}]{2012Robitaille}
{Robitaille}, T., \& {Bressert}, E. 2012, {APLpy: Astronomical Plotting Library
  in Python}, Astrophysics Source Code Library, record ascl:1208.017.
\newblock \doeprint{1208.017}

\bibitem[{{Roediger} \& {Hensler}(2008)}]{2008Roediger}
{Roediger}, E., \& {Hensler}, G. 2008, \aap, 483, 121,
  \dodoi{10.1051/0004-6361:200809438}

\bibitem[{{Roman-Oliveira} {et~al.}(2019){Roman-Oliveira}, {Chies-Santos},
  {Rodr{\'\i}guez del Pino}, {Arag{\'o}n-Salamanca}, {Gray}, \&
  {Bamford}}]{2019RomainOliveira}
{Roman-Oliveira}, F.~V., {Chies-Santos}, A.~L., {Rodr{\'\i}guez del Pino}, B.,
  {et~al.} 2019, \mnras, 484, 892, \dodoi{10.1093/mnras/stz007}

\bibitem[{{Rosas-Guevara} {et~al.}(2020){Rosas-Guevara}, {Bonoli}, {Dotti},
  {Zana}, {Nelson}, {Pillepich}, {Ho}, {Izquierdo-Villalba}, {Hernquist}, \&
  {Pakmor}}]{2020RosasGuevara}
{Rosas-Guevara}, Y., {Bonoli}, S., {Dotti}, M., {et~al.} 2020, \mnras, 491,
  2547, \dodoi{10.1093/mnras/stz3180}

\bibitem[{{Sanchez-Garcia} {et~al.}(2023){Sanchez-Garcia}, {Cervantes Sodi},
  {Fritz}, {Moretti}, {Poggianti}, {George}, {Gullieuszik}, {Vulcani},
  {Fasano}, \& {Tawfeek}}]{2023SanchezGarcia}
{Sanchez-Garcia}, O., {Cervantes Sodi}, B., {Fritz}, J., {et~al.} 2023, arXiv
  e-prints, arXiv:2301.06612, \dodoi{10.48550/arXiv.2301.06612}

\bibitem[{{Sancisi} {et~al.}(1979){Sancisi}, {Allen}, \&
  {Sullivan}}]{1979Sancisi}
{Sancisi}, R., {Allen}, R.~J., \& {Sullivan}, W.~T., I. 1979, \aap, 78, 217

\bibitem[{{Schoenmakers}(1999)}]{1999PhDThesis_Schoenmakers}
{Schoenmakers}, R. H.~M. 1999, PhD thesis, University of Groningen, Netherlands

\bibitem[{{Schr{\"o}der} {et~al.}(2001){Schr{\"o}der}, {Drinkwater}, \&
  {Richter}}]{2001Schroder}
{Schr{\"o}der}, A., {Drinkwater}, M.~J., \& {Richter}, O.~G. 2001, \aap, 376,
  98, \dodoi{10.1051/0004-6361:20010997}

\bibitem[{{Scott} {et~al.}(2010){Scott}, {Bravo-Alfaro}, {Brinks}, {Caretta},
  {Cortese}, {Boselli}, {Hardcastle}, {Croston}, \& {Plauchu}}]{2010Scott}
{Scott}, T.~C., {Bravo-Alfaro}, H., {Brinks}, E., {et~al.} 2010, \mnras, 403,
  1175, \dodoi{10.1111/j.1365-2966.2009.16204.x}

\bibitem[{{Sellwood}(2014)}]{2014Sellwood}
{Sellwood}, J.~A. 2014, Reviews of Modern Physics, 86, 1,
  \dodoi{10.1103/RevModPhys.86.1}

\bibitem[{{Sellwood} \& {S{\'a}nchez}(2010)}]{2010Sellwood}
{Sellwood}, J.~A., \& {S{\'a}nchez}, R.~Z. 2010, \mnras, 404, 1733,
  \dodoi{10.1111/j.1365-2966.2010.16430.x}

\bibitem[{{Sellwood} \& {Wilkinson}(1993)}]{1993SellwoodWilkinson}
{Sellwood}, J.~A., \& {Wilkinson}, A. 1993, Reports on Progress in Physics, 56,
  173, \dodoi{10.1088/0034-4885/56/2/001}

\bibitem[{{Shafi} {et~al.}(2015){Shafi}, {Oosterloo}, {Morganti},
  {Colafrancesco}, \& {Booth}}]{2015Shafi}
{Shafi}, N., {Oosterloo}, T.~A., {Morganti}, R., {Colafrancesco}, S., \&
  {Booth}, R. 2015, \mnras, 454, 1404, \dodoi{10.1093/mnras/stv2034}

\bibitem[{{Sheth} {et~al.}(2005){Sheth}, {Vogel}, {Regan}, {Thornley}, \&
  {Teuben}}]{2005Sheth}
{Sheth}, K., {Vogel}, S.~N., {Regan}, M.~W., {Thornley}, M.~D., \& {Teuben},
  P.~J. 2005, \apj, 632, 217, \dodoi{10.1086/432409}

\bibitem[{{Silva-Lima} {et~al.}(2022){Silva-Lima}, {Martins}, {Coelho}, \&
  {Gadotti}}]{2022SilvaLima}
{Silva-Lima}, L.~A., {Martins}, L.~P., {Coelho}, P. R.~T., \& {Gadotti}, D.~A.
  2022, \aap, 661, A105, \dodoi{10.1051/0004-6361/202142432}

\bibitem[{{Solanes} {et~al.}(2001){Solanes}, {Manrique},
  {Garc{\'\i}a-G{\'o}mez}, {Gonz{\'a}lez-Casado}, {Giovanelli}, \&
  {Haynes}}]{2001Solanes}
{Solanes}, J.~M., {Manrique}, A., {Garc{\'\i}a-G{\'o}mez}, C., {et~al.} 2001,
  \apj, 548, 97, \dodoi{10.1086/318672}

\bibitem[{{Sorgho} {et~al.}(2017){Sorgho}, {Hess}, {Carignan}, \&
  {Oosterloo}}]{2017Sorgho}
{Sorgho}, A., {Hess}, K., {Carignan}, C., \& {Oosterloo}, T.~A. 2017, \mnras,
  464, 530, \dodoi{10.1093/mnras/stw2341}

\bibitem[{{Spekkens} \& {Sellwood}(2007)}]{2007Spekkens}
{Spekkens}, K., \& {Sellwood}, J.~A. 2007, \apj, 664, 204,
  \dodoi{10.1086/518471}

\bibitem[{{Stevens} {et~al.}(2019){Stevens}, {Diemer}, {Lagos}, {Nelson},
  {Obreschkow}, {Wang}, \& {Marinacci}}]{2019Stevens}
{Stevens}, A. R.~H., {Diemer}, B., {Lagos}, C. d.~P., {et~al.} 2019, \mnras,
  490, 96, \dodoi{10.1093/mnras/stz2513}

\bibitem[{{Stuber} {et~al.}(2021){Stuber}, {Saito}, {Schinnerer}, {Emsellem},
  {Querejeta}, {Williams}, {Barnes}, {Bigiel}, {Blanc}, {Dale}, {Grasha},
  {Klessen}, {Kruijssen}, {Leroy}, {Meidt}, {Pan}, {Rosolowsky}, {Schruba},
  {Sun}, \& {Usero}}]{2021Stuber}
{Stuber}, S.~K., {Saito}, T., {Schinnerer}, E., {et~al.} 2021, \aap, 653, A172,
  \dodoi{10.1051/0004-6361/202141093}

\bibitem[{{Swaters} {et~al.}(1999){Swaters}, {Schoenmakers}, {Sancisi}, \& {van
  Albada}}]{1999Swaters}
{Swaters}, R.~A., {Schoenmakers}, R.~H.~M., {Sancisi}, R., \& {van Albada},
  T.~S. 1999, \mnras, 304, 330, \dodoi{10.1046/j.1365-8711.1999.02332.x}

\bibitem[{{Tabor} {et~al.}(2017){Tabor}, {Merrifield}, {Arag{\'o}n-Salamanca},
  {Cappellari}, {Bamford}, \& {Johnston}}]{2017Tabor}
{Tabor}, M., {Merrifield}, M., {Arag{\'o}n-Salamanca}, A., {et~al.} 2017,
  \mnras, 466, 2024, \dodoi{10.1093/mnras/stw3183}

\bibitem[{{Tawfeek} {et~al.}(2022){Tawfeek}, {Cervantes Sodi}, {Fritz},
  {Moretti}, {P{\'e}rez-Mill{\'a}n}, {Gullieuszik}, {Poggianti}, {Vulcani}, \&
  {Bettoni}}]{2022Tawfeek}
{Tawfeek}, A.~A., {Cervantes Sodi}, B., {Fritz}, J., {et~al.} 2022, \apj, 940,
  1, \dodoi{10.3847/1538-4357/ac9976}

\bibitem[{{Tonnesen} \& {Bryan}(2009)}]{2009Tonnesen}
{Tonnesen}, S., \& {Bryan}, G.~L. 2009, \apj, 694, 789,
  \dodoi{10.1088/0004-637X/694/2/789}

\bibitem[{{Vauterin} \& {Dejonghe}(1997)}]{1997Vauterin}
{Vauterin}, P., \& {Dejonghe}, H. 1997, \mnras, 286, 812,
  \dodoi{10.1093/mnras/286.4.812}

\bibitem[{{Vazdekis} {et~al.}(2010){Vazdekis}, {S{\'a}nchez-Bl{\'a}zquez},
  {Falc{\'o}n-Barroso}, {Cenarro}, {Beasley}, {Cardiel}, {Gorgas}, \&
  {Peletier}}]{2010Vazdekis}
{Vazdekis}, A., {S{\'a}nchez-Bl{\'a}zquez}, P., {Falc{\'o}n-Barroso}, J.,
  {et~al.} 2010, \mnras, 404, 1639, \dodoi{10.1111/j.1365-2966.2010.16407.x}

\bibitem[{{Verheijen}(2001)}]{2001Verheijen}
{Verheijen}, M. A.~W. 2001, \apj, 563, 694, \dodoi{10.1086/323887}

\bibitem[{{Verheijen} \& {Sancisi}(2001)}]{2001VerheijenSancisi}
{Verheijen}, M.~A.~W., \& {Sancisi}, R. 2001, \aap, 370, 765,
  \dodoi{10.1051/0004-6361:20010090}

\bibitem[{{Vollmer} {et~al.}(2008){Vollmer}, {Braine}, {Pappalardo}, \&
  {Hily-Blant}}]{2008Vollmer}
{Vollmer}, B., {Braine}, J., {Pappalardo}, C., \& {Hily-Blant}, P. 2008, \aap,
  491, 455, \dodoi{10.1051/0004-6361:200810432}

\bibitem[{{Vulcani} {et~al.}(2018){Vulcani}, {Poggianti}, {Gullieuszik},
  {Moretti}, {Tonnesen}, {Jaff{\'e}}, {Fritz}, {Fasano}, \&
  {Bettoni}}]{2018Vulcani_b}
{Vulcani}, B., {Poggianti}, B.~M., {Gullieuszik}, M., {et~al.} 2018, \apjl,
  866, L25, \dodoi{10.3847/2041-8213/aae68b}

\bibitem[{{Wang} {et~al.}(2016){Wang}, {Koribalski}, {Serra}, {van der Hulst},
  {Roychowdhury}, {Kamphuis}, \& {Chengalur}}]{2016Wang}
{Wang}, J., {Koribalski}, B.~S., {Serra}, P., {et~al.} 2016, \mnras, 460, 2143,
  \dodoi{10.1093/mnras/stw1099}

\bibitem[{{Waugh} {et~al.}(2002){Waugh}, {Drinkwater}, {Webster},
  {Staveley-Smith}, {Kilborn}, {Barnes}, {Bhathal}, {de Blok}, {Boyce},
  {Disney}, {Ekers}, {Freeman}, {Gibson}, {Henning}, {Jerjen}, {Knezek},
  {Koribalski}, {Marquarding}, {Minchin}, {Price}, {Putman}, {Ryder}, {Sadler},
  {Stootman}, \& {Zwaan}}]{2002Waugh}
{Waugh}, M., {Drinkwater}, M.~J., {Webster}, R.~L., {et~al.} 2002, \mnras, 337,
  641, \dodoi{10.1046/j.1365-8711.2002.05942.x}

\bibitem[{{Werle} {et~al.}(2022){Werle}, {Poggianti}, {Moretti}, {Bellhouse},
  {Vulcani}, {Gullieuszik}, {Radovich}, {Fritz}, {Ignesti}, {Richard},
  {Soucail}, {Bruzual}, {Charlot}, {Mingozzi}, {Bacchini}, {Tomicic}, {Smith},
  {Kulier}, {Peluso}, \& {Franchetto}}]{2022Werle}
{Werle}, A., {Poggianti}, B., {Moretti}, A., {et~al.} 2022, \apj, 930, 43,
  \dodoi{10.3847/1538-4357/ac5f06}

\bibitem[{{Yim} {et~al.}(2011){Yim}, {Wong}, {Howk}, \& {van der
  Hulst}}]{2011Yim}
{Yim}, K., {Wong}, T., {Howk}, J.~C., \& {van der Hulst}, J.~M. 2011, \aj, 141,
  48, \dodoi{10.1088/0004-6256/141/2/48}

\bibitem[{{Yim} {et~al.}(2014){Yim}, {Wong}, {Xue}, {Rand}, {Rosolowsky}, {van
  der Hulst}, {Benjamin}, \& {Murphy}}]{2014Yim}
{Yim}, K., {Wong}, T., {Xue}, R., {et~al.} 2014, \aj, 148, 127,
  \dodoi{10.1088/0004-6256/148/6/127}

\bibitem[{{Yoon} {et~al.}(2017){Yoon}, {Chung}, {Smith}, \&
  {Jaff{\'e}}}]{2017Yoon}
{Yoon}, H., {Chung}, A., {Smith}, R., \& {Jaff{\'e}}, Y.~L. 2017, \apj, 838,
  81, \dodoi{10.3847/1538-4357/aa6579}

\bibitem[{{Yu} {et~al.}(2022){Yu}, {Kalinova}, {Colombo}, {Bolatto}, {Wong},
  {Levy}, {Villanueva}, {S{\'a}nchez}, {Ho}, {Vogel}, {Teuben}, \&
  {Rubio}}]{2022Yu}
{Yu}, S.-Y., {Kalinova}, V., {Colombo}, D., {et~al.} 2022, \aap, 666, A175,
  \dodoi{10.1051/0004-6361/202244306}

\bibitem[{{Zabel} {et~al.}(2019){Zabel}, {Davis}, {Smith}, {Maddox}, {Bendo},
  {Peletier}, {Iodice}, {Venhola}, {Baes}, {Davies}, {de Looze}, {Gomez},
  {Grossi}, {Kenney}, {Serra}, {van de Voort}, {Vlahakis}, \&
  {Young}}]{2019Zabel}
{Zabel}, N., {Davis}, T.~A., {Smith}, M. W.~L., {et~al.} 2019, \mnras, 483,
  2251, \dodoi{10.1093/mnras/sty3234}

\bibitem[{{Zabel} {et~al.}(2022){Zabel}, {Brown}, {Wilson}, {Davis}, {Cortese},
  {Parker}, {Boselli}, {Catinella}, {Chown}, {Chung}, {Deb}, {Ellison},
  {Jim{\'e}nez-Donaire}, {Lee}, {Roberts}, {Spekkens}, {Stevens}, {Thorp},
  {Tonnesen}, \& {Villanueva}}]{2022Zabel}
{Zabel}, N., {Brown}, T., {Wilson}, C.~D., {et~al.} 2022, arXiv e-prints,
  arXiv:2205.05698.
\newblock \doarXiv{2205.05698}

\end{thebibliography}
\bibliographystyle{aasjournal}



\end{document}